\documentclass[paper]{JHEP}
\usepackage{epsfig}
\usepackage{multirow,booktabs}
\usepackage{amssymb}
\def\beq{\begin{equation}}
\def\beqn{\begin{eqnarray}}
\def\eeq{\end{equation}}
\def\eeqn{\end{eqnarray}}
\def\abs#1{\left|#1\right|}

\def\remove#1#2{#1\hspace{-#2truecm}\backslash}

\def\red{}

\newcommand\sss{}
\newcommand\mydot{\!\cdot\!}
\newcommand\ep{\epsilon}
\newcommand\half{\frac{1}{2}}
\newcommand\quarter{\frac{1}{4}}
\newcommand\qb{\bar{q}}
\newcommand\ub{\bar{u}}
\newcommand\db{\bar{d}}
\newcommand\bb{\bar{b}}
\newcommand\tb{\bar{t}}
\newcommand\sqs{\sqrt{s}}
\newcommand\epem{e^+e^-}
\newcommand\mpmm{\mu^+\mu^-}

\newcommand\isubrmv{\remove{i}{0.125}}

\newcommand\FKSpairs{{\cal P}_{\sss\rm FKS}}
\newcommand\FKSpairsred{\overline{{\cal P}}_{\sss\rm FKS}}
\newcommand\FKSelem{N_{\sss\rm FKS}}
\newcommand\FKSelemred{\overline{N}_{\sss\rm FKS}}
\newcommand\nchannels{N_{\rm ch}}
\newcommand\proc{r}
\newcommand\procB{r_{\sss B}}
\newcommand\procR{r_{\sss R}}
\newcommand\allproc{{\cal R}}
\newcommand\allprocnpo{\allproc_{n+1}}
\newcommand\allprocn{\allproc_{n}}
\newcommand\BornME{{\cal B}}
\newcommand\nini{n_{\sss I}}
\newcommand\nlight{n_{\sss L}}
\newcommand\nlightB{\nlight^{\sss (B)}}
\newcommand\nlightR{\nlight^{\sss (R)}}
\newcommand\nlightBorR{\nlight^{\sss (B/R)}}
\newcommand\nheavy{n_{\sss H}}
\newcommand\nzero{n_\emptyset}
\newcommand\ident{{\cal I}}
\newcommand\numofgr{N_d}
\newcommand\amp{{\cal A}}

\newcommand\ampnt{\amp^{(n,0)}}
\newcommand\ampnpot{\amp^{(n+1,0)}}
\newcommand\ampnl{\amp^{(n,1)}}
\newcommand\ampsq{{\cal M}}

\newcommand\ampsqnt{\ampsq^{(n,0)}}
\newcommand\ampsqnpot{\ampsq^{(n+1,0)}}
\newcommand\ampsqnl{\ampsq^{(n,1)}}
\newcommand\vampsqnl{{\cal V}^{(n,1)}}

\newcommand\vampsqnlF{{\cal V}^{(n,1)}_{\sss FIN}}
\newcommand\hvampsqnlF{\hat{\cal V}^{(n,1)}_{\sss FIN}}
\newcommand\tampsq{\widetilde{\cal M}}
\newcommand\tampsqnt{\tampsq^{(n,0)}}

\newcommand\Ione{\ident_1}
\newcommand\Itwo{\ident_2}
\newcommand\xii{\xi_i}
\newcommand\yij{y_{ij}}
\newcommand\phii{\varphi_i}
\newcommand\yi{y_i}
\newcommand\xic{\left(\frac{1}{\xii}\right)_c}
\newcommand\lxic{\left(\frac{\log\xii}{\xii}\right)_c}
\newcommand\omyijd{\left(\frac{1}{1-\yij}\right)_\delta}
\newcommand\omyid{\left(\frac{1}{1-\yi}\right)_\delta}
\newcommand\opyid{\left(\frac{1}{1+\yi}\right)_\delta}
\newcommand\Dfun{{\cal D}}
\newcommand\Sfun{{\cal S}}
\newcommand\Sfunij{\Sfun_{ij}}
\newcommand\asfun{a_{\Sfun}}
\newcommand\bsfun{b_{\Sfun}}
\newcommand\stepf{\Theta}
\newcommand\phsp{d\phi}
\newcommand\phspn{\phsp_{n}}
\newcommand\phspnpo{\phsp_{n+1}}
\newcommand\tphsp{d\widetilde{\phi}}
\newcommand\tphspn{\tphsp_{n}}
\newcommand\tphspnij{\tphsp_{n}^{ij}}
\newcommand\as{\alpha_{\sss S}}
\newcommand\asotwopi{\frac{\as}{2\pi}}
\newcommand\gs{g_{\sss S}}
\newcommand\xicut{\xi_{cut}}
\newcommand\ximax{\xi_{\rm max}}
\newcommand\deltaO{\delta_{\sss O}}
\newcommand\deltaI{\delta_{\sss I}}
\newcommand\NC{N_{\sss c}}
\newcommand\CA{C_{\sss A}}
\newcommand\CF{C_{\sss F}}
\newcommand\TF{T_{\sss F}}
\newcommand\DA{D_{\sss A}}
\newcommand\eikint{{\cal E}}
\newcommand\eikintD{\hat{\cal E}}
\newcommand\APdamp{\overline{P}}
\newcommand\Qdamp{\overline{Q}}
\newcommand\Qop{\vec{Q}}
\newcommand\JetsB{J^{\nlightB}}
\newcommand\JetsR{J^{\nlightB+1}}

\newcommand\velkl{v_{kl}}
\newcommand\alkl{\alpha_{kl}}
\newcommand\avg{{\cal N}}
\newcommand\symm{\varsigma}

\newcommand\symmnpoij{\symm_{ij}^{(n+1)}}
\newcommand\veck{\vec{k}}
\newcommand\kbar{\bar{k}}
\newcommand\kkdotkl{k_k\mydot k_l}
\newcommand\polv{\varepsilon}
\newcommand\MadFKS{{\tt MadFKS}}
\newcommand\polP{{\cal T}}
\newcommand\polQ{{\cal W}}
\newcommand\muF{\mu_{\sss F}}
\newcommand\muR{\mu_{\sss R}}

\newcommand\clS{{\mathbb S}}

\preprint{
 CERN-TH/2009-157\\
 CP3-09-39
 }
\title{Automation of next-to-leading order computations in QCD:
the FKS subtraction}
\author{Rikkert Frederix%
 \thanks{On leave of absence from CP3, Universit\'{e} catholique de Louvain.}\\
  PH Department, TH Unit, CERN, CH-1211 Geneva 23, Switzerland\\
  E-mail: \email{Rikkert.Frederix@cern.ch}}
\author{Stefano Frixione%
  \thanks{On leave of absence from INFN, Sez. di Genova, Italy.}\\
  PH Department, TH Unit, CERN, CH-1211 Geneva 23, Switzerland\\
  ITPP, EPFL, CH-1015 Lausanne, Switzerland\\
  E-mail: \email{Stefano.Frixione@cern.ch}}
\author{Fabio Maltoni\\
  Centre for Particle Physics and Phenomenology (CP3)\\
  Universit\'{e} catholique de Louvain\\
  Chemin du Cyclotron 2, B-1348 Louvain-la-Neuve, Belgium\\
  E-mail: \email{fabio.maltoni@uclouvain.be}}
\author{Tim Stelzer\\
  Department of Physics, University of Illinois at Urbana-Champaign\\
  1110 West Green Street, Urbana, IL\ \ 61801, USA\\
  E-mail: \email{tstelzer@uiuc.edu}}
\abstract{We present the complete automation of the universal 
subtraction formalism proposed by Frixione, Kunszt, and Signer 
for the computation of any cross section at the next-to-leading
order in QCD. Given a process, the only ingredient to be 
provided externally is the infrared- and ultraviolet-finite contribution 
of virtual origin. Our implementation, currently restricted to the
case of $\epem$ collisions, is built upon and works in the 
same way as MadGraph. It is particularly 
suited to parallel computation, and it can deal with any physical 
process resulting from a theory implemented in MadGraph, 
thus including the Standard Model as well as Beyond the Standard Model
theories. We give results for some sample processes that document
the performances of the implementation, and show in particular 
how the number of subtraction terms has an extremely mild growth
with final-state multiplicity.
}
\enlargethispage*{1000pt}

\renewcommand\arraystretch{1.1}

\begin{document}

\section{Introduction\label{sec:intro}}
The vast majority of new-physics searches at the LHC, and some 
of those at the Tevatron, are based on final states signatures
that feature the presence of a large number of jets (typically, 
from four to ten), generically denoted as multi-jet configurations.
The fairly difficult problem of giving theoretical predictions
for multi-jet processes has been completely solved in the past
few years, but the solutions are only accurate to the leading
order (LO) in QCD. All approaches are thus based on the computations of
tree-level, multi-parton matrix elements, performed by dedicated 
computer programmes (see e.g.~ref.~\cite{Dobbs:2004qw} for a review 
and a list of references); the LO accuracy implies that the computations
can be carried out in four dimensions, and that all divergences resulting
from integration over the phase space are avoided by means of 
kinematic cuts. Through the computation of the matrix elements, one
obtains final states composed of quarks, gluons, and other accompanying
particles (e.g.~$W$'s or $Z$'s). At this point, one may choose to
identify each light quark and gluon with a jet, thus making use of
the local hadron-parton duality, and compare the theoretical results 
to data; we shall call these results {\em matrix-element predictions}.
A more realistic approach, which is widely used by experimental
collaborations, is that of feeding parton-level final states obtained
from matrix element computations to event generators, which by showering
them eventually result in hadron-level final states, where any jet may contain
several tens of particles; we shall call these {\em showered predictions}.
As is known, showered predictions are only meaningful in the context
of a proper matching formalism, which avoids the double counting of 
configurations that can be obtained both from the matrix elements and 
from the showers. Several solutions are available for the problem 
of tree-level matching~\cite{Catani:2001cc,Krauss:2002up,Lonnblad:2001iq,
Alwall:2007fs}, which involve the simultaneous treatment of final states 
with different multiplicities.

While fairly successful phenomenologically, the approaches discussed
above suffer from the usual limitations of LO computations. 
Scale uncertainties, which are typically large for processes with 
high-multiplicity final states, render it unreliable the predictions of
the absolute values of the cross sections. A common strategy is 
therefore that of fixing the normalization equal to the data for a 
given jet multiplicity, and of predicting the cross sections for larger
and smaller multiplicities. It is therefore desirable to extend
the accuracy of multi-jet computations to the next-to-leading order (NLO)
in QCD. It is in fact worth recalling that NLO results, so far available 
only for small-multiplicity jet final states, have been a very important 
ingredient in establishing QCD and the SM as the correct theories of
strong and electroweak interactions at LEP, SLD, and the Tevatron,
as well as to exclude the presence of Beyond the SM signals in the data so far.
We also recall that NLO calculations, apart from resulting in smaller 
theoretical uncertainties that allow one to trust absolute predictions,
induce non-trivial effects in particle spectra that cannot be obtained
in general by a simple rescaling of LO results.

The problem of the computation of multi-jet cross sections at the NLO
has attracted considerable attention in the recent past. There are two
main obstacles that must be cleared. Firstly, one needs to compute the
one-loop corrections. There are now several different solutions to this 
problem, that are reasonably automated, and are based on unitarity
methods: BlackHat, CutTools, and
Rocket (see refs.~\cite{Berger:2009ep,vanHameren:2009dr,Ellis:2009zw}
for recent results, and for a more complete list of references).
A more traditional approach, based on Feynman diagrammatics,
is that of GOLEM~\cite{Binoth:2008uq}. It is reasonable to expect that most 
of these codes will attain larger flexibility and speed in the near future. 
Secondly, one must compute the real-emission contributions, and combine them 
with the one-loop ones in order to obtain the physical cross sections.

The aim of this paper is that of addressing the second of the problems
mentioned above. We assume that ultraviolet-renormalized one-loop corrections 
are given, and we fully automate the following steps: 
\begin{itemize}
\item[{\em a)}] computations of real-emission matrix elements and of their
local counterterms; 
\item[{\em b)}] their combinations with the one-loop matrix elements and, 
if relevant, with initial-state collinear counterterms; and the subsequent 
definition of finite short-distance cross sections;
\item[{\em c)}] integration over the phase space of the cross sections
obtained in {\em b)}; 
\item[{\em d)}] output of the results on an event-by-event basis, in the 
form of a set of four-momenta and a weight, that one can use to 
construct (infrared-safe) observables.
\end{itemize}
We stress that item {\em b)} involves the cancellation of infrared
poles as prescribed by the KLN theorem. Even if the explicit form of 
the one-loop contributions is not available, this task can be carried 
out analytically, thanks to the fact that the general
structure of the poles is known for any process; the residues
are proportional to tree-level matrix elements that can be 
computed automatically.

There are two main ingredients in our work: a programme that computes
tree-level matrix elements, and a universal formalism for the analytical
cancellation of infrared singularities, that allows one to implement in a 
computer codes short-distance cross sections free of any singularities.
For the former, we use MadGraph~\cite{Stelzer:1994ta,Alwall:2007st}.
For the latter, we adopt the formalism of Frixione, Kunszt, and Signer,
presented originally in refs.~\cite{Frixione:1995ms,Frixione:1997np}, which
we shall refer to as FKS in the rest of this paper. Through the implementation
of items {\em a)}--{\em d)} above, we basically achieve an NLO version
of MadGraph/MadEvent, up to a missing (infrared- and ultraviolet-finite) 
piece of one-loop origin, in which the only operation required from a user 
is that of typing in the process to be computed. We call this programme
\MadFKS. The current implementation of \MadFKS, and the results
given in this paper, are relevant only to the case of $\epem$ collisions, 
which is sufficient to highlight the capabilities of the code.
On the other hand, the formalism is written in full generality, and
can therefore be immediately applied to any other kind of collisions.
We defer to a forthcoming paper the complete implementation of the cases 
where initial-state hadrons are present. We stress, however, that what is
done here will apply without any modifications (except for the trivial
multiplication by the relevant parton density functions) to those cases
as well. Thus, the only new piece of implementation required by initial-state
hadrons will concern the emission of a parton from an initial-state parton.
We finally point out that the results of \MadFKS\ are matrix-element
predictions, but also give the necessary building blocks to arrive
at NLO showered predictions in the context of matching techniques
such as MC@NLO~\cite{Frixione:2002ik} or 
POWHEG~\cite{Nason:2004rx,Frixione:2007vw}.

This paper is organized as follows. In sect.~\ref{sec:NLOxsec} we give
a general overview of the issues in NLO computations of multi-jet
cross sections, and of the defining features of the FKS subtraction
method. In sect.~\ref{sec:nots} our notations are introduced. In
sect.~\ref{sec:FKSxsecs} we summarize the most relevant formulae 
of the FKS subtraction for any type of collisions, 
including hadron-hadron ones. In sect.~\ref{sec:impl} we explain how
these formulae can be implemented in a computer code.
A few refinements which we use in \MadFKS\ to optimize
the performances of the code are discussed in sect.~\ref{sec:opt}.
In sect.~\ref{sec:res} we present selected results of \MadFKS, 
for the case of $\epem$ collisions. Section~\ref{sec:concl}
reports our conclusions. Further technical information such as
longer formulae, and possible variants of the implementation,
are given in the appendices.

\section{Computations of multi-jet cross sections 
at the NLO\label{sec:NLOxsec}}
Given a production process, the computation of its NLO corrections
in QCD implies the evaluation of the corresponding one-loop and
real-emission corrections (for a pedagogical introduction to these
concepts, see e.g.~sect.~4 of ref.~\cite{Dobbs:2004qw}), which are
eventually added up to get the physical cross section.
Although from the principle point of view there is no difference 
between the computation of a small-multiplicity and that of a
large-multiplicity jet cross section, in practice the situation is 
entirely different. One-loop computations based on straightforward 
Feynman diagram evaluation rapidly become too involved to be carried 
out\footnote{It is therefore quite remarkable that the authors of
ref.~\cite{Bredenstein:2009aj} have computed the NLO corrections
to the hadroproduction of $t\tb b\bb$ by means of standard techniques.}, 
and new techniques have been developed recently to bypass this problem 
(see e.g.~refs.~\cite{Ossola:2006us,Bern:2007dw,Giele:2008ve,Berger:2008sj}).
On the other hand, the computation of real-emission and Born (i.e.,
tree-level) diagrams poses no problem, and is limited only by CPU.

There are two issues when summing one-loop and real-emission contributions.
The first is that of proving (analytically) the cancellation of the
infrared singularities that arise in the intermediate steps of the
computation, and to write the leftovers as finite contributions that
can be implemented in computer codes. The second is the actual implementation
of these finite terms, which crucially includes an integration over
the phase space. The former issue has been fully 
solved~\cite{Frixione:1995ms,Catani:1996vz,Giele:1991vf,Giele:1993dj} 
in the 90's,
for any jet multiplicity (in other words, the jet multiplicity is simply
a parameter in the resulting formulae), in the context of the so-called
universal cancellation formalisms, which use either the (approximate) 
slicing method, or the (exact) subtraction method; it is nowadays 
acknowledged that the slicing method is unsuited for describing 
complicated final states, such as those in multi-jet production.

The latter of the two issues mentioned above
has actually not even been considered, for multi-jet
cross sections, until recently. There are now several 
proposals~\cite{Seymour:2008mu,Frederix:2008hu,Hasegawa:2008ae}
that aim at constructing local counterterms for any given real-emission 
matrix elements (item {\em a)} above). What is outlined in items 
{\em a)}--{\em d)} is achieved in ref.~\cite{Gleisberg:2007md}
(restricted to massless SM particles), and very recently in
ref.~\cite{Czakon:2009ss} (which includes the treatment
of massive SM particles). All of these approaches have tackled the problem 
using the universal subtraction method originally introduced in 
ref.~\cite{Catani:1996vz} (referred to as dipole subtraction henceforth).

As was discussed in the introduction, in this paper we use
the FKS subtraction formalism. Although dipole subtraction is the 
most widely used method for NLO matrix-element computations,
FKS is the only technique used so far in practice for performing NLO 
showered computations, in the MC@NLO and POWHEG frameworks. However, 
as we shall show in this paper, FKS subtraction has also excellent performances
in the context of matrix-element computations. While dipole and FKS
subtraction methods are formally equivalent as shown in 
ref.~\cite{Frixione:2004is},
differences between the two arise because of the way in which
the building blocks for the local counterterms of the real-emission
matrix elements are combined (these building blocks corresponds to the
soft, collinear, and soft-collinear singularities). In the dipole method,
one emphasizes the role of soft emissions. A structure arises where one
sums over an ``emitted'' parton, and two colour partners that exchange
the emitted parton; the sum over dipoles is thus a sum over three indices.
In the FKS method, the emphasis is on collinear emissions, and the structure
that emerges is therefore one where there is a sum over parton pairs,
i.e.~a sum over two indices. The consequence of this is that, in general, 
the number of independent subtraction terms is smaller in the FKS formalism
than in the dipole formalism.

One of the key features of FKS subtraction is that, in the 
integration of the real-emission matrix elements, one effectively
defines partonic processes with at most one soft and one collinear 
singularities, which results in a much simpler subtraction structure 
than that of the original matrix elements.
These processes are furthermore fully independent from each
other, and can therefore be computed separately. This is a genuine
parallelization, more efficient than that of computing the same production
process several times with different seeds for random number generation,
especially in view of the use of adaptive-integration routines
(the latter type of parallelization can obviously still be set up
in the context of FKS subtraction, if need be).

The original FKS papers~\cite{Frixione:1995ms,Frixione:1997np}
addressed the case of processes with final states composed only of
massless quarks and gluons. The extension to generic processes,
with the presence of both strongly-interacting massive particles and of
non strongly-interacting particles, is almost trivial. Several of these
cases have in fact been implemented in MC@NLO using FKS subtraction
(see e.g.~ref.~\cite{Frixione:2005vw}). Here, we shall present a complete
summary of the relevant formulae, which can also be applied to computing 
cross sections to NLO accuracy for the production of non-SM particles.

In summary, we advocate the use of FKS subtraction for
matrix-element predictions for a variety of reasons: structure 
identical to that of collinear emissions, thus lending itself naturally
to matching with parton shower Monte Carlos not based on colour dipoles;
ease of importance sampling; small number of subtraction terms, 
and the modest growth of their number with the multiplicity, with
beneficial effects on numerical stability; organization of the calculation 
in a way which is parallel in nature; and the availability of all
results necessary for the treatment of fully-polarized processes,
which allows one to performing sums over helicity states with
Monte Carlo methods. 

In the rest of this paper, all formulae for 
NLO cross sections will make use of FKS subtraction.

\section{Notation\label{sec:nots}}
\subsection{Partonic processes\label{sec:nots:proc}}
For any given partonic process, we shall denote by $n$ the number 
of final-state particles at the Born level. The contributions to the 
NLO partonic cross sections will therefore either have a $2\to n$
or a $2\to (n+1)$ kinematics, which we shall call Born (or $n$-body) or 
real-emission (or $(n+1)$-body) kinematics respectively\footnote{The decay
of a particle can also be described by using eqs.~(\ref{Bkin}) 
and~(\ref{Rkin}).}. They will be denoted as follows:
\beqn
&&k_1+k_2\;\longrightarrow\; k_3+\cdots +k_{n+2}
\phantom{aaaa}{\rm Born~kinematics}\,;
\label{Bkin}
\\
&&k_1+k_2\;\longrightarrow\; k_3+\cdots +k_{n+3}
\phantom{aaaa}{\rm real~emission~kinematics}\,.
\label{Rkin}
\eeqn
We shall write the corresponding phase spaces as
\beq
\phspn\,,\;\;\;\;\;\phspnpo
\eeq
respectively.
We shall denote by $\nzero$ the number of final-state particles which
are not strongly interacting (e.g.~leptons); $\nheavy$ will denote 
the number of massive, strongly-interacting particles (e.g.~heavy quarks).
Depending on the process, some or all of the $\nzero+\nheavy$ particles
will not belong to the Standard Model. $n$- and $(n+1)$-body processes
will be characterized by the same values of $\nzero$ and of $\nheavy$;
the identities of the corresponding particles will also be the same
in the two classes of processes. On the other hand, the number of
light quarks and gluons in the $(n+1)$-body processes (denoted
by $\nlightR$) will be equal to that in $n$-body processes (denoted
by $\nlightB$), plus one. We have therefore:
\beqn
\nlightR&=&\nlightB+1\,,
\\
n&=&\nlightB+\nheavy+\nzero\,,
\\
n+1&=&\nlightR+\nheavy+\nzero\,.
\eeqn
If $1\le k\le n+2$ or $1\le k\le n+3$ is an index which runs over all
particles, we shall adopt the following labeling scheme:
\beqn
1\le&k&\le 2\;\;\;\Longrightarrow\;\;\; 
\nonumber\\*&&\phantom{aa}
{\small\rm initial~state}\,;
\\
3\le &k&\le\nlightBorR+2\;\;\;\Longrightarrow\;\;\;
\nonumber\\*&&\phantom{aa}
{\small\rm massless~quarks~and~gluons}\,;
\\
\nlightBorR+3\le &k&\le\nlightBorR+\nheavy+2\;\;\;\Longrightarrow 
\nonumber\\*&&\phantom{aa}
\textrm{strongly-interacting~massive~particles}\,;
\\
\nlightBorR+\nheavy+3\le &k&\le \nlightBorR+\nheavy+\nzero+2
\;\;\;\Longrightarrow 
\nonumber\\*&&\phantom{aa}
\textrm{non~strongly-interacting~particles}\,.
\eeqn
In order to deal with the various types of colliding particles, it will
also be useful to introduce the quantity $\nini$, which can take the
following values
\beqn
&&\nini=1\;\;\;\Longleftrightarrow\;\;\;HH~~{\rm collisions}\,,
\\
&&\nini=2\;\;\;\Longleftrightarrow\;\;\;eH~~{\rm collisions}\,,
\\
&&\nini=3\;\;\;\Longleftrightarrow\;\;\;ee~~{\rm collisions}\,,
\eeqn
where $e$ and $H$ generically denote a non-hadronic and a hadronic
particle respectively; $\nini$ is therefore the smallest value that
a label can assume when running over light quarks and gluons
(without loss of generality, we have conventionally given label 1 to the
non-hadronic particle in an $eH$ collision).

The identity of particle $k$ will be denoted by $\ident_k$; it is
also convenient to denote the anti-particle of $\ident_k$ 
by $\ident_{\bar{k}}$, although $\overline{\ident}_k$ can be used as well.
A process will be unambiguously identified by giving the list of the 
identities of the particles involved, and we shall denote it by
\beq
\proc=\left(\ident_1,\ldots\,\ident_{n+2}\right)
\phantom{aaa}{\rm or}\phantom{aaa}
\proc=\left(\ident_1,\ldots\,\ident_{n+3}\right)\,.
\label{procdef}
\eeq
The sets of all $n$- and $(n+1)$-body processes will be denoted
by $\allprocn$ and by $\allprocnpo$ respectively\footnote{If two
processes in the form of eq.~(\ref{procdef}) are related by a permutation 
over the identities of final-state particles, only one of them will be
included in $\allprocn$ or $\allprocnpo$.}. In the following,
we shall make use of the fact that a given $n$-body process can
be related to at least one $(n+1)$-body process. We therefore
introduce the following notations, where $i$ is a final-state 
light quark or a gluon \mbox{($3\le i\le\nlightR+2$)},
and $j$ is a strongly-interacting particle
\mbox{($\nini\le j\le\nlightR+\nheavy+2$)}:
\beqn
&&\proc=\left(\ident_1,\ldots\ident_i,\ldots\ident_j,
\ldots\ident_{n+3}\right)\in\allprocnpo
\\*&&\phantom{aaaaaaaaaaa}\Longrightarrow\phantom{aaa}
\proc^{\isubrmv},\proc^{j\oplus i,\isubrmv}\in\allprocn
~~~~~~{\rm if}~~~\proc^{\isubrmv},\proc^{j\oplus i,\isubrmv}~~{\rm exist}\,.
\eeqn
We have defined
\beqn
\proc^{\isubrmv}&=&
\left(\ident_1,\ldots\remove{\ident}{0.2}_i,\ldots\ident_j,
\ldots\ident_{n+3}\right)\,,
\label{sfproc}
\\
\proc^{j\oplus i,\isubrmv}&=&
\left(\ident_1,\ldots\remove{\ident}{0.2}_i,
\ldots\ident_{j\oplus i},\ldots\ident_{n+3}\right)\,.
\label{clproc}
\eeqn
Thus, process $\proc^{\isubrmv}$ in eq.~(\ref{sfproc}) is constructed
by simply removing parton $i$ from the original list. It is clear that such 
an operation results in a physical process (i.e., which exists) only when 
$i$ is a gluon. Process $\proc^{j\oplus i,\isubrmv}$ in eq.~(\ref{clproc}) 
is constructed by removing parton $i$ from the original list, and by 
replacing particle $j$ with one whose identity is that of the parton 
entering the $(\ident_{j\oplus i},\ident_j,\ident_i)$ QCD vertex (thus, 
if $\ident_i=g$ and $\ident_j=g$, then $\ident_{j\oplus i}=g$;
if $\ident_i=g$ and $\ident_j=q$, then $\ident_{j\oplus i}=q$,
and so forth). If this QCD
vertex does not exist (as e.g.~in the case $\ident_i=u$ and
$\ident_j=\db$), process $\proc^{j\oplus i,\isubrmv}$ is non physical,
i.e.~it does not exist\footnote{If $j=1$ or $j=2$, one needs to
use $\ident_{\bar i}$ instead of $\ident_i$ when constructing
$\ident_{j\oplus i}$; in the following, we may therefore use
the notation $\ident_{1\oplus\bar{i}}$ or $\ident_{2\oplus\bar{i}}$.}.
The definitions of $\proc^{\isubrmv}$ and of 
$\proc^{j\oplus i,\isubrmv}$ are obviously motivated by the relevant
soft and collinear limits respectively, but will be used in several 
different contexts\footnote{Note that eq.~(\ref{sfproc}) is a particular
case of eq.~(\ref{clproc}), since the two coincide when $\ident_i=g$.
However, it will turn out to be convenient to have the two different 
notations.}.

\subsection{Matrix elements\label{sec:nots:ME}}
NLO cross sections receive contributions from both tree-level
and one-loop amplitudes, which we shall denote as follows:
\beqn
\ampnpot(\proc)\;\; &\longrightarrow& \;\;
\textrm{real-emission~tree~amplitude}~~~
(\proc\in\allprocnpo)\,,
\\
\ampnt(\proc)\;\; &\longrightarrow& \;\;
{\small\rm Born~tree~amplitude}~~~
(\proc\in\allprocn)\,,
\\
\ampnl(\proc)\;\; &\longrightarrow& \;\;
\textrm{one-loop~amplitude}~~~
(\proc\in\allprocn)\,.
\eeqn
The amplitudes above are typically relevant to a given spin and
colour configuration; they include all the coupling constant factors.
Starting from amplitudes, one needs to construct several quantities 
which will enter short-distance cross sections. We denote them as follows:
\beqn
\ampsqnpot(\proc)&=&\frac{1}{2s}\frac{1}{\omega(\Ione)\omega(\Itwo)}
\mathop{\sum_{\rm colour}}_{\rm spin}\abs{\ampnpot(\proc)}^2,
\label{Mtreenpo}
\\
\ampsqnt(\proc)&=&\frac{1}{2s}\frac{1}{\omega(\Ione)\omega(\Itwo)}
\mathop{\sum_{\rm colour}}_{\rm spin}\abs{\ampnt(\proc)}^2,
\label{Mtreen}
\\
\ampsqnt_{kl}(\proc)&=&-\frac{1}{2s}
\frac{2-\delta_{kl}}{\omega(\Ione)\omega(\Itwo)}
\mathop{\sum_{\rm colour}}_{\rm spin}
\ampnt(\proc) \Qop(\ident_k)\mydot\Qop(\ident_l)
{\ampnt(\proc)}^{\star},\phantom{aa}
\label{Mlinked}
\\
\ampsqnl(\proc)&=&\frac{1}{2s}\frac{1}{\omega(\Ione)\omega(\Itwo)}
\mathop{\sum_{\rm colour}}_{\rm spin}
2\Re\left\{\ampnt(\proc){\ampnl(\proc)}^{\star}\right\}.
\label{Moneloop}
\eeqn
In these equations, $s=(k_1+k_2)^2=2k_1\cdot k_2$, 
and $\omega(\ident)$ is the product of spin and colour degrees of 
freedom for particle $\ident$. In $4-2\ep$ dimensions, we have
$\omega(q)=2\NC\equiv 6$ and $\omega(g)=2(1-\ep)\DA\equiv 16(1-\ep)$.
These average factors serve the sole
purpose of fully specifying the divergent part of the one-loop
contribution in the CDR scheme, which is not used in numerical
calculations (see app.~\ref{app:virt} for more details).
We stress that the formulae which give the physical cross sections
in FKS, and that are implemented in computer codes, are finite
in four dimensions. 
Note that symmetry factors accounting for the presence of identical
particles in the final state are not included in the matrix elements
defined above; they will be inserted later in the expressions for 
the short-distance cross sections.
As the notation suggests, $\proc\in\allprocnpo$ in eq.~(\ref{Mtreenpo}),
while $\proc\in\allprocn$ in eqs.~(\ref{Mtreen})--(\ref{Moneloop}).
The quantities defined in eqs.~(\ref{Mtreenpo})--(\ref{Moneloop})
are sufficient to perform all of our computations, with the exception
of an azimuthal contribution to the collinear limit of $\ampsqnpot$,
which requires a reduced matrix element that we shall denote by $\tampsqnt$.
The precise form of this quantity is irrelevant in this section; it will
be used in sect.~\ref{sec:impl:RME}, and its definition will be given
in app.~\ref{app:Qfun}.

The colour operators $\Qop(\ident)$ that enter the definition of
the so-called colour-linked Born's, eq.~(\ref{Mlinked}), give a
representation of the colour algebra, and can be defined as follows:
\beq
\Qop(\ident)=\left\{t^a\right\}_{a=1}^8\,,\;
\left\{-t^{a{\rm T}}\right\}_{a=1}^8\,,\;
\left\{T^a\right\}_{a=1}^8\,,\phantom{aaa}
\ident\in {\bf 3},{\bf \bar{3}},{\bf 8},
\label{Qopdef}
\eeq
with $t^a$ and $T^a$ the SU(3) generators in the fundamental
and adjoint representations respectively. Clearly, eq.~(\ref{Qopdef})
can be extended to higher-dimensional representations if need be.
It is trivial to show that
\beqn
&&\Qop(\ident_1)\cdot \Qop(\ident_2)=
\Qop(\ident_2)\cdot \Qop(\ident_1)\,,
\label{Qid1}
\\
&&\Qop(\ident)\cdot \Qop(\ident)\equiv Q^2(\ident) = C(\ident)I,
\label{Qid2}
\eeqn
where
\beqn
&&C(\ident)=\CF=\frac{\NC^2-1}{2\NC}\phantom{aaa}{\rm for}~~~
\ident\in {\bf 3},{\bf \bar{3}}\,,
\label{Cq}
\\
&&C(\ident)=\CA=\NC\phantom{aaaaaaa}{\rm for}~~~
\ident\in {\bf 8}\,.
\label{Cg}
\eeqn
Using the colour-singlet condition 
\beq
\sum_{k=\nini}^{\nlightB+\nheavy+2}\Qop(\ident_k)=\vec{0}
\eeq
and eqs.~(\ref{Qid1})--(\ref{Cg}), it is easy to prove that
\beqn
\ampsqnt_{kl} &=& \ampsqnt_{lk}\,,
\label{Mklident0}
\\
\mathop{\sum_{k\ne l}}_{k=\nini}^{\nlightB+\nheavy+2}
\ampsqnt_{kl} &=& 2C(\ident_l)\ampsqnt\,,
\label{Mklident1}
\\
\ampsqnt_{kk} &=& - C(\ident_k)\ampsqnt\,,
\label{Mklident2}
\eeqn
with
\beq
\nini\le k,l\le \nlightB+\nheavy+2\,.
\eeq
Note that $\ampsqnt_{kl}$ is an interference term related to the
exchange of a soft gluon between particles $k$ and $l$, which appears
in the soft limit of the real matrix elements. This term has a coefficient
proportional to an eikonal factor, which therefore vanishes identically
when $k=l$ and $k$ is massless (see app.~\ref{app:eik}). 

The final formulae of the FKS subtraction method, to be given below,
will only deal with non-divergent quantities. The interested reader
can find in the original paper (ref.~\cite{Frixione:1995ms}) the proof
of the cancellation of the infrared and collinear singularities that
arise in the intermediate steps of the computation. As in all computations
at the NLO, the finiteness of the partonic short-distance cross sections 
is in general a consequence
of imposing kinematic cuts on final-state particles. Without loss of
generality, we can always assume these cuts to be equivalent to the
request of having either $\nlightB$ or $\nlightB+1$ jets in the final
state, the jets being reconstructed with an arbitrary algorithm. This
condition will be symbolically denoted by
\beq
\JetsB\,.
\label{jetcuts}
\eeq The condition in $\JetsB$ is sufficient to prevent the appearance
of phase-space singularities\footnote{It is understood that the condition 
in $\JetsB$ also prevents all divergences due to final-state leptons and 
photons. This is necessary if the subtraction of the corresponding
QED singularities is not carried out.} 
in $n$-body quantities such as $\ampsqnt$, $\ampsqnt_{kl}$, 
and $\ampsqnl$. On the other hand, the $(n+1)$-body tree-level
matrix elements will still diverge in some regions of the phase space;
these divergences are subtracted (in all subtraction methods) by means
of suitable counterterms.  In order to classify these divergences and
to eventually subtract them, for any given process $\proc\in\allprocnpo$ we
introduce the following set of ordered pairs (called the set of FKS pairs) 
\beqn
\FKSpairs(\proc)&=&\Big\{(i,j)\;\Big|\;3\le i\le\nlightR+\nheavy+2\,, 
\nini\le j\le\nlightR+\nheavy+2\,, 
i\ne j\,,
\nonumber\\*&&\phantom{aaa}
\ampsqnpot(\proc)\JetsB\to\infty~~{\rm if}~~k_i^0\to 0~~
{\rm or}~~k_j^0\to 0~~{\rm or}~~\veck_i\parallel \veck_j\Big\}.
\phantom{aaaaa}
\label{PFKSdef}
\eeqn
In words, a pair of particles belongs to the set of FKS pairs if
they induce soft or collinear singularities (or both) in the $(n+1)$-body
matrix elements, even in the presence of jet cuts. The first element of 
the pair will be called the FKS parton, and the second element will be 
its sister\footnote{Although in the definition of $\FKSpairs$ the role
of the FKS partons and of their sisters is symmetric if they both belong
to the final state, this will not be the case when the subtraction of
singularities will be performed: hence, the convenience of distinguishing
them.}. Conversely, by definition $\FKSpairs$ takes into account
{\em all} phase-space singularities of the $(n+1)$-body tree-level 
matrix elements after jet cuts. We note that the condition $k_j\to 0$
is simply not relevant if $j=1,2$, since in such a case the momentum
$k_j$ is fixed.
In the computation of an NLO cross section according to the FKS
formalism, each pair belonging to $\FKSpairs$ will correspond
to a set of subtractions of soft and collinear singularities
which, when combined with the real-emission matrix element,
will result into a finite contribution to physical observables.
Each of these contributions is separately finite, and can therefore
be computed independently from the others, and integrated with 
numerical methods.

It is immediately obvious that some of the pairs \mbox{$(i,j)$} 
with $i,j\le\nlightR+\nheavy+2$ will not
belong to $\FKSpairs$ for any physical process. This is the case 
e.g.~when $i$ and $j$ are both massive and strongly interacting; 
or when $i$ and $j$ are a quark and an antiquark of different flavours;
or when one of the members of the pair is massive and strongly 
interacting, and the other member is not a gluon. In general, however,
for large multiplicities the number of elements in $\FKSpairs$
will scale approximately as \mbox{$\nlightR(\nlightR+\nheavy)$}.
On the other hand, depending on the identities of the particles
in $\proc$, some of the contributions due to the various FKS pairs
will actually be identical, because of the symmetry properties of 
matrix elements and phase spaces. This implies a drastic
reduction of the number of independent terms actually needed in 
the computation. We shall return to this issue in sect.~\ref{sec:opt:red}.

\newpage
\section{Cross sections\label{sec:FKSxsecs}}
We write an NLO cross section in the collision of two
particles $P_1$ and $P_2$ with momenta $K_1$ and $K_2$ using
the factorization theorem
\beqn
&&d\sigma_{\sss P_1P_2}(K_1,K_2)=
\nonumber \\*&&\phantom{aa}
\sum_{\proc_{\sss R}\in\allprocnpo}\int dx_1 dx_2
f_{\Ione}^{(P_1)}(x_1) f_{\Itwo}^{(P_2)}(x_2)
\nonumber \\*&&\phantom{aaaaaa}\times
\Big(d\sigma^{(n+1)}(\proc_{\sss R};x_1K_1,x_2K_2)+
d\bar{\sigma}^{(n+1)}(\proc_{\sss R};x_1K_1,x_2K_2)\Big) 
\nonumber\\*&&\phantom{a}+
\sum_{\proc_{\sss B}\in\allprocn}\int dx_1 dx_2
f_{\Ione}^{(P_1)}(x_1) f_{\Itwo}^{(P_2)}(x_2)
d\sigma^{(n)}(\proc_{\sss B};x_1K_1,x_2K_2)\,.
\label{factTH}
\eeqn
As the notation suggests, we have (see eqs.~(\ref{Bkin}) and~(\ref{Rkin}))
\beq
k_1=x_1K_1\,,\;\;\;\;\;\;k_2=x_2K_2\,.
\eeq
If $P_\alpha$ is a hadron, $f_{\ident_\alpha}^{(P_\alpha)}$ will be a
parton density function (PDF). Otherwise, it may simply be equal to 
$\delta(1-x_\alpha)$ (e.g.~when describing an electron beam at a
fixed energy), or to a more complicated function that effectively
describes the energy loss when the incoming particle $P_\alpha$ turns
into the particle $\ident_\alpha$ that enters the hard reaction
(e.g., $f_{\ident_\alpha}^{(P_\alpha)}$ may be the Weizs\"acker-Williams
function with $P_\alpha=e$ and $\ident_\alpha=\gamma$). These details
are in any case irrelevant in what follows, where we shall deal
only with the short-distance partonic cross sections $d\sigma^{(n)}$,
$d\sigma^{(n+1)}$, and $d\bar{\sigma}^{(n+1)}$; if necessary, we shall
refer to functions $f_{\ident_\alpha}^{(P_\alpha)}$ generically as PDFs.

In this paper, we shall denote by $\mu$ the common value of the
factorization and renormalization scales, $\mu=\muF=\muR$, which
simplifies the writing of the formulae. At the end, one 
is always able to recover the separate dependence upon $\muF$ and $\muR$,
by exploiting the renormalization group invariance with respect to these
two scales. All the relevant formulae are given in app.~\ref{app:scales}.

The short-distance cross sections $d\sigma^{(n)}$, $d\sigma^{(n+1)}$,
and $d\bar{\sigma}^{(n+1)}$, have an $n$-body, an $(n+1)$-body,
and a degenerate $(n+1)$-body kinematics respectively. They will be 
discussed in the next three subsections in turn.

\subsection{$n$-body contributions\label{sec:dsign}}
We decompose the $n$-body cross section into four terms:
\beq
d\sigma^{(n)}=d\sigma^{(B,n)}+d\sigma^{(C,n)}+
d\sigma^{(S,n)}+d\sigma^{(V,n)}\,.
\label{nbody}
\eeq
The first term on the r.h.s.~of eq.~(\ref{nbody}) is the Born 
contribution ($\proc\in\allprocn$)
\beq
d\sigma^{(B,n)}(\proc)=\ampsqnt(\proc)\frac{\JetsB}{\avg(\proc)}\phspn\,,
\label{dsignB}
\eeq
with $\avg(\proc)$ the symmetry factor that takes into account the 
presence of identical particles in the final state.
The remaining three terms on the r.h.s.~of eq.~(\ref{nbody})
have, roughly speaking, a collinear, soft, and one-loop origin 
respectively; they are separately finite, and we understand the 
kinematics given in eq.~(\ref{Bkin}). We have:
\beq
d\sigma^{(C,n)}(\proc)=\asotwopi{\cal Q}(\proc)
\ampsqnt(\proc)\frac{\JetsB}{\avg(\proc)}\phspn\,,
\label{dsignC}
\eeq
where
\beqn
{\cal Q}(\proc)&=&-\log\frac{\mu^2}{Q^2}
\Bigg(\gamma(\ident_1)+2C(\ident_1)\log\xicut
+\gamma(\ident_2)+2C(\ident_2)\log\xicut\Bigg)
\nonumber \\*&+&
\sum_{k=3}^{\nlightB+2}\Bigg[\gamma^\prime(\ident_k)
-\log\frac{s\deltaO}{2Q^2}\left(\gamma(\ident_k)
-2C(\ident_k)\log\frac{2E_k}{\xicut\sqrt{s}}\right)
\nonumber \\*&&\phantom{\sum_{k=3}^{\nlightB+2}}
+2C(\ident_k)\left(\log^2\frac{2E_k}{\sqrt{s}}-\log^2\xicut\right)
-2\gamma(\ident_k)\log\frac{2E_k}{\sqrt{s}}\Bigg].
\label{Qdef}
\eeqn
Note that, in this equation, only light quarks and gluons are involved. 
We have $s=2k_1\mydot k_2$, and $E_k$ is the energy of parton
$k$ in the c.m.~frame of the incoming particles $\ident_1$ and $\ident_2$
(i.e., the partonic c.m.~frame in the case of hadronic collisions).
The Casimir's $C(\ident)$ have been defined in eqs.~(\ref{Cq}) and~(\ref{Cg}); 
the other colour factors are:
\beqn
\gamma(g)&=&\frac{11}{6}\CA-\frac{2}{3}\TF N_f,,
\label{gammag}
\\
\gamma(q)&=&\frac{3}{2}\CF\,,
\label{gammaq}
\\
\gamma^\prime(g)&=&\left(\frac{67}{9}-\frac{2\pi^2}{3}\right)\CA
-\frac{23}{9}\TF N_f\,,
\label{gmmprimeglu}
\\
\gamma^\prime(q)&=&\left(\frac{13}{2}-\frac{2\pi^2}{3}\right)\CF\,.
\label{gmmprimeqrk}
\eeqn
Equation~(\ref{Qdef}) contains two free parameters, $\xicut$ and 
$\deltaO$. The same parameters will be used in the subtraction of
the soft and final-state collinear divergences, respectively, that 
affect the real-emission contribution, as it will be described below.
Thus, although eq.~(\ref{dsignC}) does depend on $\xicut$ and
$\deltaO$, the physical cross section does not. This constitutes
a powerful check of the correctness of the implementation of the
formalism; we shall return to this point in sect.~\ref{sec:res}.
In eq.~(\ref{Qdef}) we have denoted by $Q$ the Ellis-Sexton 
scale~\cite{Ellis:1985er}, which is an arbitrary mass scale 
that one may use to write the one-loop results; more details 
on its role will be given in app.~\ref{app:virt} and in app.~\ref{app:scales}.
Here we just stress that the NLO cross section
is exactly independent of $Q$.
Finally, we remind the reader that $\mu=\muF=\muR$.
Note that the contribution in round brackets in the first line on the
r.h.s.~of eq.~(\ref{Qdef}) is non trivial only if at least one of the incoming
particles is a hadron. Still, the form in eq.~(\ref{Qdef}) applies to 
any kind of collisions, provided that one defines
\beq
C(\ident)=\gamma(\ident)\equiv 0\;\;\;{\rm if}~~\ident~~
{\rm is~a~colour~singlet}.
\eeq

The $n$-body contribution of soft origin reads:
\beq
d\sigma^{(S,n)}(\proc)=
\asotwopi\sum_{k=\nini}^{\nlightB+\nheavy+2}\,\,\sum_{l=k}^{\nlightB+\nheavy+2}
\eikint_{kl}^{(m_k,m_l)}\ampsqnt_{kl}(\proc)\frac{\JetsB}{\avg(\proc)}\phspn\,.
\label{dsignS}
\eeq
The quantities $\eikint_{kl}^{(m_k,m_l)}$ are the finite parts of the 
integrated eikonal factors; their precise definitions and explicit forms 
are given in app.~\ref{app:eik}. We have denoted by
\beq
m_l=\sqrt{k_l^2}
\eeq
the mass of particle $l$.
We point out that the contributions with $k=l$ and $k\le\nlightB+2$
to eq.~(\ref{dsignS}) are identically equal to zero, owing to 
eq.~(\ref{eik00self}).

Finally, the $n$-body contribution of one-loop origin is:
\beq
d\sigma^{(V,n)}(\proc)=
\asotwopi\vampsqnl_{\sss FIN}(\proc)\frac{\JetsB}{\avg(\proc)}\phspn\,.
\label{dsignV}
\eeq
The quantity $\vampsqnl_{\sss FIN}$ is the non-divergent part of
$\ampsqnl$ introduced in eq.~(\ref{Moneloop}). Its definition is
therefore unambiguous only in a definite scheme, and the one
to be used in eq.~(\ref{dsignV}) is the Conventional Dimensional
Regularization (CDR) scheme. The extraction of $\vampsqnl_{\sss FIN}$
from a complete one-loop computation, and its relation to the popular
Dimensional Reduction scheme, is discussed in app.~\ref{app:virt}.

\subsection{$(n+1)$-body contributions\label{sec:dsignpo}}
The short distance cross sections $d\sigma^{(n+1)}$ in eq.~(\ref{factTH})
are due the contributions of the real-emission matrix elements $\ampsqnpot$,
with their phase-space singularities suitably subtracted. As is known,
the integration of the subtracted matrix elements is the most involved
from the numerical viewpoint, and the difficulty increases with the
number of external legs, because of the proliferation of singularities.
In the FKS formalism, this problem is simplified by effectively partitioning 
the phase space, in such a way that in each of the kinematic regions 
resulting from the partition at most one soft and one collinear 
singularity are present. This partition is achieved by introducing
a set of positive-definite functions\footnote{In the original papers,
refs.~\cite{Frixione:1995ms,Frixione:1997np}, the $\Sfun$ functions were
constructed using Heaviside $\stepf$'s. Smoother $\Sfun$ functions, resulting
in an improved numerical behaviour, were introduced in 
ref.~\cite{Frixione:2005vw}. With $\stepf$ functions, one achieves an
exact phase-space partition, in which the various regions do not overlap,
while with the present form the partition regions can overlap.}
\beq
\Sfunij(\proc)\,,\;\;\;\;\;\;(i,j)\in\FKSpairs(\proc)\,,
\eeq
where the argument $\proc\in\allprocnpo$ stresses the fact that different 
choices for $\Sfun$ can be made for different processes; this property is 
useful to reduce the computing time. In the following, we shall often 
understand the argument $\proc$ in the $\Sfun$ functions.

There is ample freedom in the definition of the $\Sfun$ functions, but
the following constraint must be obeyed:
\beq
\sum_{(i,j)\in\FKSpairs}\!\!\!\!\Sfunij = 1\,.
\label{Sfununit}
\eeq
Furthermore, these functions must have the following behaviours
in the kinematic configurations associated with the singularities
of $\ampsqnpot$:
\beqn
&&\lim_{\veck_i\parallel\veck_j}\Sfunij=
h_{ij}\!\!\left(\frac{E_i}{E_i+E_j}\right)~~~~~~~{\rm if}~~m_i=m_j=0\,,
\label{SfunC1}
\\
&&\lim_{k_i^0\to 0}\Sfunij=c_{ij}\phantom{aaaaaaaaaaaaaa}\;
{\rm if}~~~\ident_i=g\,,~~~{\rm with}
\nonumber\\*&&\phantom{aaaaa}
0<c_{ij}\le 1\quad{\rm and}\,\,
\mathop{\sum_{j}}_{(i,j)\in\FKSpairs}\!\!\!\!\!c_{ij}=1\,.
\label{SfunC2}
\\
&&\lim_{\veck_k\parallel\veck_l}\Sfunij=0~~~~~\forall\,
\{k,l\}\ne\{i,j\}~~{\rm with}~~(k,l)\in\FKSpairs~~{\rm and}~~
m_k=m_l=0,\phantom{aaa}
\label{SfunC3}
\\
&&\lim_{k_k^0\to 0}\Sfunij=0~~~~~\forall k~~{\rm with}~~\ident_k=g
~~{\rm and}~~\exists l~~{\rm with}~~(k,l)\in\FKSpairs~~{\rm or}~~
(l,k)\in\FKSpairs\,.
\nonumber\\*&&
\label{SfunC4}
\eeqn
In words, $\Sfunij$ goes to zero in all regions of phase space
where the real emission matrix elements diverge, except if this
involves particle $i$ being soft, or particles $i$ and $j$ being
collinear. Note that eqs.~(\ref{SfunC1}) and/or~(\ref{SfunC3})
need not be satisfied if one of the partons in the relevant 
FKS pair is massive; in such cases, the limits can assume arbitrary 
values. Also, eq.~(\ref{SfunC4}) implies in particular that the limit 
of $\Sfunij$ is zero when $j$ is soft; thus, although according to
eq.~(\ref{PFKSdef}) there may be soft singularities associated with
FKS sisters, they are damped by using $\Sfunij$.
The functions $h_{ij}(z)$ introduced in eq.~(\ref{SfunC1}) 
are defined in \mbox{$0\le z\le 1$}, and have the following properties:
\beqn
h_{ij}(z)&=&1~~~~~~~~~{\rm if}~~\nini\le j\le 2\,,
\label{hdef1}
\\
h_{ij}(z)&=&h(z)~~~~~{\rm if}~~3\le j\le \nlightR+\nheavy+2\,,
\label{hdef2}
\eeqn
with $h(z)$ a positive-definite function such that
\beq
\lim_{z\to 0}h(z)=1\,,\;\;\;\;\;\;
\lim_{z\to 1}h(z)=0\,,\;\;\;\;\;\;
h(z)+h(1-z)=1\,.
\label{hdef3}
\eeq
The physical cross section is independent of the choice of $h(z)$; we
shall return to this point in sect.~\ref{sec:res}. 
As stressed above, the particular form of the
$\Sfun$ functions is not important in what follows; we shall give an
explicit construction in sect.~\ref{sec:impl:Sfun}. 
As a final remark, we point out that eqs.~(\ref{hdef1}) and~(\ref{hdef2}) 
imply that the notation $h_{ij}$ is redundant here; it will however
become useful in the context of the optimization we shall carry
out in sect.~\ref{sec:opt}.

Using eq.~(\ref{Sfununit}), one proceeds by writing
\beq
\ampsqnpot(\proc)=\sum_{(i,j)\in\FKSpairs}\Sfunij(\proc)\ampsqnpot(\proc)\,.
\label{Realident}
\eeq
Thanks to eqs.~(\ref{SfunC3}) and~(\ref{SfunC4}), each term on the r.h.s.~of
eq.~(\ref{Realident}) will only be singular when particle $i$ is soft,
and/or particles $i$ and $j$ are collinear\footnote{We remind the reader
that multiple soft and collinear singularities, i.e.~configurations that 
would contribute to NNLO, are cut by $\JetsB$.}. 
Therefore, each term can actually
be regarded as describing a production process with the simplest possible
structure of phase-space divergences. Furthermore, these contributions 
are fully independent from each other.

We shall now give the expressions of the subtracted real-emission cross 
sections. In order to do so, we work in the c.m.~frame of the incoming 
particles:
\beq
k_1=\frac{\sqs}{2}(1,0,0,1)\,,\;\;\;\;\;
k_2=\frac{\sqs}{2}(1,0,0,-1)\,.
\label{cmsframe}
\eeq
In this frame, for each pair $(i,j)\in\FKSpairs$ we introduce the variables
$\xii$ and $\yij$, where
\beqn
E_i&=&\frac{\sqs}{2}\xii\,,
\label{xiidef}
\\
\veck_i\mydot\veck_j&=&\abs{\veck_i}\abs{\veck_j}\yij\,.
\label{yijdef}
\eeqn
In other words, $\xii$ is the rescaled energy of the FKS parton,
and $\yij$ is the cosine of the angle between the FKS parton and
its sister. Using these variables, the soft and collinear singularities
of $\Sfunij\ampsqnpot$ correspond to $\xii=0$ and to $\yij=1$ respectively
(note, however, that the matrix element may not be singular in either 
or both of these limits, depending on the identities of particles $i$ 
and $j$). We have (see ref.~\cite{Frixione:1995ms})
\beq
d\sigma^{(n+1)}(\proc)=
\sum_{(i,j)\in\FKSpairs}d\sigma_{ij}^{(n+1)}(\proc)\,,
\label{dsignpo}
\eeq
where
\beq
d\sigma_{ij}^{(n+1)}(\proc)=\xic\omyijd
\Big((1-\yij)\xii^2\ampsqnpot(\proc)\Big)
\Sfunij(\proc)\frac{\JetsB}{\avg(\proc)}\, 
d\xii d\yij d\phii\tphspnij\,.
\label{dsigijnpo}
\eeq
The variable $\phii$ parametrizes the azimuthal direction of the FKS parton,
but its precise definition is irrelevant here. The quantity $\tphspnij$
is related to the $n$-body phase space, and is implicitly defined as
follows:
\beq
\phspnpo = \xii d\xii d\yij d\phii \tphspnij\,.
\label{tphspdef}
\eeq
The actual form for $\tphspnij$ depends on the specific parametrization
adopted for $\phspnpo$, but the following equations must always hold:
\beqn
\lim_{\xii\to 0}\tphspnij&=&\frac{s}{(4\pi)^3}\phspn~~~~~{\rm if}~~m_i=0\,,
\label{tphisoft}
\\
\lim_{\yij\to 1}\tphspnij&=&\frac{s}{(4\pi)^3}\phspn~~~~~{\rm if}~~m_i=m_j=0\,.
\label{tphicoll}
\eeqn
The distributions entering eq.~(\ref{dsigijnpo}) are defined as 
follows\footnote{If $m_i\ne 0$, the lower limit of the $\xii$ integration
range is larger than zero. This case is however trivial, and we shall return
to it at the end of this section.}:
\beqn
\int_0^{\ximax} d\xii f(\xii)\xic&=&
\int_0^{\ximax} d\xii \frac{f(\xii)-f(0)\stepf(\xicut-\xii)}{\xii}\,,
\label{distrxii}
\\
\int_{-1}^1 d\yij g(\yij)\omyijd&=&
\int_{-1}^1 d\yij \frac{g(\yij)-g(1)\stepf(\yij-1+\delta)}{1-\yij}\,,
\label{distryij}
\eeqn
where
\beq
\ximax=1-\frac{1}{s}\left(\sum_{k=3}^{n+3}m_k\right)^2\,.
\eeq
In eq.~(\ref{distrxii}) and~(\ref{distryij}) $\xicut$ and $\delta$ are 
free parameters, that can be chosen in the ranges
\beq
0<\xicut\le\ximax\,,\;\;\;\;\;
0<\delta\le 2\,.
\eeq
It would be possible to choose different values of $\xicut$ and $\delta$
for each of the cross sections $d\sigma_{ij}^{(n+1)}$ contributing
to eq.~(\ref{dsignpo}). In practice, we shall choose the same value
of $\xicut$ for all $(i,j)$ pairs, the same value of $\delta$ for
all pairs $(i,j)$ with $j\ge 3$ (which we shall denote by $\deltaO$),
and the same value of $\delta$ for all pairs $(i,j)$ with $j\le 2$ 
(which we shall denote by $\deltaI$). The parameters $\xicut$, $\deltaO$, 
and $\deltaI$ are therefore associated with soft, final-state collinear,
and initial-state collinear singularities respectively. As was discussed 
in sect.~\ref{sec:dsign}, the physical cross section on the l.h.s.~of
eq.~(\ref{factTH}) is strictly independent of $\xicut$, $\deltaO$,
and $\deltaI$, while this is not the case for the short-distance
cross sections $d\sigma^{(n)}$, $d\sigma^{(n+1)}$, and 
$d\bar{\sigma}^{(n+1)}$. 

For the reader unfamiliar with the use of plus distributions in
the context of cross section calculations, it is useful to explicitly
show the structure of eq.~(\ref{dsigijnpo}). We start by rewriting
such an equation with the following shorthand notation:
\beq
d\sigma_{ij}^{(n+1)}=\xic\omyijd\Sigma_{ij}(\xii,\yij)
d\xii d\yij\,.
\label{dsigijnpo2}
\eeq
We can now expand the plus distributions using 
their definitions, eqs.~(\ref{distrxii}) and~(\ref{distryij}):
\beqn
d\sigma_{ij}^{(n+1)}&=&\int_0^{\ximax} d\xii \int_{-1}^1 d\yij
\frac{1}{\xii(1-\yij)}
\Big[\Sigma_{ij}(\xii,\yij)-\Sigma_{ij}(\xii,1)\stepf(\yij-1+\delta)
\nonumber\\*&&\phantom{\int_0^{\ximax} }
-\Sigma_{ij}(0,\yij)\stepf(\xicut-\xii)
+\Sigma_{ij}(0,1)\stepf(\xicut-\xii)\stepf(\yij-1+\delta)
\Big].
\nonumber\\*&&
\label{dsigijnpoE}
\eeqn
There are four terms in the integrand in eq.~(\ref{dsigijnpoE}).
We shall call them ``event'' ($\Sigma_{ij}(\xii,\yij)$),
``collinear counterevent'' ($\Sigma_{ij}(\xii,1)$),
``soft counterevent'' ($\Sigma_{ij}(0,\yij)$), and
``soft-collinear counterevent'' ($\Sigma_{ij}(0,1)$). In general, they may
have four different kinematic configurations, but one can actually
reduce them to two, as discussed in sect.~\ref{sec:impl:kin}. Note that, in
the case of a massive sister ($m_j\ne 0$), the collinear and soft-collinear
counterevents vanish identically, because of the damping factor
\mbox{$(1-\yij)$} in eq.~(\ref{dsigijnpo}).

Finally, consider the case $m_i\ne 0$. According to eq.~(\ref{PFKSdef}),
$(i,j)\in\FKSpairs$ only if $\ident_j=g$. The only possible
singularity arising from the FKS pair $(i,j)$ is the soft one $k_j^0\to 0$;
this is however damped by $\Sfunij$. It follows that all counterevents
in eq.~(\ref{dsigijnpoE}) vanish, and the distributions in 
eq.~(\ref{dsigijnpo2}) coincide with ordinary functions.
Therefore, using eq.~(\ref{tphspdef}), one simply gets
\beq
d\sigma_{ij}^{(n+1)}(\proc)=\ampsqnpot(\proc)
\Sfunij(\proc)\frac{\JetsB}{\avg(\proc)}\,\phspnpo\,,
\;\;\;\;\;
m_i\ne 0~~{\rm and}~~\ident_j=g\,.
\label{dsigijMi}
\eeq
Since there are no subtractions involved, one is not bound to use the
variables $\xii$, $\yij$, and $\phii$ in the parametrization of $\phspnpo$
in this equation. On the other hand, there is nothing that prevents one from
doing so; in such a case, note that the smallest value $\xii$ can assume 
is \mbox{$2m_i/\sqs$}, and we understand in eq.~(\ref{dsigijnpoE})
that \mbox{$\Sigma_{ij}(\xii,\yij)$} vanish identically if
\mbox{$\xii\le 2m_i/\sqs$}. The fact that when $m_i\ne 0$ one does
not need to subtract any singularities suggests that from the viewpoint
of the efficiency of the calculation it is not convenient to define
the quantity $d\sigma_{ij}^{(n+1)}$ in eq.~(\ref{dsigijMi}) as an independent 
contribution; we shall discuss this issue in sect.~\ref{sec:opt}.

\subsection{Degenerate $(n+1)$-body contributions\label{sec:dsigcoll}}
These contributions, denoted by $d\bar{\sigma}^{(n+1)}$ in 
eq.~(\ref{factTH}), are the finite remainders left by the subtraction
of initial-state collinear counterterms. We write
\beqn
d\bar{\sigma}^{(n+1)}(\proc)=
\sum_{(i,1)\in\FKSpairs}d\bar{\sigma}_{i1}^{(n+1)}(\proc)+
\sum_{(i,2)\in\FKSpairs}d\bar{\sigma}_{i2}^{(n+1)}(\proc)\,.
\label{collrem}
\eeqn
The condition $(i,j)\in\FKSpairs$ (with $j=1,2$) in eq.~(\ref{collrem})
implies that the real-emission matrix element is singular when
\mbox{$\veck_i\parallel\veck_j$}. This typically means that, 
if $P_j$ is not a hadron, the corresponding contribution
$d\bar{\sigma}_{ij}^{(n+1)}$ need not be included. A notable exception is
$\ident_i=q$ and $\ident_j=\gamma$, when the matrix element has 
a (QED) singularity, which is cancelled by the counterterm due to
the inhomogeneous part of the (Altarelli-Parisi) evolution equation 
for the hadronic photon. In any case, by writing \mbox{$(i,1)\in\FKSpairs$}
and \mbox{$(i,2)\in\FKSpairs$} in eq.~(\ref{collrem}) we can use this
equation for any kind of collisions.
As shown in ref.~\cite{Frixione:1995ms}, one obtains
\beqn
&&d\bar{\sigma}_{i1}^{(n+1)}(\proc;k_1,k_2)=
\asotwopi\Bigg\{\APdamp_{\ident_{1\oplus\bar{i}}\ident_1}^{(0)}(1-\xii)
\left[\xic\log\frac{s\deltaI}{2\mu^2}+2\lxic\right]
\nonumber \\*&&
\phantom{d\bar{\sigma}_{i1}^{(n+1)}(\proc;k_1,k_2)=\asotwopi}
-\APdamp_{\ident_{1\oplus\bar{i}}\ident_1}^{(1)}(1-\xii)\xic
-K_{\ident_{1\oplus\bar{i}}\ident_1}(1-\xii)\Bigg\}
\nonumber \\*&&\phantom{aaaaaa}\times
\ampsqnt\left(\proc^{1\oplus\bar{i},\isubrmv};(1-\xii)k_1,k_2\right)
\frac{\JetsB}{\avg(\proc)}\phspn\Big((1-\xii)k_1,k_2\Big)d\xii\,,
\label{sigcp}
\\
&&d\bar{\sigma}_{i2}^{(n+1)}(\proc;k_1,k_2)=
\asotwopi\Bigg\{\APdamp_{\ident_{2\oplus\bar{i}}\ident_2}^{(0)}(1-\xii)
\left[\xic\log\frac{s\deltaI}{2\mu^2}+2\lxic\right]
\nonumber \\*&&
\phantom{d\bar{\sigma}_{i2}^{(n+1)}(\proc;k_1,k_2)=\asotwopi}
-\APdamp_{\ident_{2\oplus\bar{i}}\ident_2}^{(1)}(1-\xii)\xic
-K_{\ident_{2\oplus\bar{i}}\ident_2}(1-\xii)\Bigg\}
\nonumber \\*&&\phantom{aaaaaa}\times
\ampsqnt\left(\proc^{2\oplus\bar{i},\isubrmv};k_1,(1-\xii)k_2\right)
\frac{\JetsB}{\avg(\proc)}\phspn\Big(k_1,(1-\xii)k_2\Big)d\xii\,,
\label{sigcm}
\eeqn
where, in analogy with eq.~(\ref{distrxii}), we have introduced the 
distribution
\beq
\int_0^{\ximax} d\xii f(\xii)\lxic=
\int_0^{\ximax} d\xii \Big(f(\xii)-f(0)\stepf(\xicut-\xii)\Big)
\frac{\log\xii}{\xii}\,.
\label{distrlxii}
\eeq
We have also defined the functions
\beqn
\APdamp_{ab}(z,\ep)=(1-z)P_{ab}(z,\ep)\equiv
\APdamp_{ab}^{(0)}(z)+\ep\APdamp_{ab}^{(1)}(z)+{\cal O}(\ep^2)\,,
\label{APdampdef}
\eeqn
i.e.~the unregularized Altarelli-Parisi splitting functions times 
a $(1-z)$ factor that damps the $z\to 1$ soft singularity. The explicit
forms of $P_{ab}^{(0)}$ and $P_{ab}^{(1)}$ are given in 
app.~\ref{app:Qfun}. The functions 
$K_{ab}$ are related to the choice of the PDF scheme; for all practical
purposes they are trivial, since in the $\overline{\rm MS}$ scheme
we have $K_{ab}=0$. We point out that in the special case alluded to
at the beginning of this section, i.e.~$\ident_i=q$ and $\ident_j=\gamma$,
the relevant $d\bar{\sigma}_{ij}^{(n+1)}$ must be proportional to 
$\alpha_{em}$. We can therefore still use eqs.~(\ref{sigcp}) and~(\ref{sigcm}),
provided that we include a factor \mbox{$\alpha_{em}/\as$} in the
kernels $P_{q\gamma}$ and $K_{q\gamma}$ (see ref.~\cite{Frixione:1997np}).

Consider $d\bar{\sigma}_{i1}^{(n+1)}$ in
eq.~(\ref{sigcp}) (the case of $d\bar{\sigma}_{i2}^{(n+1)}$ is
fully analogous). The fact that $(i,1)\in\FKSpairs$ implies that
$\proc^{1\oplus\bar{i},\isubrmv}$ exists and belongs to $\allprocn$,
and that $\ident_{1\oplus\bar{i}}$ is either a light quark or a gluon.
The notation embodies the fact that the matrix element and phase space
must be computed with $2\to n$ kinematic configurations, where
incoming partons have momenta 
\beq
k_1^\prime=(1-\xii)k_1\,,\;\;\;\;\;\;
k_2^\prime=k_2\,.
\eeq

We conclude by stressing that the symmetry factor in eqs.~(\ref{sigcp})
and~(\ref{sigcm}) is $\avg(\proc)$, and not 
$\avg(\proc^{1\oplus\bar{i},\isubrmv})$ 
or $\avg(\proc^{2\oplus\bar{i},\isubrmv})$. This can be easily
understood as follows.
Suppose that there are $p$ identical particles in the final state 
of $\proc$, with labels $i_1,\ldots i_p$; $\avg(\proc)$ will thus 
contain a factor $p!$. On the other hand, we shall also have
$d\bar{\sigma}_{i_\alpha j}^{(n+1)}=d\bar{\sigma}_{i_\beta j}^{(n+1)}$,
for $j=1,2$, and $1\le\alpha,\beta\le p$. Hence, 
\beq
\sum_{\alpha=1}^p d\bar{\sigma}_{i_\alpha j}^{(n+1)}=
p\,d\bar{\sigma}_{i_1 j}^{(n+1)}\,.
\eeq
Apart from allowing one to reduce the computing time by a factor of $p$,
this equation also shows that symmetry factors 
$\avg(\proc^{1\oplus\bar{i}_1,\isubrmv_1})$ and 
$\avg(\proc^{2\oplus\bar{i}_1,\isubrmv_1})$
will emerge naturally, since
\beq
\frac{p}{\avg(\proc)}=\frac{p}{p!\,\overline{\avg}(\proc)}=
\frac{1}{(p-1)!\,\overline{\avg}(\proc)}=
\frac{1}{\avg(\proc^{1\oplus\bar{i}_1,\isubrmv_1})}=
\frac{1}{\avg(\proc^{2\oplus\bar{i}_1,\isubrmv_1})}\,.
\eeq
This equation obviously holds also for $p=1$.

\section{Implementation\label{sec:impl}}
The most involved part in the computation of an NLO cross section is
the integration of the subtracted real-emission (i.e., $(n+1)$-body) 
matrix elements. 
In this section we shall therefore mainly present technical details
on the ingredients that will enter such an integration in our implementation.
In particular, we shall discuss the kinematics relevant to the generation
of counterevents, give the formulae relevant to the evaluation of the
damped matrix elements in the singular limits, and present an explicit
construction of the $\Sfun$ functions. We stress that the latter is
fully general, and easy to automate, but nothing prevents one from
using alternative forms, provided they fulfill the conditions given in
eqs.~(\ref{Sfununit})--(\ref{SfunC4}).
Although the practical applications in this paper are limited to the 
case of $\epem$ collisions, the formulae we give in the following
can be applied without modifications to any kind of incoming particles.

\subsection{Kinematics\label{sec:impl:kin}}
The integration of the subtracted real-emission matrix elements 
over the $(n+1)$-body phase space entails 
the generation of $3n-1$ random numbers, of which three correspond to the
integration variables $\xii$, $\yij$, and $\phii$, that have been singled 
out since they are directly associated with the FKS parton. We denote the
remaining $3n-4$ random numbers by $x_\alpha$; we can write a generic 
$2\to n+1$ kinematic configuration as follows:
\beq
\Big\{k_l\Big\}_{l=1}^{n+3}\left(\xii,\yij,\phii,x_\alpha\right)\,.
\label{momEV}
\eeq
As implied by eq.~(\ref{dsigijnpoE}), this configuration will be that
of the event. Starting from it, we construct the configurations corresponding 
to the soft, collinear, and soft-collinear counterevents\footnote{In the
case in which particle $j$ is massive, there is no need to construct
collinear and soft-collinear counterevents. If $i$ is massive, all
counterevents are identically equal to zero.}:
\beqn
\Big\{k_l^{\sss (S)}\Big\}_{l=1}^{n+3}&=&
\Big\{k_l\Big\}_{l=1}^{n+3}\left(0,\yij,\phii,x_\alpha\right)\,,
\label{momCS}
\\
\Big\{k_l^{\sss (C)}\Big\}_{l=1}^{n+3}&=&
\Big\{k_l\Big\}_{l=1}^{n+3}\left(\xii,1,\phii,x_\alpha\right)\,,
\label{momCC}
\\
\Big\{k_l^{\sss (SC)}\Big\}_{l=1}^{n+3}&=&
\Big\{k_l\Big\}_{l=1}^{n+3}\left(0,1,\phii,x_\alpha\right)\,,
\label{momCSC}
\eeqn
with $\phii$ and $x_\alpha$ in eqs.~(\ref{momCS})--(\ref{momCSC})
the same as those used in the generation of the kinematics of
the event, eq.~(\ref{momEV}). For a given choice of the $3n-1$ 
independent variables, the momenta $k_l$ depend on the phase-space
parametrization adopted. We have considered three different such
parametrizations, in order to test the convergence behaviour
and numerical stability of the numerical integration of short-distance
cross sections. All obey the following equations:
\beqn
&&k_l^{\sss (S)}=k_l^{\sss (C)}=k_l^{\sss (SC)}\,,
\phantom{aaaaaaaaaaaaaaaaaaaa}
\forall\;l\ne i,j\,,
\label{kinK1}
\\&&
k_i^{\sss (S)}+k_j^{\sss (S)}=
k_i^{\sss (C)}+k_j^{\sss (C)}=
k_i^{\sss (SC)}+k_j^{\sss (SC)}\,.
\label{kinK2}
\eeqn
These properties trivially follow from the observation that the kinematic
configurations of the counterevents are degenerate, and effectively
correspond to a $2\to n$ configuration. This configuration, whose momenta
we denote~\cite{Frixione:2002ik,Frixione:2007vw} by $\kbar$, can simply be
defined with a relabeling of the soft counterevent kinematics:
\beq
\Big\{\kbar_l\Big\}_{l=1}^{n+2}\equiv
\Big\{\kbar_{\sigma(l)}\Big\}_{l=1,l\ne i}^{n+3}=
\Big\{k_l^{\sss (S)}\Big\}_{l=1,l\ne i}^{n+3}\,,
\label{kbardef}
\eeq
with $\sigma$ denoting the relabeling.

We stress that, although eqs.~(\ref{kinK1}) and~(\ref{kinK2}) need not
be fulfilled by any phase-space parametrization, and are not mandatory
for the implementation of FKS subtraction, they are quite intuitive from
the physical point of view, and help improve the numerical stability
of the results for differential distributions, while also
reducing the computing time.

We finally point out that the results of this section
apply to the case $j\ge 3$, i.e.~to final-state emissions. The
situation is more involved in the case of $j=1,2$, i.e.~of initial-state
emissions, but similar simplifications are possible 
(see e.g.~app.~A.4 of ref.~\cite{Frixione:2002ik}). We postpone
a discussion on this issue to a forthcoming publication.

\subsection{Choice of $\Sfun$ functions\label{sec:impl:Sfun}}
As already mentioned in sect.~\ref{sec:dsignpo}, in
ref.~\cite{Frixione:2005vw} a smooth form of the $\Sfun$ function
was adopted, which was seen to improve the numerical integration
with respect to the form of refs.~\cite{Frixione:1995ms,Frixione:1997np},
where Heaviside $\stepf$'s had been used. In ref.~\cite{Frixione:2005vw}, 
the $\Sfun$ functions were constructed using invariants $k_k\mydot k_l$,
but given that FKS subtraction uses energy and angles as integration
variables, in a further generalization (see e.g.~ref.~\cite{Frixione:2007vw})
one can also define:
\beq
\Sfunij=\frac{1}{\Dfun\,d_{ij}}\,
h_{ij}(z_{ij}),\;\;\;\;\;
(i,j)\in\FKSpairs\,,
\label{Sijdef}
\eeq
where the functions $h_{ij}$ have been introduced in 
eqs.~(\ref{hdef1})--(\ref{hdef3}), and
\beqn
z_{ij}&=&\frac{E_i}{E_i+E_j}\,,\;\;\;\;\;
z_{ji}=\frac{E_j}{E_i+E_j}=1-z_{ij}\,,
\label{zijdef}
\\
\Dfun&=&\sum_{(k,l)\in\FKSpairs}\,\frac{1}{d_{kl}}h_{kl}(z_{kl})\,,
\label{Ddef}
\\
d_{kl}&=&\left(\frac{2E_k}{\sqs}\right)^{\asfun}
\left(\frac{2E_l}{\sqs}\right)^{\asfun}
\left(1-\beta_k\beta_l\cos\theta_{kl}\right)^{\bsfun}\,,
\label{ddef}
\\
\beta_k&=&\sqrt{1-\frac{m_k^2}{E_k^2}}\,.
\label{betadef}
\eeqn
The parameters $\asfun$ and $\bsfun$ introduced in eq.~(\ref{ddef})
are real, positive, and arbitrary. The fact that the physical cross section
is independent of their choices constitutes another check of the correctness
of the implementation of FKS subtraction. We shall return to this point
in sect.~\ref{sec:res}. As shown in eqs.~(\ref{hdef1}) and~(\ref{hdef2}),
the functions $h_{kl}$ are given in terms of a function $h(z)$, that
we have chosen as follows:
\beq
h(z)=\frac{(1-z)^{2a_h}}{z^{2a_h}+(1-z)^{2a_h}}\,,\;\;\;\;\;\;
a_h=1\,.
\eeq
In practice, the specific form of $h(z)$ has only a modest impact on
numerical computations; we shall briefly comment on this point in
sect.~\ref{sec:res}. The role of $h_{kl}$ in eq.~(\ref{Ddef}) is
simply that of avoiding to count twice the contribution of $1/d_{kl}$
when both $(k,l)$ and $(l,k)$ belong to $\FKSpairs$. In fact, in this case
one can immediately obtain
\beq
\frac{1}{d_{kl}}h_{kl}(z_{kl})+
\frac{1}{d_{lk}}h_{lk}(z_{lk})=
\frac{1}{d_{kl}}\Big(h_{kl}(z_{kl})+h_{kl}(1-z_{kl})\Big)=
\frac{1}{d_{kl}}\,,
\label{hident}
\eeq
having used eqs.~(\ref{hdef3}) and~(\ref{zijdef}). Alternatively, one
could have dropped the $h_{kl}$ factor in eq.~(\ref{Ddef}), and summed
over unordered pairs.

We point out that the limits of the $\Sfun$ functions given in
eqs.~(\ref{SfunC1})--(\ref{SfunC4}) are related to the vanishing of
$d_{ij}$ in the collinear ($\theta_{ij}\to 0$ with $m_i=m_j=0$) and 
soft ($E_i\to 0$) limits. When parton $i$ is not a gluon, the soft
limit is simply not relevant (the matrix element is non singular),
and the $\Sfun$ functions can assume arbitrary values. This implies
that, in such a case, we may replace \mbox{$2E_i/\sqs$} with 1 in the
definition of $d_{ij}$. Along the same lines, one may always replace
$\beta_i$ and/or $\beta_j$ with 1 in $d_{ij}$ when particle $i$ and/or 
$j$ is massive; in doing so, $d_{ij}$ will not be proportional any longer 
to $k_i\mydot k_j$ when $\asfun=\bsfun$, but this is perfectly acceptable.
We have tested these options, and will comment on this point 
in sect.~\ref{sec:res}.

The $\Sfun$ functions defined in eq.~(\ref{Sijdef}) are suitable for
numerical implementation, since they are well behaved in the whole
$(n+1)$-body phase space, and in particular in the soft and collinear
limits. In fact, after trivial algebra (also using eq.~(\ref{hident}))
one gets
\beq
\Dfun d_{ij}=1+
\mathop{\sum_{(i,l)\in\FKSpairs}}_{l\ne j}\frac{d_{ij}}{d_{il}}+
\mathop{\sum_{(k,l)\in\FKSpairs}}_{k\notin\{i,j\}~{\rm or}~l\notin\{i,j\}}
\frac{d_{ij}}{d_{kl}}\,h_{kl}\!\left(z_{kl}\right).
\label{Ddsimp}
\eeq
The $E_i$ dependence can be analytically removed in the second term on
the r.h.s of eq.~(\ref{Ddsimp}), and therefore \mbox{$\Dfun d_{ij}$}
can be computed numerically both in the soft and collinear limits,
which are needed for the evaluations of the soft, collinear, and
soft-collinear counterevents.

\subsection{$(n+1)$-body matrix elements\label{sec:impl:RME}}
In the previous section we have seen how to compute the $\Sfun$ functions
in eq.~(\ref{dsigijnpo}) in the various singular 
limits. The other major ingredient for the evaluation of 
eq.~(\ref{dsigijnpo}) is the damped matrix element
\beq
\ampsqnpot_{ij}=(1-\yij)\xii^2\ampsqnpot\,.
\label{Mnpodamp}
\eeq
We shall now give explicit, finite, expressions for $\ampsqnpot_{ij}$
in the soft and collinear limits. Using the notation introduced in
eq.~(\ref{kbardef}), we have
\beqn
\lim_{\xii\to 0}\ampsqnpot_{ij}\left(\proc;\Big\{k\Big\}\right)&=&
\delta_{g\ident_i}\,\gs^2
\mathop{\sum_{p=\nini}}_{p\ne i}^{\nlightR+\nheavy+2}\,\,
\mathop{\sum_{l=p}}_{l\ne i}^{\nlightR+\nheavy+2}
\frac{(1-\yij)\xii^2(\kbar_{\sigma(p)}\mydot \kbar_{\sigma(l)})}
{(\kbar_{\sigma(p)}\mydot k_i)(\kbar_{\sigma(l)}\mydot k_i)}
\Bigg|_{\xii=0}
\nonumber\\*&&\phantom{
\delta_{g\ident_i}\,\gs^2
\mathop{\sum_{p=\nini}}_{p\ne i}^{\nlightR+\nheavy+2}\,\,
\mathop{\sum_{l=p}}_{l\ne i}^{\nlightR+\nheavy+2}
}
\times\ampsqnt_{\sigma(p)\sigma(l)}\left(\proc^{\isubrmv};
\Big\{\kbar\Big\}\right),\phantom{aaaaaa}
\label{MEsoftlim}
\eeqn
where $\sigma$ is the relabeling of the indices introduced 
in eq.~(\ref{kbardef}).
Note that the eikonals on the r.h.s.~of eq.~(\ref{MEsoftlim}) are finite
at $\xii=0$, since the factor $\xii^2$ in the numerator is cancelled by 
that resulting from $E_i^2$ in the denominator (see eq.~(\ref{xiidef})).
Furthermore, in the case $m_j=0$, the collinear limit of eq.~(\ref{MEsoftlim})
is also finite, since the factor \mbox{$1-\yij$} in the numerator cancels
an identical factor in the denominator, when $p=j$ or $l=j$ 
(see eq.~(\ref{yijdef})); this collinear limit can thus be used to construct
the soft-collinear counterterm. The cancellations of the $\xii^2$ and
\mbox{$1-\yij$} factors are carried out analytically, and 
eq.~(\ref{MEsoftlim}) is therefore well behaved numerically in the
whole $(n+1)$-body phase space.

In the case $m_i=m_j=0$, we also need to compute the collinear limit.
We have (see ref.~\cite{Frixione:1995ms})
\beqn
\lim_{\yij\to 1}\ampsqnpot_{ij}\left(\proc;\Big\{k\Big\}\right)&=&
\gs^2\frac{(1-\yij)\xii^2}{k_i\mydot k_j}
P_{\ident_j\ident_{j\oplus i}}^{(0)}(z_{ji})
\ampsqnt\left(\proc^{j\oplus i,\isubrmv};\Big\{\kbar\Big\}\right)
\nonumber\\*&+&
\gs^2\frac{(1-\yij)\xii^2}{k_i\mydot k_j}
Q_{\ident_j\ident_{j\oplus i}^\star}(z_{ji})
\tampsqnt\left(\proc;\Big\{k^{\sss (C)}\Big\}\right),\phantom{aaaa}
\label{MEcolllim1}
\eeqn
where $z_{ji}$ has been defined in eq.~(\ref{zijdef}), and 
the functions $Q_{ab^\star}(z)$ and reduced matrix elements $\tampsqnt$
are given in app.~\ref{app:Qfun}. The second term on the r.h.s.~of
eq.~(\ref{MEcolllim1}) vanishes when the integration over the azimuthal
angle $\phii$ is carried out. However, it is in general non zero 
pointwise in the phase space and,
as eq.~(\ref{MEcolllim1}) shows, it actually has the same behaviour
for $\yij\to 1$ as the first term; it must therefore be included when
constructing a {\em local} collinear counterterm.

Using eqs.~(\ref{xiidef}) and~(\ref{yijdef}), eq.~(\ref{MEcolllim1})
becomes:
\beqn
\lim_{\yij\to 1}\ampsqnpot_{ij}\left(\proc;\Big\{k\Big\}\right)&=&
\frac{4\gs^2}{z_{ji} s}\APdamp_{\ident_j\ident_{j\oplus i}}^{(0)}(z_{ji})
\ampsqnt\left(\proc^{j\oplus i,\isubrmv};\Big\{\kbar\Big\}\right)
\nonumber\\*&+&
\frac{4\gs^2}{z_{ji} s}\Qdamp_{\ident_j\ident_{j\oplus i}^\star}(z_{ji})
\tampsqnt\left(\proc;\Big\{k^{\sss (C)}\Big\}\right),
\label{MEcolllim2}
\eeqn
where, in analogy to eq.~(\ref{APdampdef}), we have defined
\beq
\Qdamp_{ab^\star}(z)=(1-z)Q_{ab^\star}(z)\,.
\label{Qdampdef}
\eeq
It is apparent that the r.h.s.~of eq.~(\ref{MEcolllim2}) can be
computed numerically also in the soft limit, $E_i\to 0$ (which is
equivalent to $z_{ji}\to 1$), owing to the regularizing factors 
\mbox{$(1-z)$} of eqs.~(\ref{APdampdef}) and~(\ref{Qdampdef}).
In the cases $(\ident_i,\ident_j)=(g,g)$ and $(\ident_i,\ident_j)=(q,g)$ 
(where $q$ is a massless quark) the Altarelli-Parisi kernels have
a soft singularity for $z_{ji}\to 0$. However, this singularity 
is eventually damped, in the computation of the cross section, by the 
factor $h_{ij}(z_{ij})$ in eq.~(\ref{Sijdef}), and thus need not be 
subtracted.

\section{Optimization\label{sec:opt}}
What we have done so far allows one to implement the FKS subtraction method 
in a computer code. We shall now tackle the problem of how to make
this straightforward implementation more efficient. As discussed
in sect.~\ref{sec:nots:ME}, the number of contributions to the sum
on the r.h.s.~of eq.~(\ref{dsignpo}) scales approximately as the
square of the number of light partons in the final state. We shall
show in the following that within the FKS formalism it is trivial
to reduce this scaling property from a quadratic form to a 
linear function at most, and actually to a constant if one
only increases the number of final-state gluons.
We shall also discuss in this section the integration of the
$n$-body contribution, and the implementation of multi-channel
integration.

\subsection{Reduction of the number of independent contributions
\label{sec:opt:red}}
Since the number of contributions to eq.~(\ref{dsignpo}) coincides with 
the number of elements in $\FKSpairs$, we start by observing that the 
definition given in eq.~(\ref{PFKSdef}) is sufficient to include all 
singularities to be eventually subtracted, but is redundant. 
As an example, consider
the case of $\ident_i=q$, $\ident_j=\qb$, with $q$ any massless quark.
According to eq.~(\ref{PFKSdef}), both $(i,j)$ and $(j,i)$ belong to
$\FKSpairs$. On the other hand, the pairs $(i,j)$ and $(j,i)$ are
responsible for the same divergence of the matrix element, the
collinear one at \mbox{$\veck_i\parallel\veck_j$}. This divergence
is subtracted twice in eq.~(\ref{dsignpo}), in the terms
\mbox{$d\sigma_{ij}^{(n+1)}$} and \mbox{$d\sigma_{ji}^{(n+1)}$},
but is not double counted thanks to the presence of 
\mbox{$h_{ij}(z_{ij})$} and of \mbox{$h_{ji}(z_{ji})$} in the
relevant $\Sfun$ functions (see eq.~(\ref{hident})). 
Furthermore, as discussed in 
sect.~\ref{sec:dsignpo}, when the FKS parton is massive, all of the
contributions \mbox{$d\sigma_{ij}^{(n+1)}$} are finite without
subtraction; it is therefore not particularly advantageous to
have a massive FKS parton. Several other examples can be given,
the upshot of which is that it is more convenient to define
$\FKSpairs$ as follows, rather than as in eq.~(\ref{PFKSdef}):
\beqn
\FKSpairs(\proc)&=&\Big\{(i,j)\;\Big|\;3\le i\le\nlightR+2\,, 
\nini\le j\le\nlightR+\nheavy+2\,, 
i\ne j\,,
\nonumber\\*&&\phantom{aaa}
\ampsqnpot(\proc)\JetsB\to\infty~~{\rm if}~~k_i^0\to 0~~{\rm or}~~
\veck_i\parallel \veck_j, 
\nonumber\\*&&
\phantom{aaa} \textrm{non-redundancy~conditions}\Big\}\,.\phantom{aa}
\label{PFKSdefopt}
\eeqn
Equation~(\ref{PFKSdefopt}) differs from eq.~(\ref{PFKSdef}) in the
upper limit for the range of $i$, which implies that FKS partons are 
now restricted to be massless (this, in turn, has the consequence
that the the condition $k_j\to 0$ need not be used any longer);
and because of the presence of non-redundancy conditions, which we specify 
as follows:
\beqn
&&\ident_i=g,\;\;\ident_j\ne g,\;\;(i,j)\in\FKSpairs
\;\;\;\Longrightarrow\;\;\;
(j,i)\notin\FKSpairs\quad{\rm if}\quad 3\le j\,,
\label{PFKScond2}
\\
&&\ident_i\ne g,\;\;\ident_j\ne g,\;\;(i,j)\in\FKSpairs
\;\;\;\Longrightarrow\;\;\;
(j,i)\notin\FKSpairs\quad{\rm if}\quad 3\le j<i\,.
\label{PFKScond3}
\eeqn 
These conditions fully remove the symmetry between $(i,j)$ and $(j,i)$
(for $i$ and $j$ both in the final state), except in 
the case in which both of these partons are gluons; for all other flavour 
combinations, either of these pairs does not belong to $\FKSpairs$
according to the new definition.

The non-redundancy conditions are easily understood. 
Equation~(\ref{PFKScond3}) implies that for those (unordered) pairs 
of particles which do not induce soft singularities (such as the $q\qb$ 
pair of the example above), of the two possible ordered pairs only one 
is included in $\FKSpairs$. On the other hand, since only the soft
singularities associated with FKS partons are subtracted (see
eq.~(\ref{dsigijnpo})), it is useless to give to particles that are not gluons 
the role of FKS partons if their sisters are gluons. Therefore, the only
case in which one needs a symmetric role for the FKS parton and its
sister is that in which both are gluons. It should be clear, however,
that the a-symmetrization carried out in 
eqs.~(\ref{PFKSdefopt})--(\ref{PFKScond3}) need be accompanied by an
analogous a-symmetrization in the definition of the $\Sfun$ functions.
This can be achieved by replacing eqs.~(\ref{hdef1}) and~(\ref{hdef2})
with the following definitions:
\beqn
h_{ij}(z)&=&1~~~~~~~~~~{\rm if}~~\ident_i\ne g~~{\rm or}~~j=1~~{\rm or}~~j=2\,,
\label{hdef1opt}
\\
h_{ij}(z)&=&h(z)~~~~~~{\rm if}~~\ident_i=g~~{\rm and}~~\ident_j=g\,.
\label{hdef2opt}
\eeqn
It is immediate to see that all formulae given in sects.~\ref{sec:nots},
\ref{sec:FKSxsecs}, and~\ref{sec:impl} still hold, provided that one 
understands eqs.~(\ref{PFKSdefopt})--(\ref{hdef2opt}) where relevant,
rather than eqs.~(\ref{PFKSdef}), (\ref{hdef1}), and~(\ref{hdef2}).
In the rest of this paper, and in the implementation of the FKS
subtraction in \MadFKS, we always use 
eqs.~(\ref{PFKSdefopt})--(\ref{hdef2opt}).

At this point, the sum on the r.h.s.~of eq.~(\ref{dsignpo}) still
scales quadratically with the number of light partons; however,
that sum obviously contains several identical contributions.
Suppose for example that one has $m$ final-state gluons 
\mbox{$i_1\ldots i_m$}. There will be \mbox{$m(m-1)$} gluon 
pairs in $\FKSpairs$, which will all contribute to $d\sigma^{(n+1)}$.
However, it is clear that \mbox{$d\sigma_{i_\alpha i_\beta}^{(n+1)}=
d\sigma_{i_\gamma i_\delta}^{(n+1)}$}, because of the symmetries of the 
matrix elements, $\Sfun$ functions, phase space, and (IR-safe) observables. 
Likewise, if $j$ is not a gluon, and is such that \mbox{$(i_1,j)\in\FKSpairs$},
then \mbox{$(i_\alpha,j)\in\FKSpairs$}, for $2\le\alpha\le m$, and
\mbox{$d\sigma_{i_1 j}^{(n+1)}=d\sigma_{i_\alpha j}^{(n+1)}$}.
In general, the determination of which elements of $\FKSpairs$
give identical contributions depends on the process considered,
and on the underlying theory (e.g., QED only or full EW interactions),
and can be implemented with relative ease in a computer code.
It is clear, however, that if several identical particles are present
in the final state, one and only one of them (which one in particular, 
is irrelevant) will be a member of the FKS pairs.
We shall denote by
\beq
\FKSpairsred\subseteq\FKSpairs\,,
\label{PFKSreddef}
\eeq
the subset of $\FKSpairs$ whose elements give non-identical contributions
to the sum in eq.~(\ref{dsignpo}). We shall then compute
\beq
d\sigma^{(n+1)}(\proc)=
\sum_{(i,j)\in\FKSpairsred}\symmnpoij(\proc)\, d\sigma_{ij}^{(n+1)}(\proc)\,,
\label{dsignpo2}
\eeq
with $\symmnpoij$ an integer symmetry factor, equal to the number of identical 
contributions (for a given $(i,j)$) to the sum in eq.~(\ref{dsignpo}).
The number of elements in $\FKSpairs(\proc)$ and $\FKSpairsred(\proc)$ will 
be denoted by $\FKSelem(\proc)$ and $\FKSelemred(\proc)$ respectively; 
the value of $\FKSelemred$ for several production processes will be 
reported in sect.~\ref{sec:res}.

\subsection{$n$-body matrix elements\label{sec:opt:BME}}
The integration of the $n$-body contribution, eq.~(\ref{nbody}),
is straightforward, and is usually carried out independently of
that of the $(n+1)$-body contribution. In this section, we give
the formulae that allow one to integrate these two contributions
at the same time. For instance, this technique is the default in MC@NLO.
We start by observing that
\beq
{\rm if}~~~\proc_\alpha\in\allprocn\;\;\Longrightarrow\;\;
\exists\proc_\beta\in\allprocnpo~~{\rm such~that}~~
\proc_\alpha=\proc_\beta^{\isubrmv}~~{\rm with}~~\ident_i=g\,.
\eeq
This states the fact that, given a set of particles that
corresponds to a physical $2\to n$ process, by adding one gluon in
the final state one gets a physically meaningful $2\to n+1$ process.
It therefore follows that
\beq
\sum_{\proc_\alpha\in\allprocn}d\sigma^{(n)}(\proc_\alpha)\equiv
\sum_{\proc_\beta\in\allprocnpo}d\sigma^{(n)}(\proc_\beta^{\isubrmv})\;\;\;\;
{\rm for~a~given}~~i~~{\rm with}~~\ident_i=g\,.
\label{sumnnpo}
\eeq
Note that, for an arbitrary $2\to n+1$ process $\proc_\beta$,
the corresponding $2\to n$ process $\proc_\beta^{\isubrmv}$ does not
necessarily exist (e.g.~if $\ident_i=q$). In such a case, we have 
\mbox{$d\sigma^{(n)}(\proc_\beta^{\isubrmv})=0$}. On the other hand,
if one fixes $i$, if $\proc_\beta^{\isubrmv}$ exists then it is
unique, and hence eq.~(\ref{sumnnpo}) holds. 

Let us now introduce the shorthand notation
\beq
d\sigma^{(n)}=\BornME\,\phspn\,.
\eeq
The precise form of $\BornME$ can be read from eqs.~(\ref{dsignB}),
(\ref{dsignC}), (\ref{dsignS}), and~(\ref{dsignV}), but is not
relevant in what follows. Fixing $i$ and assuming $\proc^{\isubrmv}$ 
is a physical $2\to n$ process, we have, using eq.~(\ref{tphisoft}),
\beq
d\sigma^{(n)}=\BornME\,\frac{16\pi^2}{\xicut s}\,
\stepf(\xicut-\xii)d\xii d\yij d\phii\tphspnij(\xii=0)\,.
\eeq
Using eq.~(\ref{SfunC2}) we can also write
\beq
d\sigma^{(n)}=\mathop{\sum_{j}}_{(i,j)\in\FKSpairs}
\BornME\,\Sfunij(\xii=0)\,\frac{16\pi^2}{\xicut s}\,
\stepf(\xicut-\xii)d\xii d\yij d\phii\tphspnij(\xii=0)\,.
\label{sumnnpo2}
\eeq
This equation holds for any given $i$ with $\ident_i=g$. It is convenient 
to give a prescription for fixing $i$ given the process. In order to do
that, we observe that there is only one gluon among the first elements
of the pairs that belong to $\FKSpairsred$, and this gluon is therefore
a natural candidate for fixing $i$. We rewrite eq.~(\ref{sumnnpo2}) 
as follows:
\beqn
d\sigma^{(n)}(\proc)&=&
\sum_{(i,j)\in\FKSpairsred}
\delta_{g\ident_i}d\sigma_{ij}^{(n)}(\proc^{\isubrmv})\,,\;\;\;\;\;\;
\proc\in\allprocnpo\,,
\label{sumnnpo3}
\\
d\sigma_{ij}^{(n)}&=&
\BornME\,\Sfunij(\xii=0)\,\frac{16\pi^2}{\xicut s}\,
\stepf(\xicut-\xii)d\xii d\yij d\phii\tphspnij(\xii=0)\,,
\label{dsign2}
\eeqn
where on the l.h.s.~of eq.~(\ref{sumnnpo3}) the use of a $2\to n+1$
process as an argument for $d\sigma^{(n)}$ is justified by the fact
that there is a one-to-one correspondence between $\proc$ and
$\proc^{\isubrmv}$ if one considers $\FKSpairsred$.

After these manipulations, the physical cross section of 
eq.~(\ref{factTH}) is written as follows:
\beqn
&&d\sigma_{\sss P_1P_2}(K_1,K_2)=
\nonumber \\*&&\phantom{aa}
\sum_{\proc\in\allprocnpo}\int dx_1 dx_2
f_{\Ione}^{(P_1)}(x_1) f_{\Itwo}^{(P_2)}(x_2)
\nonumber \\*&&\phantom{aaaa}\times
\Big(d\sigma^{(n+1)}(\proc;x_1K_1,x_2K_2)+
d\bar{\sigma}^{(n+1)}(\proc;x_1K_1,x_2K_2)+
d\sigma^{(n)}(\proc;x_1K_1,x_2K_2)\Big),
\nonumber \\*&&
\label{factTH2}
\eeqn
with $d\bar{\sigma}^{(n+1)}$ given in eq.~(\ref{collrem}), and
\beq
d\sigma^{(n+1)}(\proc)+d\sigma^{(n)}(\proc)=
\sum_{(i,j)\in\FKSpairsred}\left(
\symmnpoij(\proc)\, d\sigma_{ij}^{(n+1)}(\proc)+
\delta_{g\ident_i}d\sigma_{ij}^{(n)}(\proc^{\isubrmv})\right).
\label{dsignpopn}
\eeq
By inspection of eqs.~(\ref{dsign2}) and~(\ref{dsigijnpoE}), we
see that the second term on the r.h.s.~of eq.~(\ref{dsignpopn})
has the very same structure as the soft counterevent of the
first term. When cast in this form, it thus becomes obvious
that the $(n+1)$- and $n$-body contributions to an NLO
cross section can be integrated together.

It is also easy to show that the degenerate $(n+1)$-body contribution
$d\bar{\sigma}^{(n+1)}$ can be integrated together with the other two
contributions. This is the default strategy in MC@NLO. We refrain from
reporting the relevant derivation here, and postpone it to a forthcoming
paper, where the implementation of the FKS subtraction will be extended
to the case of hadronic collisions.

We conclude this section by observing that the logic behind 
eqs.~(\ref{factTH2}) and~(\ref{dsignpopn}) is that of getting the
underlying Born process given a real-emission process. 
This structure matches quite smoothly that of MadGraph,
which is the reason why we have chosen it in the first version
of \MadFKS. On the other hand, one may also consider the inverse
logic, namely that of getting the real-emission process(es) given
a Born process, which is especially convenient in the context
of the matching of NLO computations with parton showers.
We shall describe this option in app.~\ref{app:variants}.

\subsection{Multi-channel integration\label{sec:opt:mch}}
In the previous sections we have shown how to define a number $\FKSelemred$
of finite and independent short-distance contributions, whose sum is
the observable cross section. For large final-state multiplicities,
the integration of each of these contributions may prove difficult
from the numerical point of view.

Let us start by summarizing the strategy adopted in a tree-level
computation by MadGraph/MadEvent~\cite{Maltoni:2002qb}.
If the tree-level matrix elements are
those relevant to a $2\to n+1$ process, one can use eq.~(\ref{factTH}),
set $d\bar{\sigma}^{(n+1)}=0$ and $d\sigma^{(n)}=0$ there, and define
\beq
d\sigma^{(n+1)}(\proc)=\ampsqnpot(\proc)
\frac{\JetsR}{\avg(\proc)}\,\phspnpo\,,
\label{dsigtree}
\eeq
where $\JetsR$ denotes that one reconstructs exactly $\nlightB+1$ jets
in the final state, which is sufficient to cut off all phase-space
singularities of $\ampsqnpot$. Let us now denote by
\beq
d_\alpha(\proc),\;\;\;\;\;\alpha=1,\ldots\numofgr(\proc)
\eeq
the subset of Feynman diagrams that contribute to $\ampsqnpot$, and
that cannot be obtained from each other with a permutation of four-momenta.
In other words,
\beq
\ampnpot(\proc)=\sum_{\alpha=1}^{\numofgr(\proc)}\Big(d_\alpha(\proc)+
{\rm permutations~of~momenta}\Big)\,.
\label{Afromd}
\eeq
MadGraph then splits the integral of $d\sigma^{(n+1)}$ in eq.~(\ref{dsigtree})
into $\numofgr(\proc)$ independent contributions\footnote{We point out that
MadGraph does not include diagrams that contain four-point vertices in the
weights used for multichanneling. This fact is irrelevant for the sake of 
the present discussion.}, called {\em integration channels}, and 
defined as follows:
\beq
d\sigma^{(n+1)}(\proc)=\sum_{\alpha=1}^{\numofgr(\proc)}
d\sigma_\alpha^{(n+1)}(\proc)\,,
\label{sigsumal}
\eeq
where
\beqn
d\sigma_\alpha^{(n+1)}(\proc)&=&
\frac{\abs{d_\alpha(\proc)}^2}{D(\proc)}
\ampsqnpot(\proc)\frac{\JetsR}{\avg(\proc)}\,\phspnpo\,,
\label{dsigalpha}
\\
D(\proc)&=&\sum_{\beta=1}^{\numofgr(\proc)}\abs{d_\beta(\proc)}^2\,.
\label{Ddiagdef}
\eeqn
The formal similarity between what done here and what is done in the 
context of the FKS subtraction formalism suggests to replace
the damped $(n+1)$ matrix elements of eq.~(\ref{Mnpodamp}) with
\beqn
&&\ampsqnpot_{ij}(\proc)=(1-\yij)\xii^2\ampsqnpot(\proc)
\nonumber\\*&&\phantom{aaaaaaaaaa}\longrightarrow\;\;
\ampsqnpot_{ij,\alpha}(\proc)=(1-\yij)\xii^2\ampsqnpot(\proc)
\frac{\abs{d_\alpha(\proc)}^2}{D(\proc)}\,,
\eeqn
and apply the subtraction procedure to this quantity. Unfortunately,
this is not going to work. In the soft limit $E_i\to 0$, all diagrams
that would induce a divergence in the square of the amplitude in
eq.~(\ref{Afromd}) vanish if squared individually, as in 
eq.~(\ref{Ddiagdef}), owing to the presence of a self-eikonal of
a massless particle (an exception is therefore that of diagrams
in which a gluon is emitted by a final-state massive particle). 
It follows that the cross sections \mbox{$d\sigma_\alpha^{(n+1)}$} 
that are equal to zero are actually those that one would like to
have the largest contributions to the sum in eq.~(\ref{sigsumal}).

From the physics viewpoint, this is easy to understand. The squares 
of individual Feynman diagrams work well in tree-level computations
because these computations are basically classical physics. On the
other hand, soft singularities are inherently a quantum effect, since 
they are due to the interference of diagrams. Thus, multichannel sampling 
based on squares of diagrams cannot possibly describe all complications
due to interference. 

We can bypass this problem in the following way. We first introduce a
monotonically decreasing, smooth function $f(x)$ with the following
properties:
\beqn 
&&0\le f(x)\le 1~~~~{\rm for}~~0\le x\le 1\,,
\\
&&\lim_{x\to 0}f(x)=1\,.
\eeqn 
We then modify eq.~(\ref{Realident})
to read 
\beqn 
\ampsqnpot(\proc)&=&\sum_{(i,j)\in\FKSpairs}
f\left(\frac{2k_i\mydot k_j}{s}\right) \Sfunij(\proc)\ampsqnpot(\proc)
\nonumber\\*&+& \sum_{(i,j)\in\FKSpairs}
\left[1-f\left(\frac{2k_i\mydot k_j}{s}\right)\right]
\Sfunij(\proc)\ampsqnpot(\proc)\,.
\label{Realident2}
\eeqn
The factor in square brackets in the second term on the r.h.s.~of
eq.~(\ref{Realident2}) damps the only soft and collinear
singularities that survive the damping of $\Sfunij$. That second
term is therefore finite over the whole $(n+1)$-body phase space,
and is thus effectively a tree-level computation that can
be dealt with in the same way as in standard MadGraph.

The effect of the function $f$ in the first term on the r.h.s.~of
eq.~(\ref{Realident2}) is that of suppressing hard and
large-angle emissions of the $(i,j)$ system. We can therefore
assume that the typical kinematic configuration will always be
well approximated by the degenerate configuration obtained by
letting particles $i$ and $j$ become collinear, or when particle $i$
becomes soft. As we have seen in sect.~\ref{sec:impl:kin}, these
two limits result in the same four-momentum configurations, and we
can choose one or the other depending on which of the two will result 
in a physical $2\to n$ process as far as particle identities are concerned. 
In practice, if $\ident_i=g$ we 
shall use the soft limit, while if $\ident_i\ne g$ the collinear
limit will be used. To avoid unnecessarily complicating the notation,
we shall denote both cases by $\proc^{\isubrmv}$ in this discussion.
Starting from eq.~(\ref{Realident2}), we get
\beq
d\sigma^{(n+1)}(\proc)=
\sum_{(i,j)\in\FKSpairsred}\sum_{\alpha=1}^{\numofgr(\proc^{\isubrmv})}
\symmnpoij(\proc)d\sigma_{ij,\alpha}^{(n+1)}(\proc)+
\sum_{\beta=1}^{\numofgr(\proc)}
d\sigma_{0,\beta}^{(n+1)}(\proc)\,,
\label{dsigijnpo3}
\eeq
with
\beq
d\sigma_{0,\beta}^{(n+1)}(\proc)=
\frac{\abs{d_\beta(\proc)}^2}{D(\proc)}
\left\{\sum_{(i,j)\in\FKSpairs}
\left[1-f\left(\frac{2k_i\mydot k_j}{s}\right)\right]
\Sfunij(\proc)\right\}
\ampsqnpot(\proc)\frac{\JetsB}{\avg(\proc)}\,\phspnpo\,.
\label{dsignpozero}
\eeq
We stress again that, in spite of the presence of $\JetsB$ rather than
that of $\JetsR$, the cross section in eq.~(\ref{dsignpozero})
is finite, and needs no subtraction.
We have also defined
\beq
d\sigma_{ij,\alpha}^{(n+1)}(\proc)=
\frac{\abs{d_\alpha(\proc^{\isubrmv})}^2}{D(\proc^{\isubrmv})}
f\left(\frac{2k_i\mydot k_j}{s}\right)
d\sigma_{ij}^{(n+1)}(\proc)\,,
\label{dsigijalnpo}
\eeq
where $d\sigma_{ij}^{(n+1)}$ has been given in eq.~(\ref{dsigijnpo}),
and we applied in eq.~(\ref{dsigijnpo3})
the same symmetry factor $\symmnpoij$ as in
eq.~(\ref{dsignpo2}) for obvious reasons. As discussed above, the
kinematic configuration used to evaluate the prefactor
\mbox{$\abs{d_\alpha}^2/D$} in eq.~(\ref{dsigijalnpo}) is that of
either the soft or collinear counterevents, which coincide. Therefore,
this prefactor will have the same value in the case of the event and
of all the counterevents. On the other hand, \mbox{$f(2k_i\mydot k_j/s)$}
will in general be different when computed with event and counterevent
kinematics, but the smoothness of this function guarantees that the
cancellation of phase-space singularities will still take place.

Note that the formulae above can be applied without modifications
even if one sets $f(x)\equiv 1$. This means that $n$-body Feynman diagrams
are used for multichannel sampling in the whole $(n+1)$-body phase space.
This may imply a loss of efficiency when hard-emission regions
are integrated over. However, as we shall see in sect.~\ref{sec:res},
this is not the case for any of the processes we have considered in this paper.

From eq.~(\ref{dsigijnpo3}), we can read the total number of integration
channels of our implementation of an NLO cross section, for a given
process $\proc$:
\beq
\nchannels(\proc)=\numofgr(\proc)+
\sum_{(i,j)\in\FKSpairsred}\!\!\!\!\numofgr(\proc^{\isubrmv})\,.
\label{nchs}
\eeq
Note that the second term on the r.h.s.~of this equation is roughly
proportional to $\FKSelemred$ (it would be exactly so if all
$\numofgr(\proc^{\isubrmv})$ were equal). So while $\FKSelemred$ 
is directly related to the performance of the FKS formalism in
keeping the number of independent subtraction terms as small
as possible, the ratio \mbox{$\nchannels/\FKSelemred$} can be used
to measure the increase of the number of integration channels due
to the inherent complexity of the underlying Feynman diagram structure.
Note that, by setting $f(x)\equiv 1$, eq.~(\ref{nchs}) becomes
\beq
\nchannels(\proc)=
\sum_{(i,j)\in\FKSpairsred}\!\!\!\!\numofgr(\proc^{\isubrmv})\,.
\label{nchsfeq1}
\eeq

\section{Results\label{sec:res}}
In this section we present \MadFKS\ results for several production
processes in $\epem$ or $\mpmm$ collisions. These correspond to the full NLO
cross section, eqs.~(\ref{factTH}) or~(\ref{factTH2}), where we set
\beq
\vampsqnl_{\sss FIN}=0
\eeq
in eq.~(\ref{dsignV}). This is equivalent to assuming that the one-loop
contribution is a pure-pole term in the CDR scheme (see app.~\ref{app:virt}),
which implies that the results given in this section are non physical.
Indeed, our aim is not that of carrying out a phenomenological study,
but to document the convergence properties of the FKS subtraction in
\MadFKS\ (i.e., the statistical errors obtained with a given
number of integration points), and to make sure that our implementation
is correct.

We point out that the optimizations implemented so far are 
restricted to the FKS formalism; from the point of view of MadGraph,
the current version of \MadFKS\ must be considered as a benchmark
upon which we shall build a fully efficient program. In particular,
there are two main lines of development for the code. Firstly, the
results we present here have been obtained by running all integration
channels with the same number of points. This procedure will be
improved following the same strategy as is used in MadGraph, namely
that of performing preliminary low-statistics runs to determine which
channels give significant contributions, in order to further the runs
with larger statistics only for those. Secondly, in the current
version of \MadFKS\ all sums over helicities are performed exactly.
A much more efficient procedure is that of performing such a sum
using Monte Carlo techniques, as is typically done in the context
of tree-level computations. This is also possible at the NLO, provided
one is able to write the subtraction formalism for any given helicity
configuration. This is straightforward in the context of the FKS 
subtraction method, since indeed all necessary ingredients are already
available in the literature~\cite{Frixione:1995ms,deFlorian:1998qp};
we shall further comment on this point, and give all the relevant
details, in app.~\ref{app:hel}.
We have refrained from implementing these improvements (and several
other minor ones) in the first version of \MadFKS, since this is the
best way to gauge in a clear manner the capabilities of the code;
for example, by treating all channels on the same footing, one is
able to check that they are all computed correctly, regardless of
whether some are numerically relevant or not. 

As explained in sect.~\ref{sec:opt}, to obtain a cross section we 
find it convenient to sum over $(n+1)$-body processes, according to 
eq.~(\ref{factTH2}). In this section, therefore, when considering
a given partonic process, it will always be understood as an
$(n+1)$-body one. The corresponding $n$-body contributions will
be added according to the procedure described in sect.~\ref{sec:opt:BME}.

We have set up two different types of {\em numerical} tests to check
the correctness of the implementation. The first type of tests make 
sure that the local behaviour of our main ingredients in the computation
of the real-emission contribution, namely the matrix elements and
the $\Sfun$ functions, behave as we expect them to do in the collinear
and soft limits. For all real-emission processes $r$, and for all
pairs \mbox{$(i,j)\in\FKSpairsred$}, we check numerically that
eqs.~(\ref{MEsoftlim}), (\ref{MEcolllim1}), 
and (\ref{Sfununit})--(\ref{SfunC4})
are satisfied. This is done automatically, and if one of these tests
fails, the programme proceeds no further.

The second type of tests rely on the fact that several free parameters
are introduced in the FKS short-distance cross sections; the final results
must be independent of these parameters, but partial results
must not. To give one example, let us consider the case of $\xicut$,
introduced in sect.~\ref{sec:FKSxsecs}. This parameter enters analytically
the $n$-body cross sections (see e.g.~eq.~(\ref{Qdef}) and the results for
the integrated eikonals in app.~\ref{app:eik}), and numerically the
$(n+1)$-body cross section through the definition of the soft counterterm,
according to eq.~(\ref{distrxii}). Therefore, the quantities
$d\sigma^{(n+1)}$ and $d\sigma^{(n)}$ will
separately depend on $\xicut$, but their sum will not. The free parameters
whose variations we have considered here are $\xicut$, $\deltaO$, $\asfun$,
$\bsfun$. We have also investigated if the final results are independent
of whether one uses the factor \mbox{$2E_i/\sqs$} in the definition
of $d_{ij}$ (this option is denoted by \texttt{useenergy=.true.}), or
one uses $1$ instead (this option is denoted by \texttt{useenergy=.false.});
see sect.~\ref{sec:impl:Sfun} for more details. Each of these parameter
choices corresponds to a set of (parallel) runs. Therefore, to be 
definite, we have used the $(n+1)$-body process
\beq
e^+(1)\,e^-(2) \to Z \to u(3)\,\bar{u}(4)\,g(5)\,g(6)\,g(7)
\label{benchmark}
\eeq
as a benchmark; this is a contribution to the four-jet cross section
at the NLO. We have run at $\sqs=100\textrm{ GeV}$, and set
the factorization, renormalization and Ellis-Sexton
scales equal to the $Z$ mass, $\muF^2=\muR^2=Q^2=m_Z^2$.
This process is indeed simple enough to run in a short amount of time, but
has the required complexity to test the implementation in all its details. 
The condition in eq.~(\ref{jetcuts}), i.e.~the hard cuts, has been
imposed with the {\tt KTCLUS} routine that implements the jet-finding
algorithm of ref.~\cite{Catani:1993hr}, with 
\mbox{$Y_{\sss\rm cut}=(10~{\rm GeV})^2$}. The jet four-momenta are defined
as the sum of the four-momenta of the partons in the jets; the lowest jet
energy is therefore about 10~GeV.
\begin{table}
\resizebox{\textwidth}{!}{
\begin{tabular}{lccccc}
\toprule
$\deltaO$&$\asfun=\bsfun$&$\xicut=\ximax$&$\xicut=0.3$&
$\xicut=0.1$&$\xicut=0.01$\\
\midrule
&&\multicolumn{4}{c}{\texttt{useenergy=.true.}}\\
\multirow{3}{*}{$2$}
&$1.0$&$    3.5988\pm    0.0146$&$    3.6173\pm    0.0122$&$    3.6190\pm    0.0140$&$    3.6126\pm    0.0141$\\
&$1.5$&$    3.6085\pm    0.0126$&$    3.5942\pm    0.0143$&$    3.5956\pm    0.0115$&$    3.5989\pm    0.0133$\\\vspace{3pt}
&$2.0$&$    3.6127\pm    0.0121$&$    3.6122\pm    0.0158$&$    3.6020\pm    0.0147$&$    3.5956\pm    0.0144$\\
\multirow{3}{*}{$0.6$}
&$1.0$&$    3.6196\pm    0.0142$&$    3.6012\pm    0.0139$&$    3.5888\pm    0.0142$&$    3.5833\pm    0.0130$\\
&$1.5$&$    3.5941\pm    0.0123$&$    3.6012\pm    0.0139$&$    3.6009\pm    0.0138$&$    3.6047\pm    0.0114$\\\vspace{3pt}
&$2.0$&$    3.6066\pm    0.0120$&$    3.6111\pm    0.0117$&$    3.6053\pm    0.0110$&$    3.5950\pm    0.0150$\\
\multirow{3}{*}{$0.2$}
&$1.0$&{\red $    3.6350\pm    0.0151$} &$    3.5927\pm    0.0145$&$    3.5813\pm    0.0128$&$    3.5811\pm    0.0146$\\
&$1.5$&$    3.6020\pm    0.0119$&$    3.6086\pm    0.0133$&$    3.6104\pm    0.0127$&$    3.5993\pm    0.0119$\\\vspace{3pt}
&$2.0$&$    3.5815\pm    0.0140$&$    3.5966\pm    0.0136$&$    3.5938\pm    0.0121$&$    3.6079\pm    0.0125$\\
\multirow{3}{*}{$0.06$}
&$1.0$&$    3.6053\pm    0.0202$&$    3.5998\pm    0.0181$&$    3.5988\pm    0.0122$&$    3.6088\pm    0.0165$\\
&$1.5$&$    3.6144\pm    0.0161$&$    3.5986\pm    0.0140$&$    3.5847\pm    0.0119$&$    3.5884\pm    0.0126$\\
&$2.0$&$    3.5990\pm    0.0166$&$    3.6016\pm    0.0158$&$    3.6014\pm    0.0147$&$    3.6191\pm    0.0133$\\
\midrule
&&\multicolumn{4}{c}{\texttt{useenergy=.false.}}\\
\multirow{3}{*}{$2$}
&$1.0$&$    3.6078\pm    0.0164$&$    3.6149\pm    0.0162$&$    3.6145\pm    0.0158$&$    3.6085\pm    0.0140$\\
&$1.5$&$    3.5695\pm    0.0156$&$    3.5841\pm    0.0180$&$    3.5975\pm    0.0165$&$    3.5986\pm    0.0142$\\\vspace{3pt}
&$2.0$&$    3.5921\pm    0.0125$&$    3.6260\pm    0.0211$&$    3.6034\pm    0.0134$&$    3.6007\pm    0.0149$\\
\multirow{3}{*}{$0.6$}
&$1.0$&$    3.5891\pm    0.0199$&$    3.5786\pm    0.0164$&$    3.6084\pm    0.0232$&$    3.5956\pm    0.0151$\\
&$1.5$&$    3.6083\pm    0.0152$&$    3.5944\pm    0.0136$&$    3.6040\pm    0.0123$&$    3.6018\pm    0.0147$\\\vspace{3pt}
&$2.0$&$    3.5838\pm    0.0141$&{\red $    3.5633\pm    0.0154$}&$    3.5964\pm    0.0129$&$    3.5920\pm    0.0158$\\
\multirow{3}{*}{$0.2$}
&$1.0$&$    3.5976\pm    0.0171$&$    3.5790\pm    0.0166$&$    3.5702\pm    0.0155$&$    3.6155\pm    0.0132$\\
&$1.5$&$    3.5804\pm    0.0163$&$    3.5925\pm    0.0136$&$    3.6012\pm    0.0137$&$    3.6091\pm    0.0138$\\\vspace{3pt}
&$2.0$&$    3.5978\pm    0.0148$&$    3.5749\pm    0.0144$&$    3.5825\pm    0.0128$&$    3.5902\pm    0.0145$\\
\multirow{3}{*}{$0.06$}
&$1.0$&$    3.6122\pm    0.0170$&$    3.5942\pm    0.0158$&$    3.5743\pm    0.0146$&$    3.5962\pm    0.0167$\\
&$1.5$&$    3.6064\pm    0.0198$&$    3.5977\pm    0.0136$&$    3.6047\pm    0.0115$&$    3.5886\pm    0.0123$\\
&$2.0$&$    3.5971\pm    0.0169$&$    3.6018\pm    0.0136$&$    3.5991\pm    0.0148$&$    3.6040\pm    0.0148$\\
\bottomrule
\end{tabular}
}
\caption{Cross section (in pb) and Monte Carlo integration errors for the 
$(n+1)$-body process \mbox{$e^+e^-\to Z\to u\bar{u}ggg$}.
See the text for details.}
\label{tab:uuxggg}
\end{table}
The cross sections and integration errors (as returned by 
Vegas~\cite{Lepage:1977sw}) for the
process in eq.~(\ref{benchmark}) are given in table~\ref{tab:uuxggg}.
There are $\FKSelemred=3$ independent FKS pairs contributing to
this process (these can be chosen to be $(6,3)$, $(6,4)$, and $(6,7)$
according to the labels of eq.~(\ref{benchmark})), and five integration
channels for each of them.
We used 50000 integration points and 10 iterations per channel; only 
about 15\% of these points result in kinematic configurations that pass the 
jet-finding conditions; at the level of phase space, the points are 
generated flat, i.e.~no information on the hard cuts is available 
when one generates the kinematics. The results 
of table~\ref{tab:uuxggg} are clearly independent\footnote{Since in this
specific example we do not 
sum over flavours, we have to set $N_f=0$ in eqs.~(\ref{gammag}) 
and~(\ref{gmmprimeglu}); this corresponds to not allowing $g\to q\qb$
splittings at the Born level, which matches the fact that we do not
include any $(n+1)$-body matrix elements other than that in
eq.~(\ref{benchmark}).} of the free parameters, typically within 
one standard deviation. As an even tighter consistency check, 
we have picked the largest and the smallest values among those in
the table, and re-evaluated them with higher statistics. For a 
six--fold increase of the number of iterations, keeping the points 
per iteration to the same value as before, the corresponding
cross sections change as follow:
\beq
\begin{array}{ll}
3.5633\pm 0.0154\,\, &\to\,\,3.6086 \pm 0.0051\,,
\nonumber\\*
3.6350\pm 0.0151\,\, &\to\,\,3.6007 \pm 0.0053\,.
\end{array}
\label{numconv}
\eeq
It is interesting to observe that, apart from the convergence of the
mean values towards a common value, the errors in eq.~(\ref{numconv})
scale approximately as we would expect them to do if they would follow
the Gaussian law typical of integrals of ordinary functions (while 
a subtracted cross section is actually a distribution). This gives us
confidence on the fact that Vegas estimates correctly the integration errors.
We have also checked that the cross section is independent of the choice
of the function $h$ (see eqs.~(\ref{hdef2opt}) and~(\ref{hdef3})); 
since the numerical effects 
are even smaller than those reported in table~\ref{tab:uuxggg}, we refrain 
from presenting the corresponding results here.
We have performed the runs with
\beq
f(x)=\frac{(1-x)^{2a}}{c\,x^{2a}+(1-x)^{2a}}\,,\;\;\;\;\;\;
a=1,\;\;\;\;c=1\,.
\eeq
It is remarkable that, with this choice, the contribution of 
$d\sigma_{0,\beta}^{(n+1)}$
(see eq.~(\ref{dsignpozero})) is smaller than the statistical errors
reported in table~\ref{tab:uuxggg}, and is therefore not included
in the results of the table. 

As a final comment on the process in eq.~(\ref{benchmark}), we point out
a fairly pleasant feature that could have been anticipated. Namely,
the results presented in table~\ref{tab:uuxggg} have integration
errors that are by far and large independent of the choices of the free
parameters, thus including the case of ``maximal'' subtraction,
$\xicut=\ximax$ and $\deltaO=2$, when for each event one always
subtracts the three counterevents. This is quite an useful property,
since it implies that it will not be necessary to scan the
space of the free parameters in order to find those values that
maximize the numerical stability, all choices being almost equivalent.
At the core of this property stands the fact that, for any given
FKS pair (i.e., a finite and independent contribution), when scanning
the $(n+1)$-body phase space one finds at most one soft and one
collinear singularities which, as discussed in sect.~\ref{sec:impl:kin},
are associated with a single kinematic configuration. Therefore, the
behaviour of the subtracted cross section will not change much
by varying $\xicut$ and $\delta$. 
It is instructive to compare this with 
the case of dipole subtraction. There, by changing the parameter that
is typically called $\alpha$ (the analogue of our $\xicut$ and $\delta$),
one can change drastically the number of subtraction terms included
in the computation, and the tuning of $\alpha$ may become necessary.
The capability of the FKS subtraction method to give numerically
stable results also for large values of $\xicut$ and $\delta$ is
especially promising in view of the fact that such values are 
usually the optimal choices when matching NLO computations with
parton showers in MC@NLO or POWHEG.
\begin{center}
\begin{table}[h!]
\renewcommand{\arraystretch}{1.3}
\begin{center}
\begin{tabular}{lr@{ $\pm$ }lc}
\toprule
\multirow{2}{*}{$(n+1)$-body process}&
\multicolumn{2}{c}{\multirow{2}{*}{cross section (pb)\qquad}}&
\multirow{2}{*}{$\FKSelemred$}\\
\\
\midrule
$e^+e^-\to Z\to u\bar{u}gg$  &(0.4144&0.0006 (0.15\%))$\times  
10^{2}$&3\\
$e^+e^-\to Z\to u\bar{u}ggg$ &(0.3601&0.0014 (0.38\%))$\times  
10^{1}$&3\\
$e^+e^-\to Z\to u\bar{u}gggg$&(0.8869&0.0054 (0.61\%))$\times  
10^{-1}$&3\\
\midrule
$e^+e^-\to \gamma^* / Z\to jjjj$&(0.1801&0.0002 (0.12\%))$\times  
10^{3}$&14\\
$e^+e^-\to \gamma^* / Z\to jjjjj$&(0.1529&0.0004 (0.26\%))$\times  
10^{2}$&30\\
$e^+e^-\to \gamma^* / Z\to jjjjjj$&(0.3954&0.0015 (0.38\%))$\times  
10^0$&55\\
\midrule
$e^+e^-\to Z \to t\tb gg$&(0.1219&0.0003 (0.24\%))$\times  
10^{-1}$&3\\
$e^+e^-\to Z \to t\tb ggg$&(0.1521&0.0013 (0.83\%))$\times  
10^{-2}$&3\\
$e^+e^-\to Z \to t\tb gggg$&(0.1108&0.0031  (2.76\%))$\times  
10^{-3}$&3\\
\midrule
$e^+e^-\to Z \to t\tb b\bb g$&(0.1972&0.0024 (1.23\%))$ 
\times 10^{-4}$&4\\
$e^+e^-\to Z \to t\tb b\bb gg$&(0.2157&0.0029 (1.34\%))$ 
\times 10^{-4}$&5\\
\midrule
$e^+e^-\to Z \to \tilde{t}_1\tilde{t}_1^\star ggg 
$&(0.3712&0.0037 (1.00\%))$\times 10^{-8}$&3\\
$e^+e^-\to Z \to \tilde{g}\tilde{g}ggg 
$&(0.1584&0.0019 (1.23\%))$\times 10^{-1}$&2\\
\midrule
$\mu^+\mu^-\to H\to gggg
$&(0.1404&0.0005 (0.34\%))$\times 10^{-7}$&1\\
$\mu^+\mu^-\to H\to ggggg
$&(0.2575&0.0018 (0.69\%))$\times 10^{-8}$&1\\
$\mu^+\mu^-\to H\to gggggg
$&(0.1186&0.0008 (0.70\%))$\times 10^{-9}$&1\\
\bottomrule
\end{tabular}
\end{center}
\caption{Cross sections and Monte Carlo integration errors (in absolute
value and as a fraction relative to the cross section) for various processes.}
\label{tab:2}
\renewcommand{\arraystretch}{1.1}
\end{table}
\end{center}

The independence of the results of the parameters $\xicut$ and 
$\deltaO$ allows one to check the correctness of the subtraction terms
for the real-emission matrix elements, and of the logarithmic terms
in the $n$-body contributions. In order to also check the non-logarithmic
terms that enter the $n$-body contributions, we have computed the
physical NLO cross sections for the processes \mbox{$\epem\to Z\to 2$~jets}
and \mbox{$H\to 2$~jets}, where $H$ is a SM Higgs with \mbox{$m_H=120$~GeV},
using the readily available relevant matrix elements for the virtual 
corrections. We have found complete agreement with the known results.
We point out that, by doing so, we have tested all analytical
formulae presented in this paper, except for eqs.~(\ref{sigcp})
and~(\ref{sigcm}), which only enter hadronic cross sections,
and for the integrals of the non-massless eikonals, 
eqs.~(\ref{eik0M}), (\ref{eikself}), and~(\ref{eikMM}). The latter
integrals, however, have been checked by computing them numerically 
with very high precision.

We have considered a variety of other processes, in order to prove
the flexibility of the code, and to check the convergence of the
numerical integration; the processes and the corresponding results
are reported in table~\ref{tab:2}. 
We stress that, in order to obtain the results displayed in
that table, we just had to write the relevant process
in an input card for \MadFKS; this is the same procedure
as that done in MadGraph.
All cross sections appearing in table~\ref{tab:2}
have been computed by setting $\xicut=0.1$, $\deltaO=0.6$,
$\asfun=1.5$, $\bsfun=1.5$, and \mbox{\texttt{useenergy=.true.}}.
We have run 10 Vegas iterations; for each of them, phase-space points
have been generated in order to obtain about 10000 kinematic configurations
per iteration that pass the jet-finding cuts, which corresponds to the
statistics used to generate the process in eq.~(\ref{benchmark}).
For final states with only massless partons, we set $\sqs=100\textrm{ GeV}$;
we have used the notation $j$ to indicate the sum over partons
(we have considered four flavours here);
thus, e.g.~\mbox{$e^+e^-\to jjjjjj$} is, up to the missing finite
one-loop contribution, the physical five-jet cross section.
For top-quark production processes, we set $\sqs=500\textrm{ GeV}$.
The $b$ quark is taken to be massive ($m_b=4.7$~GeV) in processes with 
$t\tb b\bb+X$ final states, and it does not enter
the jet-finding algorithm. Finally, for stop and gluino production we have
used $m_{\tilde{t}_1}=400$~GeV, $m_{\tilde{g}}=400$~GeV, and 
$\sqs=1\textrm{ TeV}$; we assume an effective $Z\tilde{g}\tilde{g}$ 
axial vertex, and set the corresponding coupling equal to one.
We have also studied the production of fully gluonic final states
arising from the decay of an (off-shell) SM Higgs with \mbox{$m_H=120$~GeV},
that contribute to three-, four-, and five-jet observables.
In order to keep this process on the same footing as the others that feature
only massless final-state particles, we have set $\sqs=100\textrm{ GeV}$,
but used a $\mu^+\mu^-$ initial state in view of the smallness of
the electron Yukawa coupling. As one can see from table~\ref{tab:2},
fully-gluonic Higgs decays have a single non-trivial contribution
from FKS pairs.
In the cases of processes involving massive particles we also tested
the option of replacing $\beta_k$ and/or $\beta_l$ with 1 
in the expressions of the
relevant $d_{kl}$ (see sect.~\ref{sec:impl:Sfun}). We did not observe
any significant differences with respect to the results obtained with
the default choice.

\begin{figure}[t]
  \begin{center}
    \epsfig{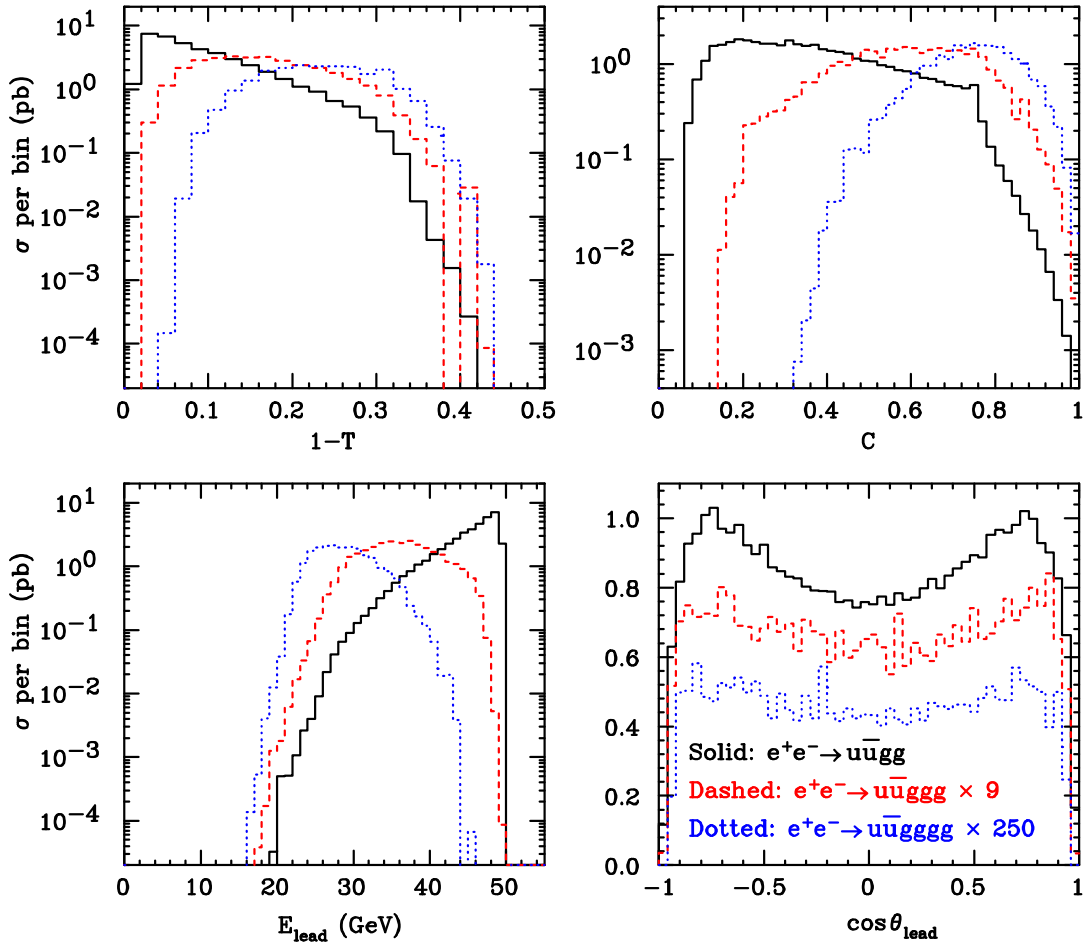}
\caption{\label{fig:compj} 
Differential spectra for the first three partonic processes
listed in table~\ref{tab:2}. The histograms for the latter 
two processes have been rescaled (by a factor of 9 and 250
respectively) in order to fit into the layout.
We present thrust, $C$ parameter, and the energy and polar angle of 
the leading jet.
}
  \end{center}
\end{figure}
As one can see from the table, the numerical errors are fairly 
small even with the limited statistics we used. The largest of them
(but still rather modest) are associated with the processes with a 
$t\tb$ pair in the final state, plus either four gluons or a 
$b\bb$ pair and extra gluons. It is clear that these processes
feature several mass scales quite different from each other -- the
c.m.~energy, the top mass, the $b$ mass, and the minimum jet
energy ($\sim 10$~GeV) -- and therefore one expects the coefficients
of the perturbative series to be plagued by several large logarithms,
that inherently increase the complexity of the numerical calculation.
We point out that, since we treat the
$b$ quark as massive, the $b\bb$ pair is not an FKS pair,
according to eq.~(\ref{PFKSdefopt}). On the other hand, given
the smallness of the $b$ mass, it may be beneficial to treat such a
pair as an FKS pair, which would also possibly imply the definition of a 
``quasi-collinear''  counterterm, whose analytical form has to 
tend to that in eq.~(\ref{MEcolllim1}) in the limit $m_b\to 0$. 
We leave the implementation of this option to a future work.

It is important to note that the growth of $\FKSelemred$ with
the final-state multiplicity is always rather modest. From
table~\ref{tab:2}, one can actually see that $\FKSelemred$ 
is a constant if the number of gluons is increased ($\FKSelemred$ 
for $t\tb b\bb gg$ is one unit larger than in the case of
$t\tb b\bb g$, since only in the former process one can form
an FKS pair with two gluons). The increase of $\FKSelemred$
in the \mbox{$\epem\to (n+1)j$} processes is mainly due to the
corresponding increase in partonic subprocesses: in the case
of four, five, and six particles in the final state of the
$(n+1)$-body processes, the number of contributing subprocesses 
is equal to 7, 7, and 17 respectively. Among individual contributions
to the five-jet cross section, those with the largest $\FKSelemred$
are the four-quark, different-flavour processes such as
\mbox{$\epem\to u\ub d\db gg$}, which have seven possible FKS pairs
(since one distinguishes between quarks and antiquarks when pairing
them with gluons, in order not to neglect possible charge asymmetries).

\begin{figure}[t]
  \begin{center}
    \epsfig{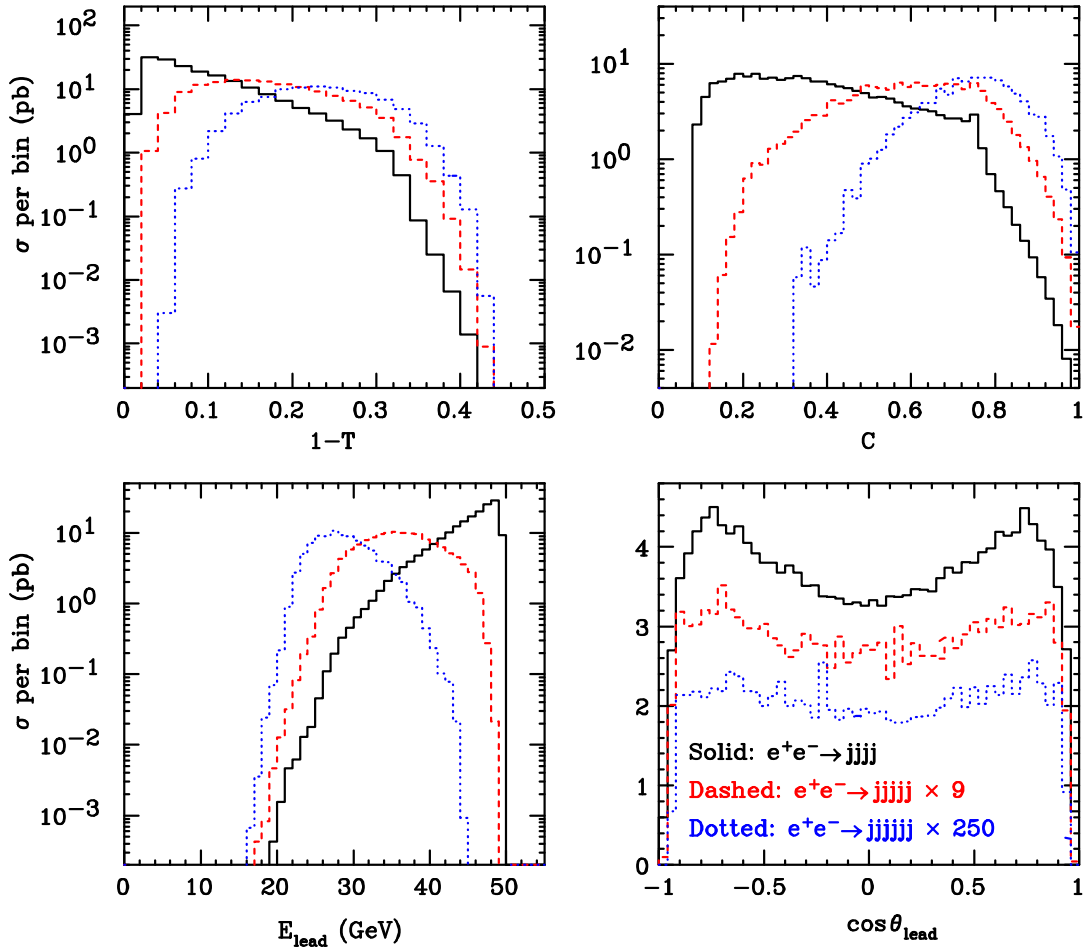}
\caption{\label{fig:lightj}
Same as in fig.~\ref{fig:compj}, for the three- (black solid), 
four- (red dashed), and five-jet (blue dotted) cross sections
(fourth to sixth processes in table~\ref{tab:2}).
}
  \end{center}
\end{figure}
As discussed in sect.~\ref{sec:opt:BME}, we integrate the $n$-body
matrix elements at the same time as the $(n+1)$-body ones. On an
event-by-event basis, we can therefore obtain both the NLO and the
LO contributions. We have checked that the latter is, for all processes,
fully consistent with the one predicted by standard MadGraph.
If one switches off in MadGraph the optimizations relevant to
the separate treatment of different integration channels, our LO
computation has the same statistical accuracy as that in standard MadGraph.
More importantly, if we only integrate the Born contributions to the processes
listed in table~\ref{tab:2} with the same number of points as that used for 
the NLO contributions (distributed equally among the possibly smaller number of
integration channels), the resulting integration uncertainties are 
a relative factor 1.9 to 4.5 smaller than those relevant to the NLO results 
presented here. Exceptions are found for the $(n+1)$-body processes 
$t\tb b\bb g$ and $t\tb gggg$, whose LO contributions have integration
accuracies better than the NLO ones by a factor 7 and 9 respectively.
Overall, these figures give us another indication of the fact that
the subtraction of singularities is achieved in very satisfactory
manner from the numerical point of view.

\begin{figure}[t]
  \begin{center}
    \epsfig{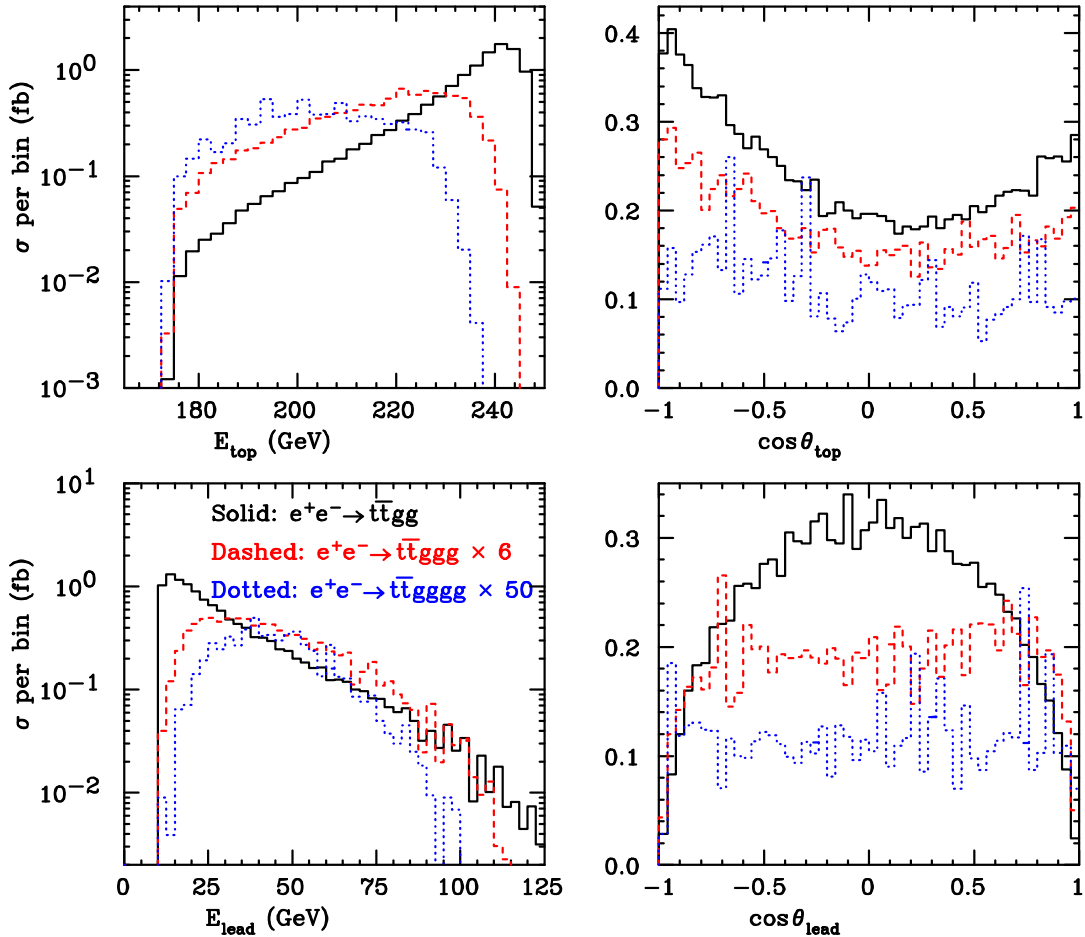}
\caption{\label{fig:comtt}
Differential spectra for the seventh to ninth partonic processes
listed in table~\ref{tab:2}. The histograms for the latter 
two processes have been rescaled (by a factor of 6 and 50
respectively) in order to fit into the layout. We present
the energy and polar angle of the top quark and of the leading
jet.
}
  \end{center}
\end{figure}
We now turn to presenting differential distributions for some of the processes
listed in table~\ref{tab:2}. As clarified in sect.~\ref{sec:impl} (see
in particular sect.~\ref{sec:impl:kin}), for a given process and
choice of \mbox{$(i,j)\in\FKSpairsred$}, to each random number
there correspond two kinematic configurations (for the event 
and for the counterevents plus $n$-body contributions), with 
associated weights. These configurations give the complete information
on all the four-momenta of the (resolved) final-state particles,
and thus one can plot as many (infrared-safe) observables as one
wants in the course of a run. The plots we present here are obtained
with the same statistics (in fact, in the same runs) as that used
for the total cross section results given in table~\ref{tab:2}.
We point out that we do not use any smoothing procedure for the
spectra we show: we limit ourselves to filling the histograms putting
the weights in the bins determined by the kinematic configurations
given by the code. This implies, in particular, that large weights
for the event and counterevents may fall into different bins.
We choose a relatively fine binning, in order to expose in a clear
manner whether this mis-binning (which is unavoidable in any
subtraction-based computation) is a severe problem or not.

In fig.~\ref{fig:compj} we show the distributions in the thrust,
$C$-parameter, and energy and polar angle of the hardest jet, for
the first three processes listed in table~\ref{tab:2}. In 
fig.~\ref{fig:lightj}, the same distributions are shown for 
three-, four-, and five-jet final states, and including the $Z/\gamma$
interference effects -- the plots in fig.~\ref{fig:lightj} would therefore 
correspond to physical jet cross sections, had we included the proper
one-loop contributions. Finally, in fig.~\ref{fig:comtt} we show the
energies and polar angles of the top quark and of the hardest jet
relevant to the \mbox{$\epem\to t\tb+(n+1)g$} processes.
All spectra are fairly smooth, except those relevant to the
polar angles, where fluctuations are marginal for light-jet-only
production processes, and more evident in the case of $t\tb$
processes. 

\section{Conclusions\label{sec:concl}}
In this paper, we have presented a complete automation of the
computation of any cross section at the next-to-leading order in
QCD, with the exception of an infrared- and ultraviolet-finite term of
one-loop origin. This is achieved by embedding into the MadGraph
framework the universal subtraction formalism of Frixione, Kunszt,
and Signer (FKS); the resulting computer code has been named \MadFKS.
The only intervention needed to run the code is the definition 
of the inputs, among which is the production process one wants to study.
The process is any physical process that results from a theory implemented 
in MadGraph, thus including the Standard Model and any user-defined Beyond the 
Standard Model theory. The current version of the implementation 
allows one to compute cross sections with colourless particles in
the initial state, but there are no difficulties
in extending it to other types of colliding particles; we defer this to 
future work. It is also worth mentioning that the formalism
presented here requires only trivial modifications to be extended
to the subtraction of QED infrared singularities. With minimal effort,
one could therefore fully automate the FKS subtraction 
for the complete electro-weak corrections; we point out, however,
that in order to actually perform such computations it will probably 
be necessary to implement a complex-mass scheme into MadGraph.

The key features of our implementation are a direct consequence of the
FKS subtraction formalism. Through a dynamic partition of the phase
space, one effectively defines partonic processes which are either
non singular, or have at most one soft and one collinear singularity, 
thus rendering the subtraction of these singularities a
fairly easy, and numerically efficient, task. These partonic 
processes are separately finite and fully independent of each 
other, and therefore the method is naturally suited to parallel 
computations. For each phase-space point, there are only two
independent kinematic configurations; one corresponds to a fully-resolved
configuration (all partons are hard and well separated from each
other), and the other one is the associated partly unresolved
configuration (where either one parton is soft, or two partons are
collinear, or both), whose weight is the sum of all subtraction
terms, of all finite remainders of the subtraction procedure, and
of the Born; this structure is exact, i.e.~does not imply any approximation.
Being able to limit the independent kinematic configurations to two
is quite beneficial for the numerical stability of the results,
and in particular for the smoothness of the predictions for
differential spectra. The latter point is easy to understand,
since in FKS one renders it small the probability that weights large in 
absolute value and opposite in sign will not fall in the same histogram 
bin (thus, one reduces the mis-binning problem that affects all 
computations based on a subtraction method).

The FKS subtraction method allows one to introduce several free parameters
in the calculation; partial results (e.g., the $(n+1)$-body and the
$n$-body contributions) depend on the values chosen for these parameters,
but the physical cross section does not. We have exploited this property
to check the correctness of our implementation. In addition, we have 
shown that the integration errors are largely independent of the choices
of the free parameters, which is a direct consequence of having
only two independent kinematic configurations for each phase space
point. This is an important fact, since it allows one to choose the
free parameters without having to worry about the numerical stability
of the results.

We have presented predictions for several production processes in $\epem$ 
or $\mpmm$ collisions. The numerical integrations converge quickly, 
requiring relatively modest statistics to achieve accurate results.
We have run in parallel, with each of the parallel 
jobs corresponding to one integration channel, and using the same 
statistics. This is certainly not the best way to perform a multi-channel
computation. The most obvious improvement, which is straightforward
to automate, is that of starting with a low statistics run for all channels, 
checking which contributions are largest in absolute value or/and
have the largest fractional errors, and further the run with larger
statistics only for those channels. We plan to release a public
web interface to \MadFKS\ as soon as the case of hadronic collisions
will also be included, and some of the optimizations discussed in
this paper will be implemented.

The current version of MadGraph, upon which \MadFKS\ is constructed,
is based on the evaluation of Feynman diagrams. We point out, however,
that \MadFKS\ can be rendered fully independent of Feynman diagram
information, thus implying the possibility of using recursively-generated
amplitudes. In fact, the only part in the present implementation which
is strictly dependent on Feynman graphs is that relevant to 
multi-channel sampling. A different criterion for sampling, based
e.g.~on topological information or on the computation of the amplitudes,
can be easily incorporated in \MadFKS\ (where multi-channel sampling
is the very last step of the computation).

\enlargethispage*{50pt}
If combined with one-loop computations, \MadFKS\ will allow one to predict
any NLO cross section, limited only by CPU time. The present bound
is actually set by the capability of MadGraph to handle the large
number of Feynman diagrams and colour configurations for very large
final-state multiplicities; tree-level amplitude computations based
on recursion relations will push this bound to larger multiplicities.
The computer programme can now use any external routine that returns
the finite one-loop contribution as defined in this paper, feeding it
with four-momenta and information on the tree-level structure.
It is clearly desirable to fully incorporate one-loop computations
into the programme, which appears to be feasible given the recent
progress made in the evaluation of virtual graphs. Either way, the
implementation presented here is the first step towards a full automation
of NLO computations matched with partons showers, as done in
MC@NLO or POWHEG.

\section*{Acknowledgments}
SF would like to thank Zoltan Kunszt and Werner Vogelsang for discussions 
during the course of this work, and for a long-time collaboration on 
related subjects; and Eric Laenen, Patrick Motylinski, and Bryan Webber
for work relevant to FKS subtraction with massive particles.
FM is grateful to CERN for hospitality during the course of this work.
RF and FM are partially funded by Technical and Cultural Affairs 
through the Interuniversity Attraction Pole P6/11.
This work was supported in part by the National Science Foundation 
under award PHY-0757889.

\appendix
\section{Eikonal integrals\label{app:eik}}
We define the integrals of the eikonal factors as follows:
\beq
\eikintD_{kl}^{(m_k,m_l)}+\eikint_{kl}^{(m_k,m_l)}=
-\frac{\xicut^{-2\ep}}{2\ep}
\frac{2^{2\ep}}{(2\pi)^{1-2\ep}}\left(\frac{s}{\mu^2}\right)^{-\ep}
\int d\Omega_i [k_k,k_l]_i\,,
\label{eikIdef}
\eeq
with
\beq
[k_k,k_l]_i=E_i^2
\frac{k_k\mydot k_l}{(k_k\mydot k_i)(k_l\mydot k_i)}\,.
\label{eikdef}
\eeq
In eq.~(\ref{eikIdef}), $\eikintD_{kl}$ collects by definition all
divergent terms (times an $\ep$-dependent pre-factor, to be given explicitly
below), while $\eikint_{kl}$ is finite. The normalization
of eq.~(\ref{eikIdef}) is conventional, and is equal to that used
in ref.~\cite{Frixione:1995ms}, times a factor $8\pi^2$ (this factor
has been compensated by a different normalization of the colour-linked
Born's $\ampsqnt_{kl}$ w.r.t.~that of ref.~\cite{Frixione:1995ms}). The 
energy of parton $i$ and the measure (in $3-2\ep$ dimensions) over
its angular variables $d\Omega_i$ are defined in the c.m.~frame of
the colliding partons (see eq.~(\ref{cmsframe})).
In ref.~\cite{Frixione:1995ms} only the massless case $m_k=0$ and
$m_l=0$ had been considered. The cases $m_k=0$, $m_l\ne 0$, and
$m_k\ne 0$ with $k=l$, have been dealt with in ref.~\cite{Frixione:2005vw},
but the analytical results for eq.~(\ref{eikIdef}) have not been published 
there. In this paper, we have computed the only missing ingredient,
i.e.~$m_k\ne 0$ and $m_l\ne 0$. We summarize here all the relevant results.

\vskip 0.3truecm
\noindent
$\bullet$~$m_k=0$, $l=k$ (massless self-eikonal).

\beqn
\eikintD_{kk}^{(0,0)}&=&0\,,
\label{eikD00self}
\\
\eikint_{kk}^{(0,0)}&=&0\,.
\label{eik00self}
\eeqn
These equations trivially follow from the mass-shell condition
$k_k^2=0$ in eq.~(\ref{eikdef}), and serve the sole purpose
of excluding the contributions of $\ampsqnt_{kk}$ with
$\nini\le k\le\nlightB+2$ from eq.~(\ref{dsignS}).

\newpage
\noindent
$\bullet$~$m_k=0$, $m_l=0$, $l\ne k$.

\beqn
\eikintD_{kl}^{(0,0)}&=&\frac{(4\pi)^\ep}{\Gamma(1-\ep)}
\left(\frac{\mu^2}{Q^2}\right)^\ep\left[
\frac{1}{\ep^2}-\frac{1}{\ep}
\left(\log\frac{2\kkdotkl}{Q^2}-
\log\frac{4E_kE_l}{\xicut^2 s}\right)\right]\,,
\label{eikD00}
\\
\eikint_{kl}^{(0,0)}&=& 
\half\log^2\frac{\xicut^2 s}{Q^2}+
\log\frac{\xicut^2 s}{Q^2}\log\frac{\kkdotkl}{2E_kE_l}
-{\rm Li}_2\left(\frac{\kkdotkl}{2E_kE_l}\right)
\nonumber\\*&+&
\half\log^2\frac{\kkdotkl}{2E_kE_l}
-\log\left(1-\frac{\kkdotkl}{2E_kE_l}\right)\log\frac{\kkdotkl}{2E_kE_l}\,.
\label{eik00}
\eeqn

\vskip 0.3truecm
\noindent
$\bullet$~$m_k=0$, $m_l\ne 0$.

\beqn
\eikintD_{kl}^{(0,m_l)}&=&\frac{(4\pi)^\ep}{\Gamma(1-\ep)}
\left(\frac{\mu^2}{Q^2}\right)^\ep\left[
\frac{1}{2\ep^2}-\frac{1}{\ep}
\left(\log\frac{2\kkdotkl}{Q^2}-
\half\log\frac{4m_l^2E_k^2}{\xicut^2 s Q^2}\right)\right]\,,
\label{eikD0M}
\\
\eikint_{kl}^{(0,m_l)}&=& 
\log\xicut\left(\log\frac{\xicut s}{Q^2}+2\log\frac{\kkdotkl}{m_lE_k}\right)
-\frac{\pi^2}{12}+\quarter\log^2\frac{s}{Q^2}
\nonumber\\*&-&
\quarter\log^2\frac{1+\beta_l}{1-\beta_l}
+\half\log^2\frac{\kkdotkl}{(1-\beta_l)E_k E_l}
+\log\frac{s}{Q^2}\log\frac{\kkdotkl}{m_lE_k}
\nonumber\\*&-&
{\rm Li}_2\left(1-\frac{(1+\beta_l)E_kE_l}{\kkdotkl}\right)
+{\rm Li}_2\left(1-\frac{\kkdotkl}{(1-\beta_l)E_kE_l}\right)\,.
\label{eik0M}
\eeqn
The definition of $\beta_l$ has been given in eq.~(\ref{betadef}).

\vskip 0.3truecm
\noindent
$\bullet$~$m_k\ne 0$, $l=k$ (massive self-eikonal).
\beqn
\eikintD_{kk}^{(m_k,m_k)}&=&\frac{(4\pi)^\ep}{\Gamma(1-\ep)}
\left(\frac{\mu^2}{Q^2}\right)^\ep\left(-\frac{1}{\ep}\right)\,,
\label{eikDself}
\\
\eikint_{kk}^{(m_k,m_k)}&=& 
\log\frac{\xicut^2 s}{Q^2}
-\frac{1}{\beta_k}\log\frac{1+\beta_k}{1-\beta_k}\,.
\label{eikself}
\eeqn

\vskip 0.3truecm
\noindent
$\bullet$~$m_k\ne 0$, $m_l\ne 0$, $l\ne k$.
\beqn
\eikintD_{kl}^{(m_k,m_l)}&=&\frac{(4\pi)^\ep}{\Gamma(1-\ep)}
\left(\frac{\mu^2}{Q^2}\right)^\ep\left(
-\frac{1}{2\ep}\frac{1}{\velkl}\log\frac{1+\velkl}{1-\velkl}\right)\,,
\label{eikDMM}
\\
\eikint_{kl}^{(m_k,m_l)}&=& 
\frac{1}{2\velkl}\log\frac{1+\velkl}{1-\velkl}
\log\frac{\xicut^2 s}{Q^2}
\nonumber\\*&+&
\frac{(1+\velkl)(\kkdotkl)^2}{2m_k^2}
\left({\rm J}^{(A)}\left(\alkl E_k,\alkl E_k\beta_k\right)
-{\rm J}^{(A)}\left(E_l,E_l\beta_l\right)\right)\,,
\label{eikMM}
\eeqn
where we have introduced the function
\beq
{\rm J}^{(A)}(x,y)=\frac{1}{2\lambda\nu}\left[
\log^2\frac{x-y}{x+y}+4{\rm Li}_2\left(1-\frac{x+y}{\nu}\right)
+4{\rm Li}_2\left(1-\frac{x-y}{\nu}\right)\right],
\eeq
and defined
\beqn
\velkl&=&\sqrt{1-\left(\frac{m_k m_l}{k_k\mydot k_l}\right)^2}\,,
\\
\alkl&=&\frac{1+\velkl}{m_k^2}\kkdotkl\,,
\\
\lambda&=&\alkl E_k-E_l\,,
\\
\nu&=&\frac{\alkl^2m_k^2-m_l^2}{2\lambda}\,.
\eeqn
It is straightforward to check that, if $k_l=k_k$, eqs.~(\ref{eikDMM}) 
and~(\ref{eikMM}) coincide with eqs.~(\ref{eikDself}) and~(\ref{eikself}) 
respectively.

\section{Finite one-loop contribution\label{app:virt}}
Assuming the validity of KLN theorem, the most general form of the 
divergent part of the ultraviolet-renormalized one-loop contribution 
$\ampsqnl$ defined in 
eq.~(\ref{Moneloop}) can be obtained by computing the divergent
contributions resulting from real-emission diagrams, and changing
the signs in front of the poles in $\ep$ obtained in this way.
These divergences may have collinear and/or soft origin. The 
results for the former have been computed in ref.~\cite{Frixione:1995ms}
(see sects.~4.3 and~4.4 there). The structure of the latter is 
more complicated, since it arises from the integrals of the eikonal factors,
which in turn depend on the masses of the two particles entering the
eikonal. A summary of the relevant results is given in sect.~\ref{app:eik};
here, we shall need the divergent parts, to be found in 
eqs.~(\ref{eikD00}), (\ref{eikD0M}), (\ref{eikDself}), and~(\ref{eikDMM}).
Using eqs.~(\ref{Mklident1}) and~(\ref{Mklident2}), we obtain:
\beq
\ampsqnl(\proc)=\asotwopi\frac{(4\pi)^\ep}{\Gamma(1-\ep)}
\left(\frac{\mu^2}{Q^2}\right)^\ep {\cal V}(\proc)\,,
\label{Virt1}
\eeq
with
\beqn
{\cal V}&=&-\Bigg(
\frac{1}{\ep^2}\sum_{k=\nini}^{\nlightB+2}C(\ident_k)
+\frac{1}{\ep}\sum_{k=\nini}^{\nlightB+2}\gamma(\ident_k)
+\frac{1}{\ep}\sum_{k=\nlightB+3}^{\nlightB+\nheavy+2}C(\ident_k)
\Bigg)\ampsqnt
\nonumber\\*&&
+\frac{1}{\ep}\sum_{k=\nini}^{\nlightB+2}
\sum_{l=k+1}^{\nlightB+\nheavy+2}\log\frac{2k_k\mydot k_l}{Q^2}
\ampsqnt_{kl}
\nonumber\\*&&
+\frac{1}{2\ep}\sum_{k=\nlightB+3}^{\nlightB+\nheavy+1}
\sum_{l=k+1}^{\nlightB+\nheavy+2}
\frac{1}{\velkl}\log\frac{1+\velkl}{1-\velkl}
\ampsqnt_{kl}
\nonumber\\*&&
-\frac{1}{2\ep}\sum_{k=\nlightB+3}^{\nlightB+\nheavy+2}
\log\frac{m_k^2}{Q^2}
\sum_{l=\nini}^{\nlightB+2}\ampsqnt_{kl}
+\vampsqnl_{\sss FIN}\,.
\label{Virt2}
\eeqn
Equation~(\ref{Virt2}) generalizes eq.~(3.2) of ref.~\cite{Frixione:1995ms}
to the case of arbitrary particle masses, and is consistent with the
results given in ref.~\cite{Catani:2000ef}. We stress again that
the poles in eq.~(\ref{Virt2}) have infrared origin, since the ultraviolet
divergences are assumed to have been eliminated through renormalization.

The Ellis-Sexton scale $Q$, originally introduced in ref.~\cite{Ellis:1985er}, 
is any scale that may be used in one-loop computations to express the arguments
of all the logarithms appearing there as \mbox{$s_{ij}/Q^2$} rather
than as \mbox{$s_{ij}/s_{kl}$}, where $s_{ij}$ and $s_{kl}$ are two
invariants constructed with the four-momenta of the particles that
enter the hard process (see e.g.~sect.~6 of ref.~\cite{Kunszt:1993sd}).
Clearly, it is not mandatory to introduce the Ellis-Sexton 
scale in the computations of the virtual amplitudes, and fairly often one
just uses the factorization or renormalization scale instead. On the
other hand, if $Q$ is kept distinct from both $\muF$ and $\muR$, one
sees that the NLO cross section is exactly independent of 
its choice; in other words, variations of $Q$ do not lead to any
estimate of missing higher orders in perturbation theory, but allow
one to check the internal consistency of the implementation.
Although indeed most of the available one-loop results do not use
the Ellis-Sexton scale, it is always possible to introduce it;
we show how this can be done in app.~\ref{app:scales}.

Once the virtual amplitude is computed, one can interfere it
with the Born amplitude and obtain $\ampsqnl$. Thus, eqs.~(\ref{Virt1}) 
and~(\ref{Virt2}) can be solved for the finite contribution 
$\vampsqnl_{\sss FIN}$ needed in eq.~(\ref{dsignV}). This operation
can be unambiguously carried out only after specifying a scheme.
Two typical situations may be considered: the Conventional Dimensional
Regularization (CDR) scheme, and the Dimensional Reduction (DR)
scheme. In the former, the quantities $\ampsqnt$ and $\ampsqnt_{kl}$ 
in eq.~(\ref{Virt2}) are evaluated in $4-2\ep$ dimensions, while in
the latter they are evaluated in 4 dimensions. The finite parts
$\vampsqnl_{\sss FIN}$ defined in the two schemes can be easily related
(see e.g.~ref.~\cite{Kunszt:1993sd}):
\beq
\vampsqnl_{\sss FIN}({\rm CDR})=\vampsqnl_{\sss FIN}({\rm DR})
-\ampsqnt\sum_{k=\nini}^{\nlightB+2}\tilde{\gamma}(\ident_k)\,,
\label{CDRtoDR}
\eeq
with
\beq
\tilde{\gamma}(q)=\half\CF\,,\;\;\;\;
\tilde{\gamma}(g)=\frac{1}{6}\CA\,.
\eeq
Note that the sum on the r.h.s.~of eq.~(\ref{CDRtoDR}) is extended
to light quarks and gluons only. Consistently with the fact that
the quantities in eq.~(\ref{CDRtoDR}) are ultraviolet-finite, we understand
that all scheme changes relevant to ultraviolet divergences have
already been performed; this implies that, from the ultraviolet point of view,
the term \mbox{$\vampsqnl_{\sss FIN}({\rm DR})$} is given in the
CDR scheme.

We stress again that $\vampsqnl_{\sss FIN}$ which is used 
in eq.~(\ref{dsignV}) must be computed in CDR.

\section{How to set $\muF\ne\muR$\label{app:scales}}
The formulae presented in this paper assume $\mu=\muF=\muR$, and
that another arbitrary mass scale, the Ellis-Sexton scale $Q$, is
introduced in the computation of the one-loop corrections.
In this appendix we explain how to relax the former condition,
and how to insert the dependence on the Ellis-Sexton scale
in an one-loop result that was computed without introducing it.
We perform these tasks by imposing renormalization group invariance
w.r.t. $\muF$ and $\muR$. We start by giving our convention for
the scales that appear in the various formulae of this paper:

\vskip 0.3truecm
\noindent
\begin{center}
\begin{minipage}{0.85\textwidth}
{\em If $\muF\ne\muR$, all the formulae for the short-distance
cross sections given in this paper must be computed with $\mu=\muF$, 
except for the argument of $\as$, which must be set equal to $\muR$.}
\end{minipage}
\end{center}

\vskip 0.3truecm
\noindent
The statement above implies in particular that the PDFs are computed
at $\mu=\muF$, as customary. The convention given above cannot possibly
allow us to recover terms proportional to logarithms of ratios of scales,
which we now proceed to determine. In order to do so, we rewrite 
eq.~(\ref{factTH}) or eq.~(\ref{factTH2}) symbolically as follows
\beq
d\sigma_{\sss P_1P_2}=f^{(P_1)}\star f^{(P_2)}\star\left(
d\sigma^{(n+1)}+d\bar{\sigma}^{(n+1)}+
d\sigma^{(n)}+C\log\frac{\muF^2}{\muR^2}\right),
\label{factTH3}
\eeq
with $C$ the unknown quantity that we need to determine.
In order to proceed, we have to specify the power of $\as$ that
enters our cross section formulae; we denote by $b$ an integer
such that
\beq
\ampsqnt={\cal O}\left(\as^b\right)\,,\;\;\;\;\;\;
\ampsqnpot={\cal O}\left(\as^{b+1}\right)\,.
\eeq
Note that in general we may have $b=0$ (e.g.~in the case of a purely-EW
process at the Born level). We can now impose the invariance of the
physical cross section w.r.t. to the renormalization scale
\beq
\frac{\partial}{\partial\log\muR^2}d\sigma_{\sss P_1P_2}=0\,,
\label{RGEmuR}
\eeq
where as usual this equation holds up to terms of 
${\cal O}\left(\as^{b+2}\right)$. Using the explicit forms of the
short-distance cross sections that appear on the r.h.s.~of 
eq.~(\ref{factTH3}), and the convention stated above, we find that
eq.~(\ref{RGEmuR}) is equivalent to
\beq
\frac{\partial}{\partial\log\muR^2}
\left(d\sigma^{(B,n)}+C\log\frac{\muF^2}{\muR^2}\right)=0\,.
\eeq
Therefore, since
\beq
\frac{\partial\as(\muR^2)}{\partial\log\muR^2}=-\beta_0\as^2(\muR^2)+
{\cal O}(\as^3)\,,\;\;\;\;\;\;
\beta_0=\frac{11\CA-4\TF N_f}{12\pi}\,,
\eeq
we obtain
\beq
C=-2\pi\beta_0\,b\left(\frac{\as(\muR^2)}{2\pi}\right)d\sigma^{(B,n)}\,,
\label{Csol}
\eeq
where $d\sigma^{(B,n)}$ is the Born cross section, given in
eq.~(\ref{dsignB}).

This would be the end of the story if all short-distance cross sections
could be computed according to the convention given above. This may not
be the case for the one-loop contribution. We can however always manage
to rewrite the one-loop contribution using our convention, regardless
of its original form. To do this, we impose the condition that the physical 
cross section be independent of the factorization scale:
\beq
\frac{\partial}{\partial\log\muF^2}d\sigma_{\sss P_1P_2}=0\,,
\label{RGEmuF}
\eeq
which is the analogue of eq.~(\ref{RGEmuR}). After some algebra (where
one also uses the evolution equations for the PDFs), eq.~(\ref{RGEmuF})
is found to be equivalent to the condition
\beq
\frac{\partial}{\partial\log\muF^2}d\sigma^{(V,n)}=-C\,,
\label{soldsigv}
\eeq
with $C$ given in eq.~(\ref{Csol}). In order to proceed, let us thus consider 
eq.~(\ref{dsignV}), and denote by $t_k$ all possible quantities
with mass-squared dimensions that can be constructed with four-momenta.
Following our convention, we must write
\beqn
\vampsqnlF(\muR^2,\muF^2,Q^2)&=&
\as^b(\muR^2)\,\hvampsqnlF(\muF^2,Q^2)\,,
\label{Veq1}
\\
\hvampsqnlF(\muF^2,Q^2)&=&a_V\left(\{t_k\}\right)\log\frac{\muF^2}{Q^2}+
b_V\left(Q^2,\{t_k\}\right)\,,
\label{Veq2}
\eeqn
where use has been made of the fact that, in an one-loop computation,
there will be an explicit linear dependence on the logarithm of the 
renormalization scale; according to our convention, in the argument of 
this logarithm $\muR$ must be replaced by $\muF$, hence eq.~(\ref{Veq2}).
On the other hand, the computation will in general contain terms
of the kind \mbox{$\log^p t_k/t_l$}, with $p=1,2$. The procedure of 
Ellis and Sexton implies that these must be replaced by
\mbox{$(\log t_k/Q^2-\log t_l/Q^2)^p$}, and $b_V$ in eq.~(\ref{Veq2})
may thus include up to double logarithms of $Q^2$; this is however
unimportant in what follows. Equation~(\ref{soldsigv}) can now be seen as 
a relation to solve for $a_V$; we obtain
\beq
\hvampsqnlF(\muF^2,Q^2)=2\pi\beta_0\,b
\left(\frac{\ampsqnt}{\as^b(\muR^2)}\right)\log\frac{\muF^2}{Q^2}+
\hvampsqnlF(Q^2,Q^2)\,.
\label{Veq3}
\eeq
Note that the r.h.s.~of this equation is indeed independent of $\muR$,
since $\ampsqnt$ is proportional to $\as^b(\muR^2)$. Equations~(\ref{Veq1})
and~(\ref{Veq3}) can be used to insert the renormalization, factorization,
and Ellis-Sexton scale dependences into a one-loop result originally
computed with a single scale $M$. In other words, if we are given
the quantity $v(M^2)$ such that
\beq
\vampsqnlF(M^2,M^2,M^2)=v(M^2)\,,
\eeq
then we can construct the finite contribution of one-loop origin
which complies with our convention as follows:
\beq
\vampsqnlF(\muR^2,\muF^2,Q^2)=
2\pi\beta_0\,b\,\ampsqnt\,\log\frac{\muF^2}{Q^2}+
\as^b(\muR^2)\frac{v(Q^2)}{\as^b(Q^2)}\,.
\label{VmuFneMuR}
\eeq
We stress that the quantity $v(Q^2)$ is proportional 
to $\as^b(Q^2)$, and therefore the whole r.h.s.~of eq.~(\ref{VmuFneMuR})
is proportional to $\as^b(\muR^2)$.

In summary, we shall compute the cross section with the most general
assignment of scales $\muF\ne\muR\ne Q$ using 
\beq
d\sigma_{\sss P_1P_2}=f^{(P_1)}\star f^{(P_2)}\star\left(
d\sigma^{(n+1)}+d\bar{\sigma}^{(n+1)}+
d\sigma^{(n)}-2\pi\beta_0\,b
\left(\frac{\as(\muR^2)}{2\pi}\right)d\sigma^{(B,n)}
\log\frac{\muF^2}{\muR^2}\right),
\label{factTH4}
\eeq
where the scales in the formulae for the short-distance cross sections
that appear on the r.h.s.~of this equation have to be chosen according
to the convention given at the beginning of this appendix. When following
this convention, the finite part of the one-loop contribution, 
eq.~(\ref{dsignV}), can be derived using eq.~(\ref{VmuFneMuR}) from
a result $v(M^2)$ obtained with a single scale $M$.

\section{Azimuthal terms in collinear limits\label{app:Qfun}}
When computing the collinear limit of the damped real matrix 
elements, we obtain eq.~(\ref{MEcolllim1}) (or the equivalent form
eq.~(\ref{MEcolllim2})). It is customary to write such a limit
by only keeping the first term on the r.h.s.~of eq.~(\ref{MEcolllim1})
but, as discussed in sect.~\ref{sec:impl:RME}, the second term is
equally important if one wants to construct local counterterms.
Appendix~B of ref.~\cite{Frixione:1995ms} explains in detail how to 
compute the reduced matrix element $\tampsqnt$, and the kernels 
$Q_{ab^\star}(z)$. These kernels are universal in the same sense as 
the Altarelli-Parisi ones. The notation suggests that already at the 
leading order (which is not the case for the Altarelli-Parisi kernels) 
their forms for timelike branchings are different from those for spacelike 
branchings.  From ref.~\cite{Frixione:1995ms} we obtain
\beqn
&&\tampsqnt\left(\proc;\Big\{k^{\sss (C)}\Big\}\right)=
\frac{1}{2s}
\frac{1}{\omega(\Ione)\omega(\Itwo)}
\label{Mtilda}\\*&&\phantom{aaaaaaaaaaaa}
\times
\Re\left\{\frac{\langle k_i^{\sss (C)}k_j^{\sss (C)}\rangle}
{[k_i^{\sss (C)}k_j^{\sss (C)}]}
\mathop{\sum_{\rm colour}}_{\rm spin}
\ampnt_{+}\left(\proc^{j\oplus i,\isubrmv},\Big\{\kbar\Big\}\right) 
{\ampnt_{-}\left(\proc^{j\oplus i,\isubrmv},\Big\{\kbar\Big\}\right)}^{\star}
\right\}.\phantom{aa}
\nonumber
\eeqn
The notation
\beq
\ampnt_{\pm}\left(\proc^{j\oplus i,\isubrmv}\right)
\label{ampntpm}
\eeq
that appears in eq.~(\ref{Mtilda}) understands that parton $j\oplus i$ 
has helicity equal to $\pm$. $\tampsqnt$ is therefore an interference
between the plus and minus helicity states of parton $j\oplus i$; the 
remaining helicities (i.e.~excluding that of parton $j\oplus i$) 
are summed over. 
The first term inside the curly brackets on the r.h.s.~of eq.~(\ref{Mtilda})
is the ratio of two spinor products. We follow here the notation
and conventions of Mangano and Parke, ref.~\cite{Mangano:1990by}. 
We note that this term is a pure phase, and is therefore numerically 
well defined also in the soft limit ($E_i\to 0$); we have worked out 
its form analytically for the various phase-space parametrizations 
we have considered.

For final-state (i.e.~timelike) branchings, the $Q$ kernels read as
follow~\cite{Frixione:1995ms}:
\beqn
Q_{gg^\star}(z)&=&-4\CA\,z(1-z)\,,
\label{Q1}
\\
Q_{qg^\star}(z)&=&4\TF\,z(1-z)\,,
\\
Q_{gq^\star}(z)&=&0\,,
\\
Q_{qq^\star}(z)&=&0\,.
\label{Q4}
\eeqn
The corresponding results for initial-state (i.e.~spacelike) branchings are:
\beqn
Q_{g^\star g}(z)&=&-4\CA\,\frac{1-z}{z}\,,
\label{Q1spc}
\\
Q_{q^\star g}(z)&=&0\,,
\\
Q_{g^\star q}(z)&=&-4\CF\,\frac{1-z}{z}\,,
\\
Q_{q^\star q}(z)&=&0\,.
\label{Q4spc}
\eeqn
The $^\star$ symbol allows one to remember easily which parton is 
off-shell. We note that for branchings which involve an off-shell 
quark, the azimuthal term
in the collinear limit vanishes identically. This is due to the
conservation of the helicity along a light-quark line; see
ref.~\cite{Frixione:1995ms} for details.

We point out that in eq.~(\ref{Mtilda}) one must use the same conventions
for the spinors in the computation of the amplitudes as in that of the 
pure-phase factor. As mentioned above, eq.~(\ref{Mtilda}) has been worked
out using the conventions of ref.~\cite{Mangano:1990by}. On the other
hand, in the actual numerical implementation in MadGraph the amplitudes
are computed following the conventions of HELAS~\cite{Murayama:1992gi}. 
Since eqs.~(\ref{Q1})--(\ref{Q4}) imply that we need to consider only the
cases in which parton $j\oplus i$ is a gluon, we can write
\beq
\ampnt_{\pm}=\ampnt_\mu\polv^\mu(\pm)\,.
\label{ampmu}
\eeq
Using the explicit representations of ref.~\cite{Mangano:1990by} and
of HELAS, it is a matter of algebra to arrive at
\beqn
\polv^\mu(+,k;{\rm MP})&=&
\exp\left(-i\,\varphi_k\right)
\polv^\mu(+,k;{\rm HELAS})+\alpha k\,,
\\
\polv^\mu(-,k;{\rm MP})&=&-
\exp\left(i\,\varphi_k\right)
\polv^\mu(-,k;{\rm HELAS})+\beta k\,,
\eeqn
where $k$ is the gluon (light-like) four-momentum, which in the coordinate 
system chosen by HELAS has azimuthal angle equal to $\varphi_k$, and $\alpha$,
$\beta$ are two constants, which are irrelevant in what follows, since
the corresponding contributions to eq.~(\ref{ampmu}) vanish because
of gauge invariance. Using these equations we obtain
\beq
\ampnt_{+}({\rm MP}){\ampnt_{-}({\rm MP})}^\star=
-\exp\left(-2i\,\varphi_{j\oplus i}\right)
\ampnt_{+}({\rm HELAS}){\ampnt_{-}({\rm HELAS})}^\star.
\eeq

We conclude this section by reporting the $4-2\ep$ dimensional forms
of the unpolarized Altarelli-Parisi kernels for $z<1$~\cite{Altarelli:1977zs}:
\beqn
P_{gg}(z,\ep)&=&2\CA\left(\frac{z}{1-z}+\frac{1-z}{z}+z(1-z)\right),
\label{APgg}
\\
P_{qg}(z,\ep)&=&\TF\left(z^2+(1-z)^2-2\ep\,z(1-z)\right),
\label{APqg}
\\
P_{qq}(z,\ep)&=&\CF\left(\frac{1+z^2}{1-z}-\ep\,(1-z)\right),
\label{APqq}
\\
P_{gq}(z,\ep)&=&\CF\left(\frac{1+(1-z)^2}{z}-\ep\,z\right).
\label{APgq}
\eeqn
According to eq.~(\ref{APdampdef}), the coefficients of the 
\mbox{${\cal O}(\ep^0)$} and \mbox{${\cal O}(\ep^1)$} terms in
these equations are the kernels $P_{ab}^{(0)}$ and $P_{ab}^{(1)}$
respectively, which are used in several equations of this paper.

\section{Possible variants in the implementation\label{app:variants}}
As we have discussed in the text, there are a large number of arbitrary
parameters which enter our implementation of the FKS formalism into MadGraph,
and any physical observable is independent of them. Each parameter choice
can be considered as a different way to implementing the subtraction
procedure in a computer code. In this appendix we report two variants 
of the implementation which are not parametric in nature, that we may
want to consider in the future, and which aim at further reducing the
number of independent contributions to the physical cross section.
We also present a technique alternative to that discussed
in sect.~\ref{sec:opt:BME} to integrate together the $n$- 
and $(n+1)$-body contributions, which is best suited for NLO
computations matched with parton showers.

We start by considering the contributions to the $(n+1)$-body 
cross sections due to $(i,1)$ and $(i,2)$ (for a given $i$, and
assuming that {\em both} of these two pairs belong to $\FKSpairs$).
Using the properties of the $\Sfun$ functions, it is a matter
of simple algebra to show that
\beqn
&&d\sigma_{i1}^{(n+1)}+d\sigma_{i2}^{(n+1)}=
\half\xic\left[\omyid+\opyid\right]
\nonumber\\*&&\phantom{aaaaaaaa}\times
\Big((1-\yi^2)\xii^2\ampsqnpot\Big)
\left(\Sfun_{i1}+\Sfun_{i2}\right)\frac{\JetsB}{\avg}\, 
d\xii d\yi d\phii\tphspnij\,.
\label{ISRFKS}
\eeqn
Here we have set
\beq
\yi=y_{i1}\;\;\;\;\Longrightarrow\;\;\;\;\yi=-y_{i2}\,,
\label{yidef}
\eeq
which follows from the fact that we work in the c.m.~frame of the
incoming partons, and thus we can use eq.~(\ref{cmsframe}). Eq.~(\ref{ISRFKS})
is the form used in refs.~\cite{Frixione:1995ms,Frixione:1997np},
and has the virtue of allowing one to subtract the two initial-state
collinear singularities in one single integration. On the other hand,
eq.~(\ref{dsigijnpo}) is closer to what MadGraph does when it integrates
tree-level matrix elements. We postpone the comparison between these
two possibilities to a forthcoming publication. We point out that
in refs.~\cite{Frixione:1995ms,Frixione:1997np} the notation
was used \mbox{$\Sfun_i^{(0)}=\Sfun_{i1}+\Sfun_{i2}$}, and
\mbox{$\Sfunij^{(1)}\equiv\Sfunij$} denoted the $\Sfun$ functions
relevant to both $i$ and $j$ in the final state.

We now turn to considering the contributions to the $(n+1)$-body 
cross sections due to all the pairs formed by a given FKS parton $i$ 
and its {\em massive} sisters. Since $m_j\ne 0$ for such sisters, 
the matrix elements are not 
singular in the collinear limits. This implies that the distribution in
$\yij$ defined in eq.~(\ref{distryij}) is just a regular function
in eq.~(\ref{dsigijnpo}), and therefore there is a cancellation
of the factors \mbox{$(1-\yij)$} between the numerator and the denominator.
We thus immediately obtain
\beqn
\mathop{\sum_{j=\nlightR+3}}_{(i,j)\in\FKSpairs}^{\nlightR+\nheavy+2}
d\sigma_{ij}^{(n+1)}&=&\xic\Big(\xii^2\ampsqnpot\Big)
\Sfun_i^{\sss (M)}\frac{\JetsB}{\avg}\, 
d\xii d\yi d\phii\tphspn^i\,,
\label{MassFKS}
\\
\Sfun_i^{\sss (M)}&=&
\mathop{\sum_{j=\nlightR+3}}_{(i,j)\in\FKSpairs}^{\nlightR+\nheavy+2}
\Sfunij\,.
\label{SfunM}
\eeqn
On the r.h.s.~of eq.~(\ref{MassFKS}) we have changed variables
$\yij\to\yi$ for all $j$. The variable $\yi$ can be equal to $\yij$
for a given $j$, or can be defined as in eq.~(\ref{yidef}). The
specific choice of $\yi$ should not matter much in terms of numerical
performance, given that the collinear regions $\yij=1$ are not
associated with any peaks in the matrix elements (this may not be
the case for particles with small masses, such as $b$ quarks).
Note also that multi-channel integration can be performed in
eq.~(\ref{MassFKS}) following exactly the procedure that leads
to eq.~(\ref{dsigijalnpo}).

In the case of processes with many massive strongly-interacting particles,
eq.~(\ref{MassFKS}) may be a more convenient alternative to
eq.~(\ref{dsigijnpo}), since it will reduce the number of independent
integrations. We did not explore this possibility in the present paper,
and we plan to do so in the future.

We finally reconsider the problem of the simultaneous integration 
of the $(n+1)$- and $n$-body contributions. In sect.~\ref{sec:opt:BME} we
have shown how to split the $n$-body contribution into a sum of terms,
each of which is then associated with a real-emission contribution
due to a given partonic process and a given FKS pairs. We did so by
unambiguously (but arbitrarily) ``constructing'' the Born-level
processes by removing from the real processes the only gluon 
(in the cases in which there is a gluon) that is an FKS parton
in $\FKSpairsred$. This procedure is clearly inspired by the connection
between an $(n+1)$-body process and its soft limits.

One can turn this logic around, and construct real-emission
processes starting from a given Born process. In doing so, it is
apparent that this construction can be achieved by considering
$a\to bc$ branchings, with $a$ a particle entering the Born process.
These branchings are reminiscent of an underlying collinear
kinematics, but this holds only at a formal level. For example,
if one has a top quark in the Born process, by considering the
branching $t\to tg$ ones gets a new final state with an extra gluon, 
which can be interpreted as a real-emission process, in spite of the
fact that there is no collinear singularity associated with 
the $tg$ pair.

At this point, one observes that for a given real-emission process,
pairs of particles that lead to an underlying Born structure are
in fact FKS pairs. For a given $\procB\in\allprocn$ and a given
$\procR\in\allprocnpo$, after constructing \mbox{$\FKSpairsred(\procR)$}
we introduce a generalized Kronecker symbol
\beq
\delta\left(\procB,\procR^{j\oplus i,\isubrmv}\right)=\left\{
\begin{array}{ll}
1 &\phantom{aaaa} {\rm if}~~\procB=\procR^{j\oplus i,\isubrmv}\,,\\
0 &\phantom{aaaa} {\rm if}~~\procB\ne\procR^{j\oplus i,\isubrmv}\,.
\end{array}
\right.
\eeq
We remind the reader that two processes are considered identical if
one of them can be obtained from the other with a permutation
of the identities of final-state particles. It is then a matter of 
algebra to show that
\beqn
\sum_{\procR\in\allprocnpo}d\sigma^{(n+1)}(\procR)&=&
\sum_{\procR\in\allprocnpo}\sum_{(i,j)\in\FKSpairsred(\procR)}
\symmnpoij(\procR)\, d\sigma_{ij}^{(n+1)}(\procR)
\nonumber\\*&=&
\sum_{\procB\in\allprocn}
\sum_{\procR\in\allprocnpo}\sum_{(i,j)\in\FKSpairsred(\procR)}
\delta\left(\procB,\procR^{j\oplus i,\isubrmv}\right)
\symmnpoij(\procR)\, 
d\sigma_{ij}^{(n+1)}(\procR)\,.
\nonumber\\*&&
\eeqn
This equation suggests a definition analogue to that given 
in eq.~(\ref{sumnnpo3}). There, the $n$-body cross section 
$d\sigma^{(n)}$ was given an $(n+1)$-body process as an argument.
In this case, we give the $(n+1)$-body cross section $d\sigma^{(n+1)}$
an $n$-body process as an argument:
\beq
d\sigma^{(n+1)}(\procB)=
\sum_{\procR\in\allprocnpo}\sum_{(i,j)\in\FKSpairsred(\procR)}
\delta\left(\procB,\procR^{j\oplus i,\isubrmv}\right)
\symmnpoij(\procR)\, 
d\sigma_{ij}^{(n+1)}(\procR)\,.
\label{dsigijnpoB}
\eeq
It is quite obvious that the same procedure can be applied to the
degenerate $(n+1)$-body contributions of eqs.~(\ref{sigcp})
and~(\ref{sigcm}), also in view of the fact that they have a structure
identical to that of the initial-state collinear limits of the
$(n+1)$-body cross sections. We thus define:
\beqn
d\bar{\sigma}_{1}^{(n+1)}(\procB)&=&
\sum_{\procR\in\allprocnpo}\sum_{(i,1)\in\FKSpairsred(\procR)}
\delta\left(\procB,\procR^{1\oplus\bar{i},\isubrmv}\right)
\,d\bar{\sigma}_{i1}^{(n+1)}(\procR)\,,
\label{sigcpB}
\\
d\bar{\sigma}_{2}^{(n+1)}(\procB)&=&
\sum_{\procR\in\allprocnpo}\sum_{(i,2)\in\FKSpairsred(\procR)}
\delta\left(\procB,\procR^{2\oplus\bar{i},\isubrmv}\right)
\,d\bar{\sigma}_{i2}^{(n+1)}(\procR)\,.
\label{sigcmB}
\\
d\bar{\sigma}^{(n+1)}(\procB)&=&
d\bar{\sigma}_{1}^{(n+1)}(\procB)+
d\bar{\sigma}_{2}^{(n+1)}(\procB)\,.
\label{collremB}
\eeqn
Equations~(\ref{dsigijnpoB})--(\ref{collremB}) allow us to use 
eq.~(\ref{factTH2}) with the formal replacement:
\beq
\sum_{\proc\in\allprocnpo}\;\longrightarrow\;\sum_{\proc\in\allprocn}\,.
\eeq.

Equation~(\ref{dsigijnpoB}) (or, which is equivalent for the sake of
this discussions, eqs.~(\ref{sigcpB}) and~(\ref{sigcmB})) may appear
more complicated than its analogue, eq.~(\ref{sumnnpo3}). However, the
reader should note that, owing to the presence of the Kronecker symbol,
the sum over the $(n+1)$-body processes in eq.~(\ref{dsigijnpoB})
is almost trivial, and it amounts to simply counting the 
$a\to bc$ branchings allowed by particle identities. Given that
one will have in any case to construct the sets of FKS pairs in
order to be able to perform the computations of real-emission
contributions, eq.~(\ref{dsigijnpoB}) is simple to implement
algorithmically -- in particular, we stress that the current
version of \MadFKS\ does have all necessary ingredients to
implement it.
 
We conclude this session by pointing out a couple of issues.
Firstly, the use of eqs.~(\ref{dsigijnpoB})--(\ref{collremB}) 
in place of those of sect.~\ref{sec:opt:BME} implies that
the $n$-body contributions are computed a smaller number of times.
This may lead to a significant difference in CPU time when including 
the finite term of one-loop origin $\vampsqnlF$ into \MadFKS, whose
computation for large-multiplicity final states is quite laborious. 
Obviously, the procedure of sect.~\ref{sec:opt:BME} can be improved in this
respect (for example, by not calling the function that returns
$\vampsqnlF$ every time the Born is computed), but the technique
described in this section is perhaps a simpler alternative.
Secondly, eqs.~(\ref{dsigijnpoB})--(\ref{collremB}) are required
for the matching with parton showers according to the POWHEG
formalism. Although not strictly necessary, the same technique
can also be used in the context of the generation of $\clS$
events (which are Born-like) in MC@NLO~\cite{Frixione:2002ik}.

\section{Integration of matrix elements of given helicities\label{app:hel}}
For large-multiplicity final states, the sum over all helicities
is a time-consuming operation. A successful strategy at the tree level is 
that of replacing such an exact sum with one performed with Monte Carlo
methods, which in turn requires the matrix elements be available
for any possible fixed-helicity configuration. This is indeed the
case in MadGraph, and the aim of this appendix is that of showing
that the FKS formalism presented in this paper needs only trivial
modifications to handle this situation.

We recall that we compute the physical cross section using 
eq.~(\ref{factTH2}), and let us again consider for the moment the case of
$\epem$ collisions only, which implies $d\bar{\sigma}^{(n+1)}\equiv 0$.
The \mbox{$(n+1)$}-body contribution is given in eqs.~(\ref{dsignpo})
and~(\ref{dsigijnpo}),
while the $n$-body contribution is the sum (see eq.~(\ref{nbody}))
of the quantities given in eqs.~(\ref{dsignB}), (\ref{dsignC}),
(\ref{dsignS}), and (\ref{dsignV}). In these equations, the 
matrix elements appear that are defined in 
eqs.~(\ref{Mtreenpo})--(\ref{Moneloop}) (where in the latter one
only the finite contribution is actually used, as specified
in eq.~(\ref{Virt2})). 

We now state the following fact: the equations we have referred to
in the previous paragraph apply {\em without any modifications} to the
case of a partonic process with fixed helicities, except for the
fact that in eqs.~(\ref{Mtreenpo})--(\ref{Moneloop}) 
one has to make the formal replacement
\beq
\mathop{\sum_{\rm colour}}_{\rm spin}\;\;\longrightarrow
\sum_{\rm colour}\,.
\label{nohelsum}
\eeq
In other words, the sum over helicities is now {\em not} understood
in the matrix element.

The $(n+1)$-body contribution has the subtracted structure given in
eq.~(\ref{dsigijnpoE}). In eqs.~(\ref{MEsoftlim}) and~(\ref{MEcolllim1})
we have given the necessary ingredients to construct the soft and
collinear counterterms respectively; these formulae however apply to 
the case in which the matrix elements are summed over helicities.
When one considers the soft limit of the \mbox{$(n+1)$}-body matrix 
element for a given helicity configuration, eq.~(\ref{MEsoftlim})
is basically unchanged. The colour-linked Born matrix elements 
which appear on the r.h.s.~of that equation are computed with the same 
helicities as those entering the $(n+1)$-body matrix element on the l.h.s.,
except for that of parton $i$, which is simply not present in the
reduced process, as suggested by the notation $\proc^{\isubrmv}$.
The eikonal pre-factors are also unchanged, but need be multiplied
by a factor \mbox{$1/2$} in the case in which the helicity of parton
$i$ is kept fixed; if, on the other hand, such helicity is summed
over, no modifications are needed in eq.~(\ref{MEsoftlim}).

The case of the collinear limit is slightly more complicated. Indeed,
in the reduced process $\proc^{j\oplus i,\isubrmv}$ the helicity of the
branching
parton $j\oplus i$ can in general assume any value, at fixed helicities
$h_i$ and $h_j$ of partons $i$ and $j$ respectively.
Equation~(\ref{MEcolllim1}) gets modified as follows
\beqn
\lim_{\yij\to 1}\ampsqnpot_{ij}\left(\proc;\Big\{k\Big\}\right)&=&
\gs^2\frac{(1-\yij)\xii^2}{k_i\mydot k_j}
\sum_h
\polP_{\ident_{j\oplus i}\to\ident_j\ident_i}^{hh_jh_i}(z_{ji})
\ampsqnt_h\left(\proc^{j\oplus i,\isubrmv};\Big\{\kbar\Big\}\right)
\nonumber\\*&+&
\gs^2\frac{(1-\yij)\xii^2}{k_i\mydot k_j}
\polQ_{\ident_{j\oplus i}^\star\to\ident_j\ident_i}^{h_jh_i}(z_{ji})
\tampsqnt\left(\proc;\Big\{k^{\sss (C)}\Big\}\right).\phantom{aaaaa}
\label{MEcolllimhel}
\eeqn
In the first term on the r.h.s.~of this equation, $h$ is the helicity
of parton $j\oplus i$ and, consistently with eq.~(\ref{ampntpm}), 
we have denoted
\beq
\ampsqnt_{\pm}\left(\proc^{j\oplus i,\isubrmv}\right)=
\frac{1}{2s}\frac{1}{\omega(\Ione)\omega(\Itwo)}
\sum_{\rm colour}\abs{\ampnt_{\pm}\left(\proc^{j\oplus i,\isubrmv}\right)}^2.
\label{nbodyhel}
\eeq
The quantity $\tampsqnt$ has the same definition as in eq.~(\ref{Mtilda}),
except for the sum over helicities, which is not performed here in accordance
to the general rule of this section, given in eq.~(\ref{nohelsum}).
We point out that in the second term on the r.h.s.~of eq.~(\ref{MEcolllimhel})
the sum over the helicity $h$ of parton $j\oplus i$ is not performed, since
that term results from the interference between the $h\!=\!+$ and $h\!=\!-$
states. This is also the reason why the kernels $\polQ$ do not depend on $h$.
We finally stress that all helicities other than those of partons $i$ and $j$
have the same values in the \mbox{$(n+1)$}- and $n$-body processes
that enter the matrix elements on the two sides of
eq.~(\ref{MEcolllimhel}) respectively. 

The kernels $\polP$ and $\polQ$ introduced in eq.~(\ref{MEcolllimhel}) are 
the fully polarized versions of the (four-dimensional) Altarelli-Parisi and 
$Q$ kernels respectively. 
They can be computed using the results reported in app.~B of 
ref.~\cite{Frixione:1995ms}. In the branching $a\to bc$ we use the
following momentum fraction and helicity assignments
\beq
a(1;h_a)\;\longrightarrow\;b(z;h_b)\,c(1-z;h_c)\,,
\eeq
and we obtain what follows (in four dimensions).

\noindent
$\bullet$~$g\longrightarrow gg$

\beqn
\polP_{g\to gg}^{+++}(z)&=&\CA\frac{1}{z(1-z)}\,,
\label{polPfirst}
\\
\polP_{g\to gg}^{++-}(z)&=&\CA\frac{z^3}{1-z}\,,
\\
\polP_{g\to gg}^{+-+}(z)&=&\CA\frac{(1-z)^3}{z}\,,
\\
\polP_{g\to gg}^{+--}(z)&=&0\,,
\\
\polQ_{g^\star\to gg}^{h_bh_c}(z)&=&-2\CA z(1-z)\delta_{h_b\bar{h}_c}\,,
\eeqn
where $\bar{h}_c=-h_c$.

\noindent
$\bullet$~$g\longrightarrow q\qb$

\beqn
\polP_{g\to q\qb}^{+++}(z)&=&0\,,
\\
\polP_{g\to q\qb}^{++-}(z)&=&\TF z^2\,,
\\
\polP_{g\to q\qb}^{+-+}(z)&=&\TF (1-z)^2\,,
\\
\polP_{g\to q\qb}^{+--}(z)&=&0\,,
\\
\polQ_{g^\star\to q\qb}^{h_bh_c}(z)&=&2\TF z(1-z)\delta_{h_b\bar{h}_c}\,.
\eeqn

\noindent
$\bullet$~$q\longrightarrow qg$

\beqn
\polP_{q\to qg}^{+++}(z)&=&\CF\frac{1}{1-z}\,,
\\
\polP_{q\to qg}^{++-}(z)&=&\CF\frac{z^2}{1-z}\,,
\\
\polP_{q\to qg}^{+-+}(z)&=&0\,,
\\
\polP_{q\to qg}^{+--}(z)&=&0\,,
\\
\polQ_{q^\star\to qg}^{h_bh_c}(z)&=&0\,.
\eeqn

\noindent
$\bullet$~$q\longrightarrow gq$

\beqn
\polP_{q\to gq}^{h_ah_bh_c}(z)&=&\polP_{q\to qg}^{h_ah_ch_b}(1-z)\,,
\label{polPlast}
\\
\polQ_{q^\star\to gq}^{h_bh_c}(z)&=&0\,.
\eeqn

\noindent
The $\polP$ kernels relevant to the case of the branching parton with
negative helicity are
\beq
\polP_{a\to bc}^{-h_bh_c}(z)=\polP_{a\to bc}^{+\bar{h}_b\bar{h}_c}(z)
\eeq
for all types of branchings; we have again used the notation
\mbox{$\bar{h}_b=-h_b$} and \mbox{$\bar{h}_c=-h_c$}.
The Altarelli-Parisi kernels at fixed helicities in four dimensions
can be easily obtained from the $\polP$ kernels:
\beq
P_{ab}^{(0)h_ah_b}(z)=\sum_{h_c}\polP_{b\to ac}^{h_bh_ah_c}(z)\,.
\label{APhel}
\eeq
Furthermore, we have
\beqn
\sum_{h_bh_c}\polP_{a\to bc}^{h_ah_bh_c}(z)&=&P_{ba}^{(0)}(z)\,,\;\;\;\;\;\;
{\rm for}~~h_a=\pm\,,
\\
\sum_{h_bh_c}\polQ_{a^\star\to bc}^{h_bh_c}(z)&=&Q_{ba^\star}(z)\,.
\eeqn
It is then immediate to see that eqs.~(\ref{polPfirst})--(\ref{polPlast})
lead to eqs.~(\ref{APgg})--(\ref{APgq}) (with $\ep=0$ there),
to eqs.~(\ref{Q1})--(\ref{Q4}), and that one 
recovers eq.~(\ref{MEcolllim1}) starting from eq.~(\ref{MEcolllimhel}).
In order to see this, one has to make  use of the identities that
relate Altarelli-Parisi kernels of like and unlike helicities
\beqn
&&P_{ab}^{++}(z,\ep)=P_{ab}^{--}(z,\ep)\equiv
P_{ab}^{\uparrow\uparrow}(z,\ep)\,,
\label{Pupup0}
\\
&&P_{ab}^{-+}(z,\ep)=P_{ab}^{+-}(z,\ep)\equiv
P_{ab}^{\downarrow\uparrow}(z,\ep)\,,
\label{Pupdown0}
\eeqn
which, as the notation suggests, hold in general in $4-2\ep$ dimensions. 

Although not relevant to the applications presented in this paper, 
we conclude this section by discussing the generalizations of 
eqs.~(\ref{sigcp}) and~(\ref{sigcm}) (the degenerate $(n+1)$-body
contributions) to the case of processes with given helicities.
These generalizations can be obtained from ref.~\cite{deFlorian:1998qp}, 
where the FKS subtraction method was extended to the case of polarized
collisions. We have
\beqn
&&d\bar{\sigma}_{i1}^{(n+1)}(\proc;k_1,k_2)=
\asotwopi\sum_h\Bigg\{
\APdamp_{\ident_{1\oplus\bar{i}}\ident_1}^{(0)hh_1}(1-\xii)
\left[\xic\log\frac{s\deltaI}{2\mu^2}+2\lxic\right]
\nonumber \\*&&
\phantom{d\bar{\sigma}_{i1}^{(n+1)}(\proc;k_1,k_2)=\asotwopi}
-\APdamp_{\ident_{1\oplus\bar{i}}\ident_1}^{(1)hh_1}(1-\xii)\xic
-K_{\ident_{1\oplus\bar{i}}\ident_1}^{hh_1}(1-\xii)\Bigg\}
\nonumber \\*&&\phantom{aaaaaa}\times
\ampsqnt_h\left(\proc^{1\oplus\bar{i},\isubrmv};(1-\xii)k_1,k_2\right)
\frac{\JetsB}{\avg(\proc)}\phspn\Big((1-\xii)k_1,k_2\Big)d\xii\,,
\label{sigcphel}
\\
&&d\bar{\sigma}_{i2}^{(n+1)}(\proc;k_1,k_2)=
\asotwopi\sum_h\Bigg\{
\APdamp_{\ident_{2\oplus\bar{i}}\ident_2}^{(0)hh_2}(1-\xii)
\left[\xic\log\frac{s\deltaI}{2\mu^2}+2\lxic\right]
\nonumber \\*&&
\phantom{d\bar{\sigma}_{i2}^{(n+1)}(\proc;k_1,k_2)=\asotwopi}
-\APdamp_{\ident_{2\oplus\bar{i}}\ident_2}^{(1)hh_2}(1-\xii)\xic
-K_{\ident_{2\oplus\bar{i}}\ident_2}^{hh_2}(1-\xii)\Bigg\}
\nonumber \\*&&\phantom{aaaaaa}\times
\ampsqnt_h\left(\proc^{2\oplus\bar{i},\isubrmv};k_1,(1-\xii)k_2\right)
\frac{\JetsB}{\avg(\proc)}\phspn\Big(k_1,(1-\xii)k_2\Big)d\xii\,.
\label{sigcmhel}
\eeqn
The reduced matrix elements that appear in eqs.~(\ref{sigcphel})
and~(\ref{sigcmhel}) are defined according to eq.~(\ref{nbodyhel});
as the notation suggests, $h$ is the helicity of parton 
\mbox{$1\oplus\bar{i}$} (or \mbox{$2\oplus\bar{i}$}).
We are not aware of any results for the kernels $K$ for polarized
scattering, which are non zero in PDF schemes different from 
$\overline{\rm MS}$; therefore, when considering fixed helicities 
we shall limit ourselves to using $\overline{\rm MS}$ PDFs.

The damped Altarelli-Parisi kernels at fixed helicities, 
$\APdamp_{ab}^{(k)h_ah_b}$, with $k=0,1$, have the same definition 
as in eq.~(\ref{APdampdef}):
\beq
\APdamp_{ab}^{(k)h_ah_b}=(1-z)P_{ab}^{(k)h_ah_b}(z)\,,
\eeq
with
\beqn
P_{ab}^{(0)hh}(z)+\ep P_{ab}^{(1)hh}(z)+{\cal O}(\ep^2)&=&
P_{ab}^{\uparrow\uparrow}(z,\ep)\,,
\label{Pabhh}
\\
P_{ab}^{(0)h\bar{h}}(z)+\ep P_{ab}^{(1)h\bar{h}}(z)+{\cal O}(\ep^2)&=&
P_{ab}^{\downarrow\uparrow}(z,\ep)\,,
\label{Pabhhbar}
\eeqn
where the quantities on the r.h.s.~of these equations have been
introduced in eqs.~(\ref{Pupup0}) and~(\ref{Pupdown0}), and 
we again denoted $\bar{h}=-h$. The \mbox{${\cal O}(\ep^0)$} term 
$\APdamp_{ab}^{(0)h_ah_b}$ can be obtained from eq.~(\ref{APhel}).
That equation, however, cannot be used to obtain $\APdamp_{ab}^{(1)h_ah_b}$,
since the kernels $\polP$ have been worked out only in four dimensions.
To proceed, we make use of the standard definitions:
\beqn
P_{ab}^{\uparrow\uparrow}(z,\ep)&=&\half\left(P_{ab}(z,\ep)+
\Delta P_{ab}(z,\ep)\right)\,,
\label{Pupup}
\\
P_{ab}^{\downarrow\uparrow}(z,\ep)&=&\half\left(P_{ab}(z,\ep)-
\Delta P_{ab}(z,\ep)\right)\,.
\label{Pdownup}
\eeqn
The quantities $\Delta P_{ab}$ 
are conveniently introduced in the study of collisions with 
incoming polarized beams. From refs.~\cite{Altarelli:1977zs,Vogelsang:1995vh}
we obtain
\beqn
\Delta P_{gg}(z,\ep)&=&2\CA\left(\frac{1}{1-z}-2z+1+2\ep\,(1-z)\right),
\label{DeltaPgg}
\\
\Delta P_{qg}(z,\ep)&=&\TF\left(2z-1-2\ep\,(1-z)\right),
\label{DeltaPqg}
\\
\Delta P_{qq}(z,\ep)&=&\CF\left(\frac{1+z^2}{1-z}-\ep\,(1-z)\right),
\label{DeltaPqq}
\\
\Delta P_{gq}(z,\ep)&=&\CF\left(2-z+2\ep\,(1-z)\right),
\label{DeltaPgq}
\eeqn
after taking into account a finite scheme transformation which affects
the ${\cal O}(\ep)$ term of the $qq$ kernel~\cite{Vogelsang:1995vh}.

\end{document}